\documentclass[10pt,journal,compsoc]{IEEEtran}
\ifCLASSOPTIONcompsoc
  \usepackage[nocompress]{cite}
\else
  \usepackage{cite}
\fi
\ifCLASSINFOpdf
  \usepackage[pdftex]{graphicx}
\else
\fi
\usepackage{hyperref}
\usepackage{comment}
\usepackage{multirow}
\usepackage{wrapfig}
\usepackage{color}
\usepackage{rotating}
\usepackage[table]{xcolor}
\usepackage{tikz}
\usepackage[normalem]{ulem}
\usepackage{microtype}
\usepackage{setspace}
\usepackage{todonotes}
 \usepackage{xspace}

\definecolor{red}{rgb}{0.6, 0, 0}
\definecolor{blue}{rgb}{0, 0, 0.6}
\definecolor{green}{rgb}{0.16, 0.435, 0.16}

\usepackage{subcaption}
\newcommand{\para}[1]{\textbf{#1.}\xspace}

\begin{document}
%
\title{A Survey of Perception-Based Visualization Studies by Task}
%
%
%
%

\author{Ghulam Jilani Quadri and~Paul Rosen
\IEEEcompsocitemizethanks{\IEEEcompsocthanksitem G.J.\ Quadri was with the Department
of Computer Science and Engineering, University of South Florida, Tampa,
FL, 33647.\protect\\
E-mail: ghulamjilani@usf.edu
\IEEEcompsocthanksitem P.\ Rosen was with the Department
of Computer Science and Engineering, University of South Florida, Tampa,
FL, 33647.\protect\\
E-mail: prosen@usf.edu}
\thanks{Manuscript received April 19, 2005; revised August 26, 2015.}}

%
%

\markboth{Journal of \LaTeX\ Class Files,~Vol.~14, No.~8, August~2015}%
{Shell \MakeLowercase{\textit{et al.}}: Bare Demo of IEEEtran.cls for Computer Society Journals}
%




\IEEEtitleabstractindextext{%
\begin{abstract}
    Knowledge of human perception has long been incorporated into visualizations to enhance their quality and effectiveness. The last decade, in particular, has shown an increase in perception-based visualization research studies. With all of this recent progress, the visualization community lacks a comprehensive guide to contextualize their results. In this report, we provide a systematic and comprehensive review of research studies on perception related to visualization. This survey reviews perception-focused visualization studies since 1980 and summarizes their research developments focusing on low-level tasks, further breaking techniques down by visual encoding and visualization type. In particular, we focus on how perception is used to evaluate the effectiveness of visualizations, to help readers understand and apply the principles of perception of their visualization designs through a task-optimized approach. We concluded our report with a summary of the weaknesses and open research questions in the area.
\end{abstract}

\begin{IEEEkeywords}
Visualization, perception,  graphical perception, visual analytics tasks, evaluation, survey.
\end{IEEEkeywords}}

\maketitle

\IEEEdisplaynontitleabstractindextext

%
\IEEEpeerreviewmaketitle

\setstretch{1}

\IEEEraisesectionheading{\section{Introduction}\label{sec:introduction}}

    \IEEEPARstart{V}{isualization} provides valuable assistance in data analysis and decision-making tasks. The human perceptual and cognitive systems are essential in the process of visualization, influencing visual analysis activities, e.g., data exploration, data gathering, and data manipulation. As an example, data exploration requires forming high-level analysis goals, planning actions, and evaluating results effectively, all of which are cognitive activities. Before higher-level cognitive processes analyze data, visualization passes through the human perceptual system impacting the visualization's utility~\cite{tory2004human}. The design of the visualization should make it as easy and unambiguous as possible to understand the data. Ultimately, a better understanding of human perception aids visualization design in both a quantitative and qualitative manner~\cite{ware2012information}.

\begin{figure}[!b]
        \centering
        
        \includegraphics[width=0.875\linewidth]{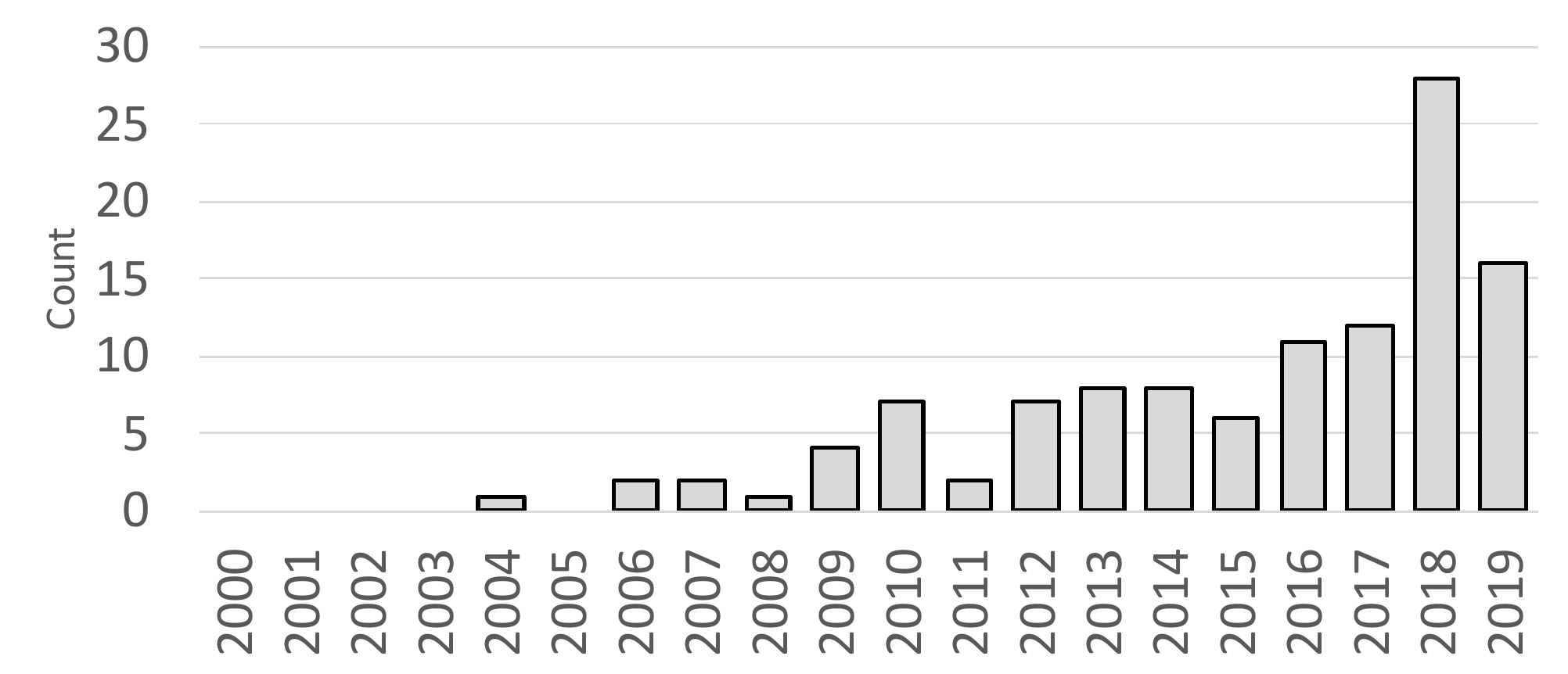}\hspace{5pt}
        
        \caption{Publications per year in our survey considering perception and visualization since 2000.}
        \label{fig:paperCountByyear}
    \end{figure}

    Numerous fields of science have studied perception, including perceptual psychology, visualization, and human-computer interaction. Much of our understanding of perception in visualization is rooted in an early work that ranked the order of \textit{visual encodings} based on their effectiveness for visual judgment~\cite{cleveland1984graphical}. The findings were pivotal in nature and a milestone in research, demonstrating that the application and understanding of perception lead to guidelines for an effective and expressive visual design. Furthermore, numerous works on improving and evaluating visualization's effectiveness have utilized the knowledge of attention, psychophysics, stimulus, judgment estimation, and perceptual laws to confirm the importance of perception throughout the process of generating a visualization~\cite{doherty2007perception, haroz2012capacity, gutwin2017peripheral, healey2012attention, rensink2006attention,  szafir2018modeling}, and we have observed a notable increase in the interest in perception-based visualization studies over recent decades (see \autoref{fig:paperCountByyear}).

    In this paper, we review perception-focused visualization studies since 1980 and summarize their research developments. Our focus is primarily on information visualization---areas such as scientific visualization, 3D perception, etc., have been surveyed elsewhere, e.g.,~\cite{preim2016survey}. Within this context, we create a practical taxonomy of prior studies based upon Amar et al.'s low-level task taxonomy~\cite{amar2005low}, further breaking techniques down by visual encoding and visualization type. In particular, we focus on how perception is used to evaluate the quality of visualizations, to help readers understand and apply the principles of perception to their visualization designs through a task-optimized approach.

    We focus our survey on studies that measured the efficacy of the visual design using \textit{graphical perception} when the user is performing various low-level tasks. Furthermore, since perception and cognition are not entirely separable, we discuss some perceptual effects on cognitive performance, e.g., completion time, accuracy, and error rate. Unless otherwise noted, throughout the remainder of this survey, we will use the term \textit{perception} to refer to \textit{graphical perception}.

    One of the key challenges with many graphical perception studies is their limited scope and reproducibility~\cite{quadri2019you}, caused in part by the difficulty of constructing human studies. Human studies are often resource-limited by the types and sizes of data, variety of visualizations or visual encodings, tasks being performed, and the size and diversity of subject pools, to name a few. This all puts transferability of results on tenuous footing. There is mounting evidence of the perceptual efficacy for many tasks and visualization types, but our survey provides a window into open research questions in many other situations. One of our survey's goals is collecting and organizing findings in such a way that the reader can make a judgment of the applicability of results to their context.

\subsection{Audience Guide for this Report}
    In this report, we target students, practitioners, and researchers, each of which will find value in different sections. First, \autoref{sec:taxonomy} establishes the structure of our taxonomy, including a discussion of tasks, as well as secondary issues, including the visual encoding and visualization types. Next, \autoref{sec.task} is the main survey of papers, divided by task, which provides background and context to aspects of perception that have been studied in visualization. \autoref{sec:discussion} concludes the work by discussing key points, limitations, and open research questions. Finally, as an appendix, \autoref{sec.background.fundamentals} provides a brief introduction to some fundamentals of perception for individuals looking to gain basic knowledge of the perceptual principles studied on visualization designs.

    An interactive list of the papers we have identified in our survey is available at {\small$<$\textcolor{blue}{{\href{ https://usfdatavisualization.github.io/VisPerceptionSurvey/}{https://usfdatavisualization. github.io/VisPerceptionSurvey/}}}$>$}.

 \begin{figure}[!b]
    \centering

    \includegraphics[width=0.825\linewidth]{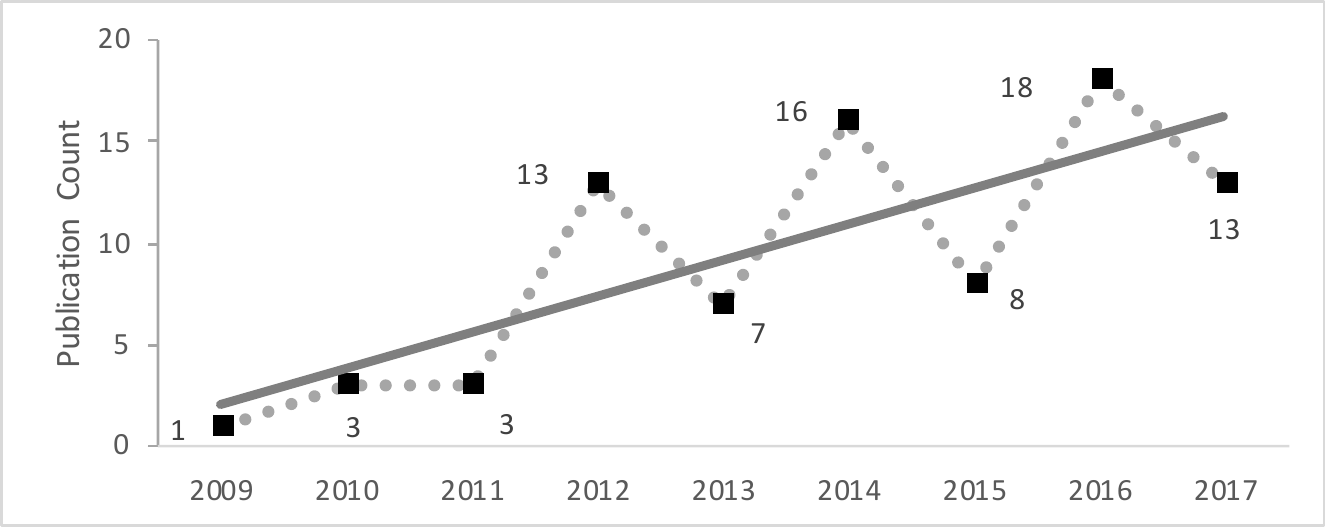}\hspace{15pt}

    \caption{Reproduction of a plot of the number of publications utilizing crowdsourcing experiments in recent years~\cite{borgo2018information}.}

    \label{fig:paperCountBycrowdsourcing}
\end{figure}

\subsection{Related Works}

    Optimizing visualization design is a perennial topic.

    \para{Design Recommendations}
    Design choices and recommendations form a critical element of effective visualization. In his book~\cite{ware2012information}, Colin Ware talked about human perception in the context of information visualization design. He aimed at broadly summarizing the design implications of research in perception and suggested explicit design guidelines. In their survey, Healey and Enns focused more specifically on the role that attention and visual memory play in the perception of visualizations~\cite{healey2012attention}. The work highlights how what we see impacts the viewer's accuracy in information judgment. Finally, VisGuides {\small$<$\textcolor{blue}{\url{https://visguides.org/}$>$}} is a web-based forum, which was established to collect practical knowledge of visualization guidelines and feedback on designs~\cite{diehl2018visguides}.

    \para{Frameworks}
    In addition to general guidelines, several researchers have focused on developing robust frameworks for optimizing design. Rensink's framework for reasoning about perceptions of visualization designs suggests using techniques from vision  science~\cite{rensink2014prospects}. The extended-vision theory asserts that a viewer and visualization system is a single system, whereas the optimal-reduction thesis postulates an optimal visualization. The work focuses on a few of the fundamental questions, e.g.: \textit{What is the best way to measure how a given visualization works?} \textit{Or, could we determine if its design is optimal?} Recent work by Elliott et al.\ introduced a design space of experimental methods for empirically investigating the perceptual processes involved in viewing data visualizations to inform visualization design guidelines~\cite{elliott2020design}. The paper provides shared design space and lexicon for facilitating empirical visualization research. Researchers can use this design space to create innovative studies and progress scientific understanding of design choices and evaluations in visualization. In contrast, our paper provides an overview of perception-based studies  by surveying papers that have evaluated the effectiveness of various visualization under a variety of tasks.

    \para{Evaluation}
    When optimizing a design, measurements of effectiveness are critical to understanding their impact. Studies often rely upon subject evaluations of visualizations. One important innovation is the introduction of crowdsourcing environments which are faster, less costly, and provide more diverse subject pools than lab-based studies (\autoref{fig:paperCountBycrowdsourcing}). Borgo et al.\ provided a detailed review of the use of crowdsourcing for evaluation in visualization research~\cite{borgo2018information}. In addition to subject evaluations, there are many methods and metrics for quantitatively evaluating the effectiveness of a visualization. Behrisch et al.~\cite{behrisch2018quality} gave an extensive audit of the state-of-art in quality metrics for various visualization techniques, along with details on a variety of implementation possibilities. The papers we discuss throughout this survey use a combination of both subject evaluations and quantitative measures to formulate their conclusions.

\subsection{Systematic Survey Literature}
\label{sec:lit_review}

Perception, being a critical part of visualization, has a wide range of applications and is included throughout different visualization-related journals and conferences. In addition, perception and its incorporation to visualization are derived from psychology-related journals, such as the Journal of Vision, Attention, Perception, and Psychophysics, and Psychonomic. With the survey's objective and the vast availability of perception-based papers, it was challenging to perform a comprehensive literature search. Our taxonomy discusses the application of perception to visualization. Therefore, we focused on visualization journals and conferences.

We identified papers from major visualization journals and conferences between 1980-2019 (see \autoref{tab:surveyed_papers}). We targeted the \textit{ACM}, \textit{IEEE}, and \textit{EG/CGF} libraries to collect the papers using a combination of keywords, including: \textit{perception}, \textit{visuali{z/s}ation}, \textit{evaluation}, \textit{design}, \textit{modeling}, \textit{visual perception}, \textit{attention}, \textit{visual task}, \textit{user study}, \textit{graphical encoding}, and \textit{effectiveness}. The ``others'' categories in \autoref{tab:surveyed_papers} are highly cited, pivotal works that were discovered during our search of the primary sources. Using the PRISMA framework (see {\small$<$\textcolor{blue}{{\url{http://www.prisma-statement.org/}}}$>$}) as a guide, we coded the scope of the survey into categories to screen the literature. We filtered the paper based on the study's objective and characteristics. We could not include all filtering categories due to space limitations, but some examples are below.

\begin{table}[!t]
    \centering
    \caption{The number of surveyed papers by source.}
    \label{tab:surveyed_papers}
    \def\arraystretch{1.025}
    \resizebox{0.925\linewidth}{!}{%
        \begin{tabular}{|p{9.05cm}|@{ }c@{ }|} 
        \hline
            \multirow{2}{*}{\textbf{Sources}}   & \textbf{Paper}   \\ 
                                                & \textbf{Count}   \\ 
        \hline\hline
            IEEE Trans.\ on Visualization and Computer Graphics (TVCG)   & \multirow{4}{*}{64}  \\ 
            IEEE Information Visualization (InfoVis)                     & \\
            IEEE Visual Analytics in Science and Technology (VAST)       & \\
            IEEE Pacific Visualization Symposium (PacificVis)            & \\
        \hline
            ACM Conf.\ on Human Factors in Information Systems (CHI)    & \multirow{3}{*}{30} \\ 
            \hspace{5pt} including Extended Abstracts                   & \\
            ACM Transaction of Graphics (TOG)                           & \\
        \hline
            Computer Graphics Forum (CGF)               & \multirow{4}{*}{17} \\
            Eurographics (EG)                           & \\
            EG/IEEE VGTC Conference on Visualization (EuroVis)  & \\ 
            \hspace{5pt} including Short Papers         & \\
        \hline
            Others---Beyond Time and Errors on Novel Evaluation Methods for Visualization (BELIV); Journal of Vision; Perception and Psychophysics; Science; Journal of the American Statistical Association; International Conference on Theory and Application of Diagram; Cartographics; Journal of Man-Machine Studies; Behaviour and Information Technology; Others
                & \begin{minipage}[t]{0.8cm}\vspace{22pt}\centering 11\end{minipage} \\ 
        \hline
        \end{tabular}
    }  
    
\end{table}
\def\arraystretch{1.0}

\para{Examples for Inclusion}
\begin{itemize}
    \item Experiments focused on graphical perception on visual tasks, e.g.,~\cite{saket2018evaluating}, or visual design, e.g.,~\cite{szafir2018modeling}.
    \item Experiments focused on the graphical perception of visualization methods, e.g.,~\cite{saket2018task}.
    \item Studies including discussion and suggestions of design guidelines, e.g.,~\cite{talbot2014four}.
    \item Experiments on modeling the visualization to improve inference-making, decision-making, or judgment estimation, based on visual channel and graphical perception, e.g.,~\cite{harrison2013influencing, yang2018correlation}.
\end{itemize}

\para{Examples for Exclusion}
\begin{itemize}
    \item User-study comparing two or more models, e.g.,~\cite{brehmer2019comparative}.
    \item Empirical studies to check on quality metrics of a system by user study, e.g.,~\cite{choo2010ivisclassifier}. 
    \item Evaluations of the performance of visualization tools or systems, e.g.,~\cite{wongsuphasawat2016voyager}. 
    \item Empirical studies on graphics or user interfaces, not the visualization, e.g.,~\cite{saket2016visualization}. 
\end{itemize}

In addition to the interactive taxonomy at {\small$<$\textcolor{blue}{{\url{https://usfdatavisualization.github.io/VisPerceptionSurvey/}}}$>$}, you will find (1) a spreadsheet of surveyed papers, (2) the systematic flow of paper collection and filtering, i.e., PRISMA, (3) a paper categorization template, and (4) a summary table with the count of studies.

\section{Structure of the Taxonomy}
\label{sec:taxonomy}

    The method of visually encoding data is usually thought to be the main component of visualization. However, the analysis task is equally, if not more, important. Several evaluation studies have suggested visualization effectiveness is task-dependent~\cite{saket2018task}, and a large body of research seeks to determine which data representations are perceptually optimal for specific low-level tasks, e.g.,~\cite{cleveland1984graphical, harrison2014ranking,talbot2014four,ware2012information}. Seeing that the vast majority of perceptual studies in visualization had a specific low-level task as the main study objective or a low-level task was used in the evaluation, we centered on that as the main category of the taxonomy. Furthermore, most papers consider the low-level tasks in the context of a limited subset of visual encodings and/or visualization types. Therefore, each of our low-level task discussions is further split by types of \textbf{visual encoding} and \textbf{visualization}.

\subsection{Low-Level Tasks} 
    We considered two existing task taxonomies as a framework for our survey. We first considered Brehmer and Munzner's taxonomy of abstracted tasks, which is a higher-level taxonomy, i.e., perceptual + cognitive~\cite{brehmer2013multi}. Since our focus was perception, it did not fit well. Despite perception and cognition not being entirely separable, an ideal task framework, would ensure as little cognition as possible occurs within the task. Ultimately, Amar et al.'s low-level task taxonomy~\cite{amar2005low} fit better, as the tasks they define require less reasoning about the data. From there, we took its ten low-level tasks: \textit{retrieve value}, \textit{filter}, \textit{compute a derived value}, \textit{find extremum}, \textit{sort}, \textit{determine range}, \textit{characterize distribution}, \textit{find anomalies}, \textit{cluster}, and \textit{correlate}, and we included one derived task, \textit{compare}, for a total of 11 low-level task groups. Each paper was placed into one or more of these task groups. We summarize the tasks using descriptions from prior work:

    \noindent
    \begin{minipage}[t]{\linewidth}
    \begin{wrapfigure}[3]{L}{0.09\linewidth}
        \vspace{-14pt}
        \begin{minipage}[t]{1.2\linewidth}
            \hbox{\includegraphics[width=28pt]{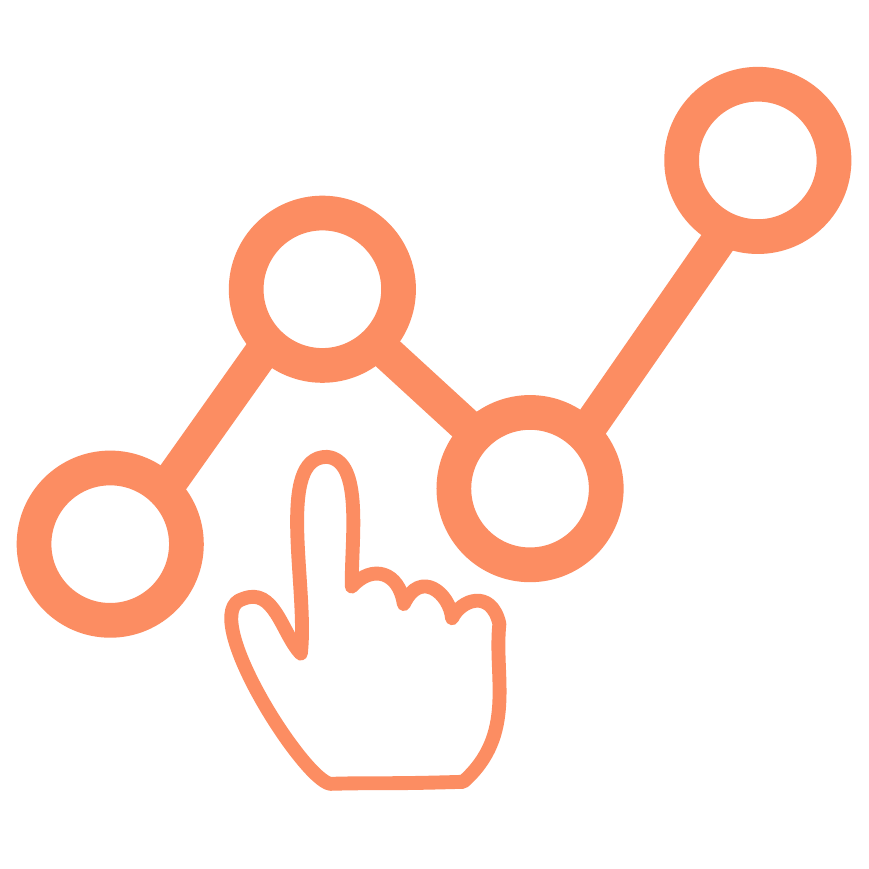}}
        \end{minipage}
    \end{wrapfigure} 
    \noindent
    \textbf{Retrieve Value} (\autoref{sec:task-retrieve}) -- \textit{``Given a set of specific cases, find attributes of those cases.''}~\cite{amar2005low}
    \end{minipage} 
    
    \vspace{6pt}
    \noindent
    \begin{minipage}[t]{\linewidth}
    \begin{wrapfigure}[3]{L}{0.09\linewidth}
        \vspace{-9pt}
        \begin{minipage}[t]{1.2\linewidth}
        \includegraphics[width=28pt]{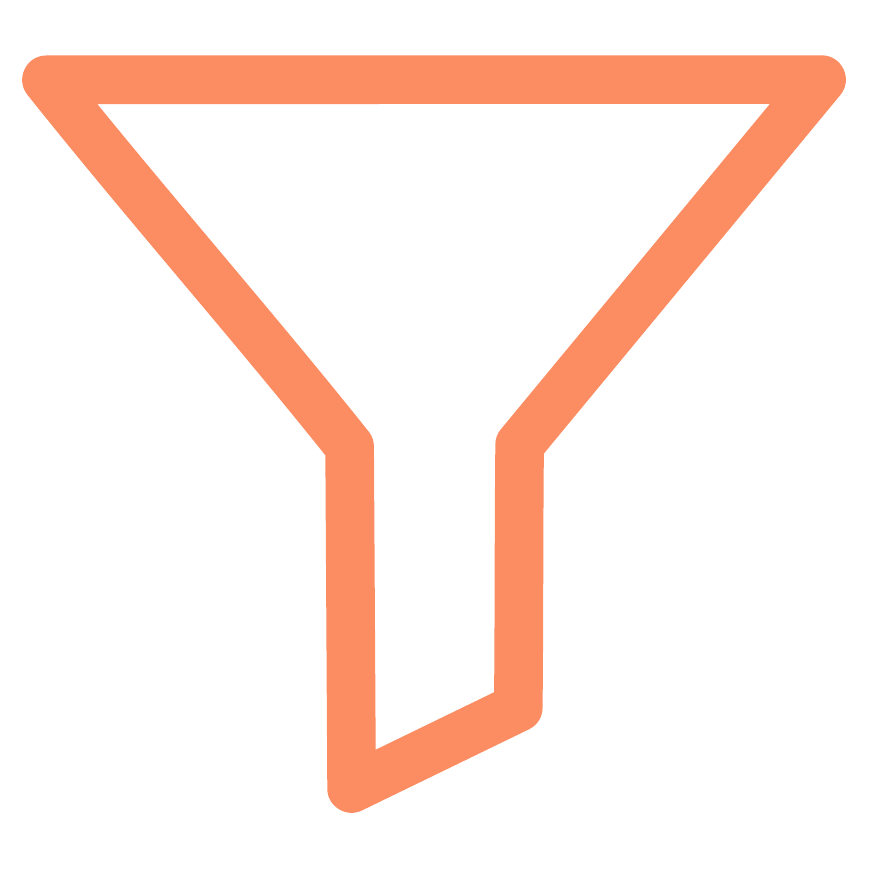}
        \end{minipage}
    \end{wrapfigure} 
    \noindent
    \textbf{Filter} (\autoref{sec:task-filter}) -- \textit{``Given some concrete conditions on attribute values, find data cases satisfying those conditions.''}~\cite{amar2005low}
    \end{minipage} 
    
    \vspace{4pt}
    \noindent
    \begin{minipage}[t]{\linewidth}
    \begin{wrapfigure}[3]{L}{0.09\linewidth}
        \vspace{-9pt}
        \begin{minipage}[t]{1.2\linewidth}
        \includegraphics[width=28pt]{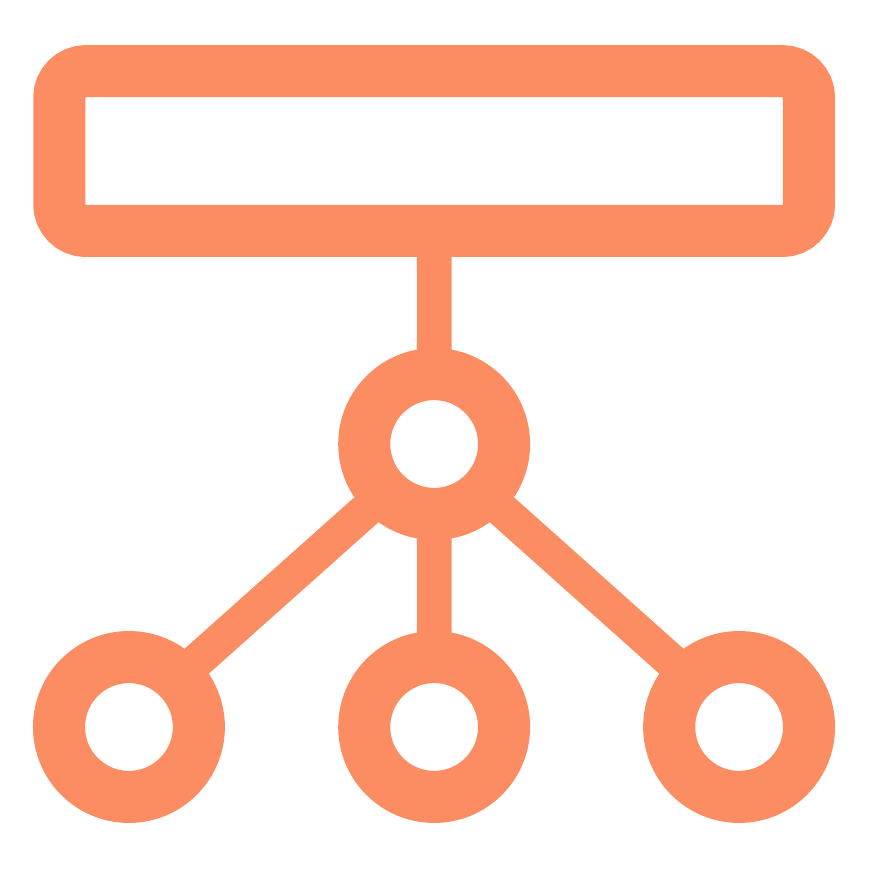}
        \end{minipage}
    \end{wrapfigure} 
    \noindent
    \textbf{Compute Derived Value} (\autoref{sec:task-compute}) -- \textit{``Given a set of data cases, compute an aggregate numeric representation of those data cases.''}~\cite{amar2005low}
    \end{minipage} 
    
    \vspace{4pt}
    \noindent
    \begin{minipage}[t]{\linewidth}
    \begin{wrapfigure}[3]{L}{0.09\linewidth}
        \vspace{-9pt}
        \begin{minipage}[t]{1.2\linewidth}
        \includegraphics[width=28pt]{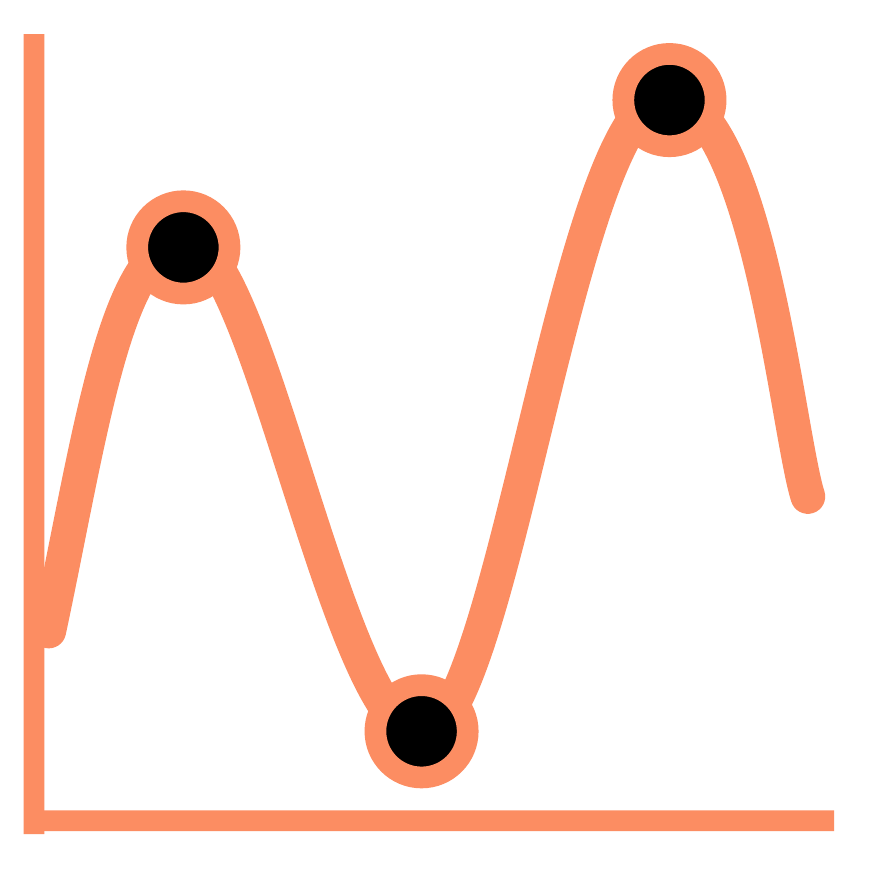}
        \end{minipage}
    \end{wrapfigure} 
    \noindent
    \textbf{Find Extremum} (\autoref{sec:task-extremum}) -- \textit{``Find data cases possessing an extreme value of an attribute over its range within the data set.''}~\cite{amar2005low}
    \end{minipage} 
    
    \vspace{6pt}
    \noindent
    \begin{minipage}[t]{\linewidth}
    \begin{wrapfigure}[3]{L}{0.09\linewidth}         \vspace{-14pt}         \begin{minipage}[t]{1.2\linewidth}
        \includegraphics[width=28pt]{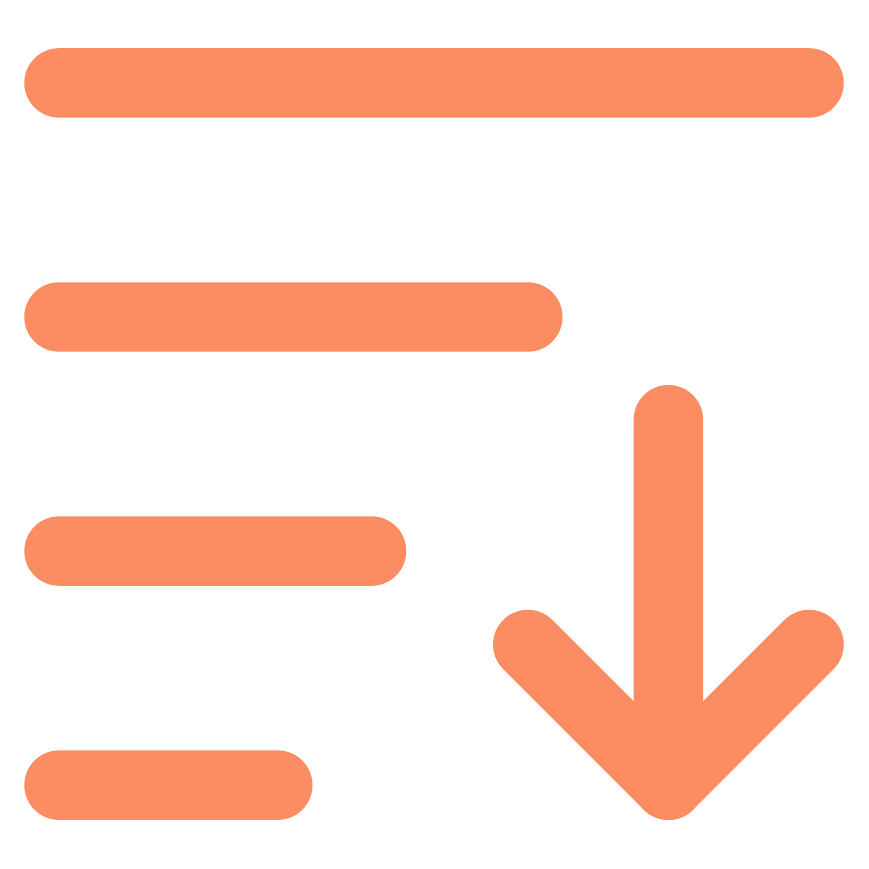}
        \end{minipage}
    \end{wrapfigure} 
    \noindent
    \textbf{Sort} (\autoref{sec:task-sort}) -- \textit{``Given a set of data cases, rank them according to some ordinal metric.''}~\cite{amar2005low}
    \end{minipage} 
    
    \vspace{6pt}
    \noindent
    \begin{minipage}[t]{\linewidth}
        \begin{wrapfigure}[3]{L}{0.09\linewidth}         \vspace{-9pt}         \begin{minipage}[t]{1.2\linewidth}
        \includegraphics[width=28pt]{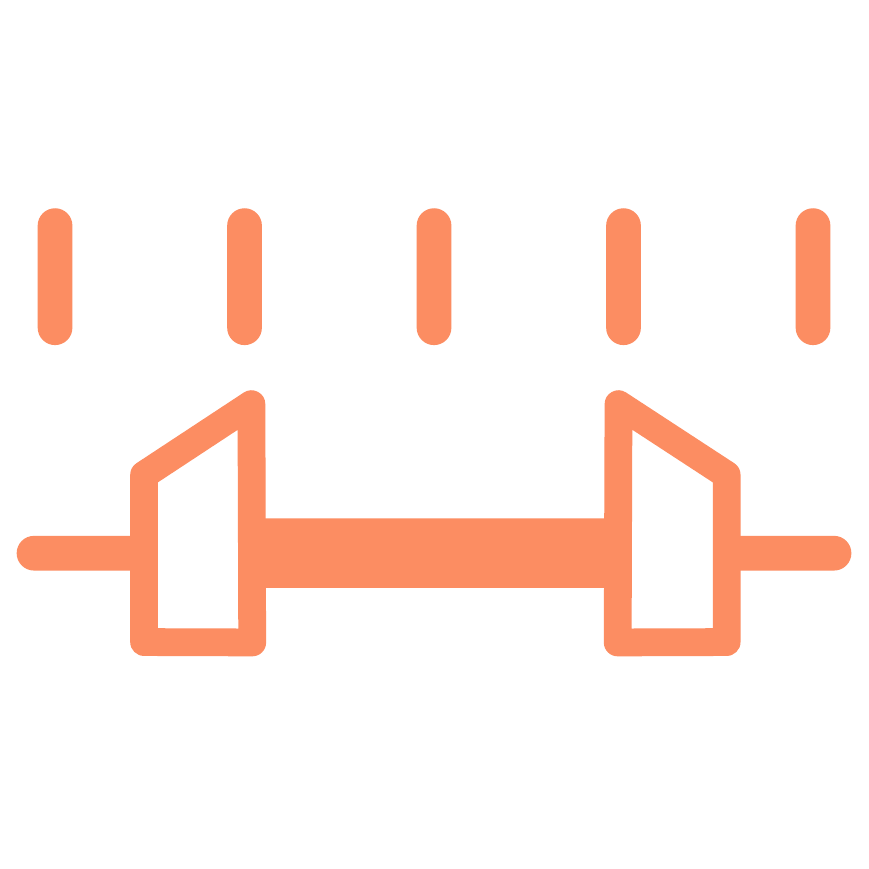}
        \end{minipage}
    \end{wrapfigure} 
    \noindent
    \textbf{Determine Range} (\autoref{sec:task-range}) -- \textit{``Given a set of data cases and an attribute of interest, find the span of values within the set.''}~\cite{amar2005low}
    \end{minipage}
    
    \vspace{4pt}
    \noindent
    \begin{minipage}[t]{\linewidth}
    \begin{wrapfigure}[3]{L}{0.09\linewidth}         \vspace{-9pt}         \begin{minipage}[t]{1.2\linewidth}
        \includegraphics[width=28pt]{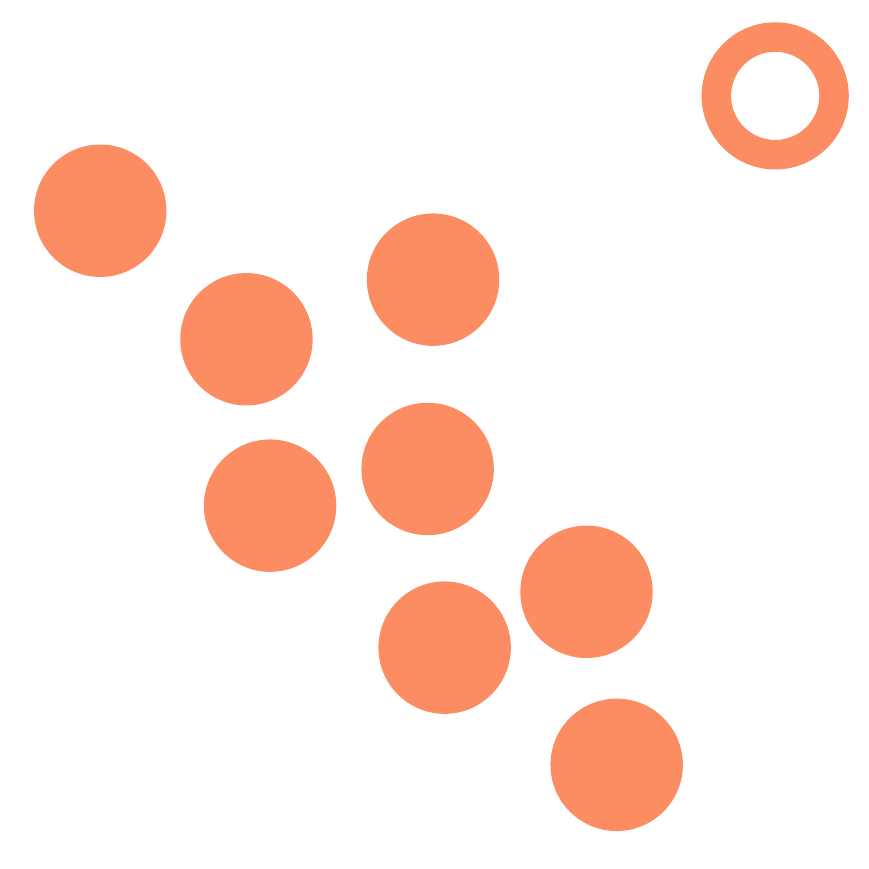}
        \end{minipage}
    \end{wrapfigure} 
    \noindent
    \textbf{Find Anomalies} (\autoref{sec:task-anomalies}) -- \textit{``Identify any anomalies within a given set of data cases concerning a given relationship or expectation, e.g., statistical outliers.''}~\cite{amar2005low}
    \end{minipage} 
    
    \vspace{4pt}
    \noindent
    \begin{minipage}[t]{\linewidth}
    \begin{wrapfigure}[4]{L}{0.09\linewidth}         \vspace{-4pt}         \begin{minipage}[t]{1.2\linewidth}
        \includegraphics[width=28pt]{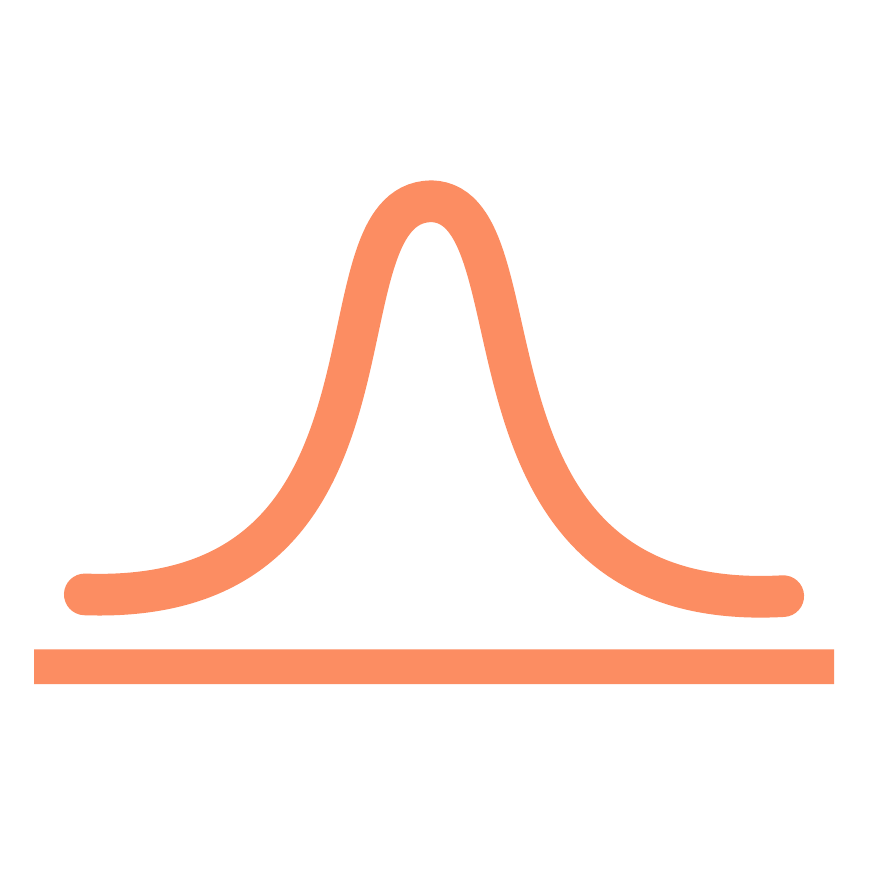}
        \end{minipage}
    \end{wrapfigure} 
    \noindent
    \textbf{Characterize Distribution} (\autoref{sec:task-distribution}) -- \textit{``Given a set of data cases and a quantitative attribute of interest, characterize the distribution of that attribute’s values over the set.''}~\cite{amar2005low}
    \end{minipage} 
    
    \vspace{6pt}
    \noindent
    \begin{minipage}[t]{\linewidth}
    \begin{wrapfigure}[3]{L}{0.09\linewidth}         \vspace{-14pt}         \begin{minipage}[t]{1.2\linewidth}
        \includegraphics[width=28pt]{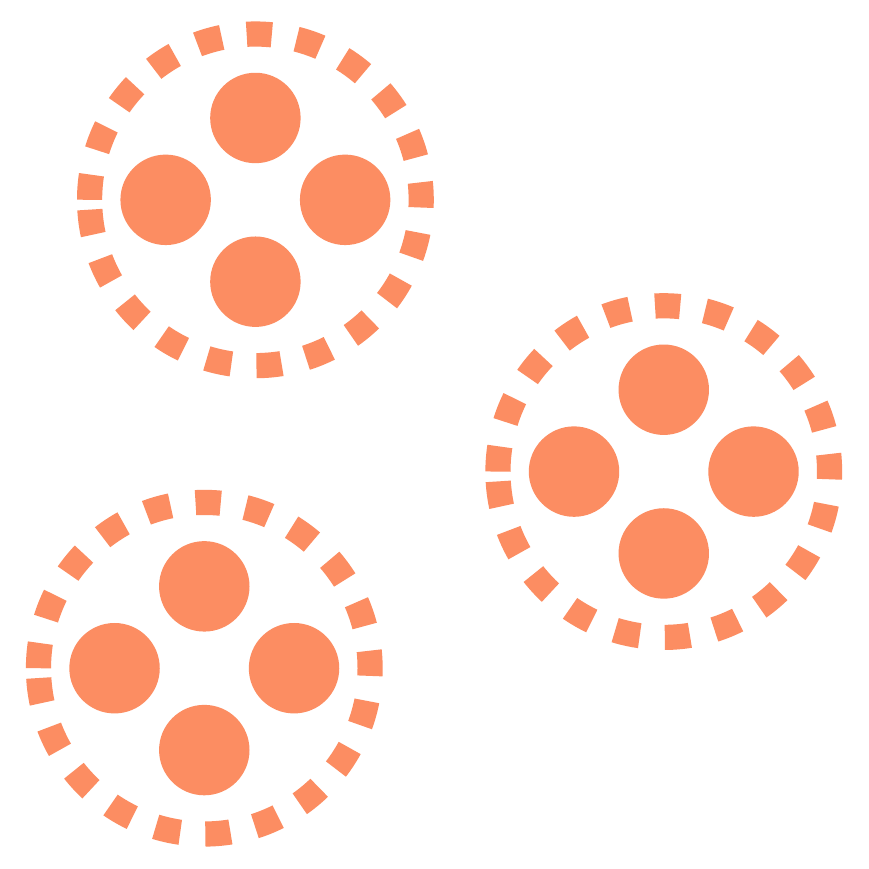}
        \end{minipage}
    \end{wrapfigure} 
    \noindent
    \textbf{Cluster} (\autoref{sec:task-cluster}) -- \textit{``Given a set of data cases, find clusters of similar attribute values.''}~\cite{amar2005low}
    \end{minipage} 
    
    \vspace{6pt}
    \noindent
    \begin{minipage}[t]{\linewidth}
    \begin{wrapfigure}[3]{L}{0.09\linewidth}         \vspace{-9pt}         \begin{minipage}[t]{1.2\linewidth}
        \includegraphics[width=28pt]{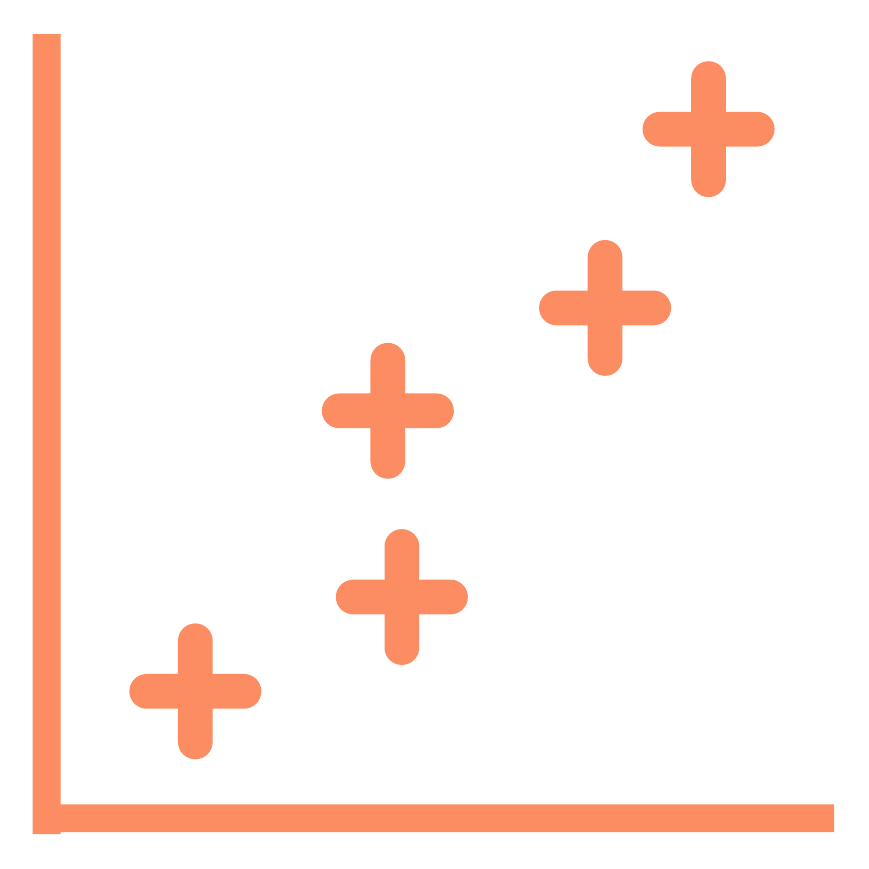}
        \end{minipage}
    \end{wrapfigure} 
    \noindent
    \textbf{Correlate} (\autoref{sec:task-correlation}) -- \textit{``Given a set of data cases and two attributes, determine useful relationships between the values of those attributes.''}~\cite{amar2005low}
    \end{minipage} 
    
    \vspace{4pt}
    \noindent
    \begin{minipage}[t]{\linewidth}
    \begin{wrapfigure}[4]{L}{0.09\linewidth}         \vspace{-4pt}         \begin{minipage}[t]{1.2\linewidth}
        \includegraphics[width=28pt]{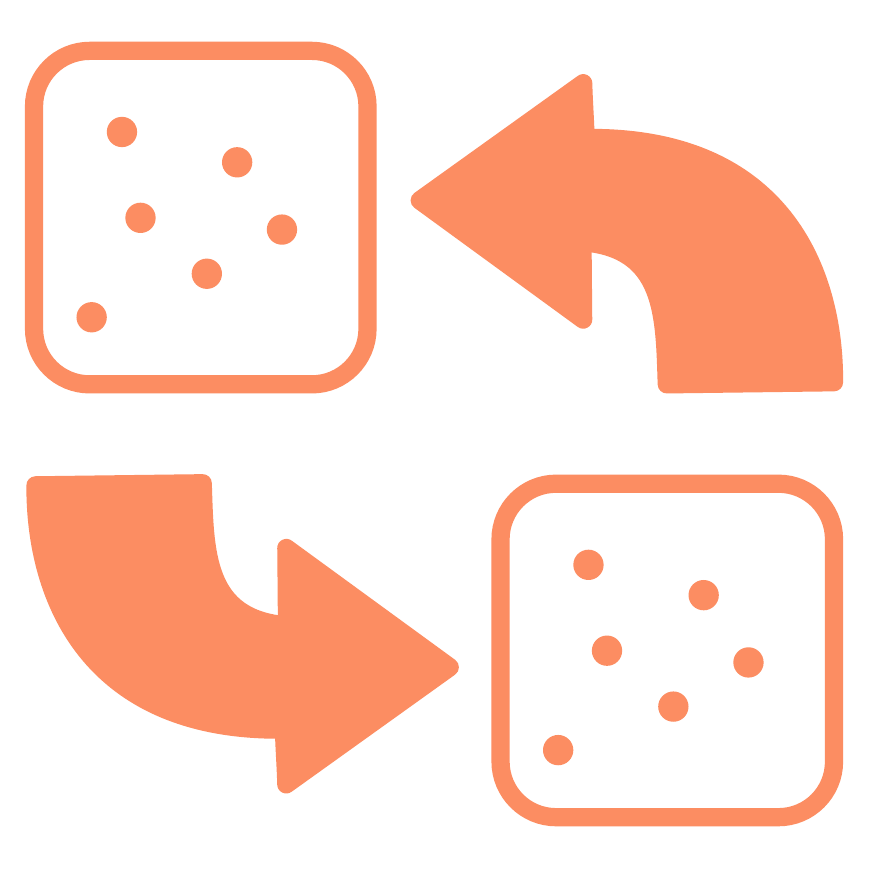}
        \end{minipage}
    \end{wrapfigure} 
    \noindent
    \textbf{Compare} (\autoref{sec:task-compare}) -- \textit{``Given a set of data cases, compare any attributes within and between relations of the given set of data cases for a given relationship condition.''}~\cite{amar2005low}
    \end{minipage}

\subsection{Visual Encoding}
\label{sec.visual_encoding}

\begin{figure}[!t]
    \centering
    \includegraphics[width=0.95\linewidth]{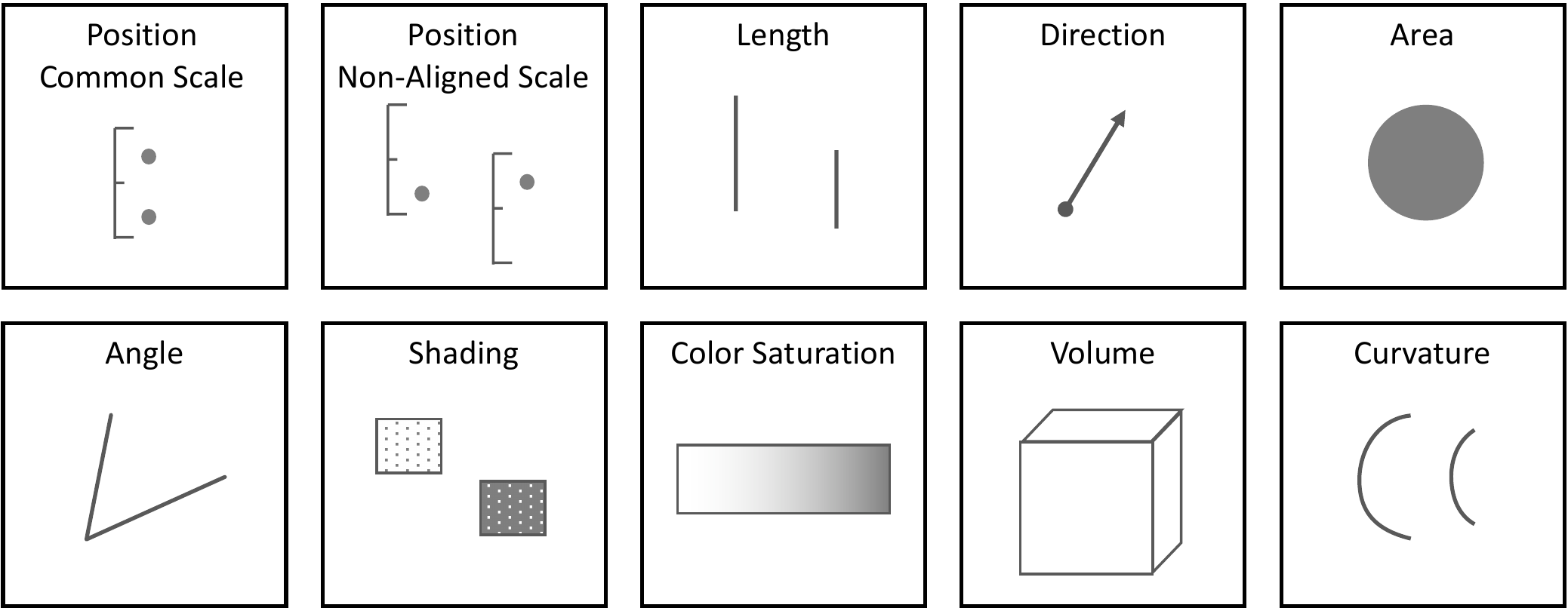}
    \caption{Reproduction of visual encodings types introduced in Cleveland and McGill's work\cite{cleveland1984graphical}.}
    \label{fig:cleveland_types}
\end{figure}

Visual encodings are properties used to encode data in a visualization, including \textit{position}, \textit{length}, \textit{angle}, \textit{area}, \textit{volume}, \textit{shading}, \textit{direction}, \textit{curvature}, and \textit{color} (see \autoref{fig:cleveland_types}). The terms graphical encoding, visual channel, visual encodings, and visual properties are often interchanged, but generally, they mean the same thing. Throughout the remainder of this paper, we refer to them as visual encodings.

\begin{figure}[!b]
    \centering
    \includegraphics[trim=0 0 40pt 0, clip, width=0.825\linewidth]{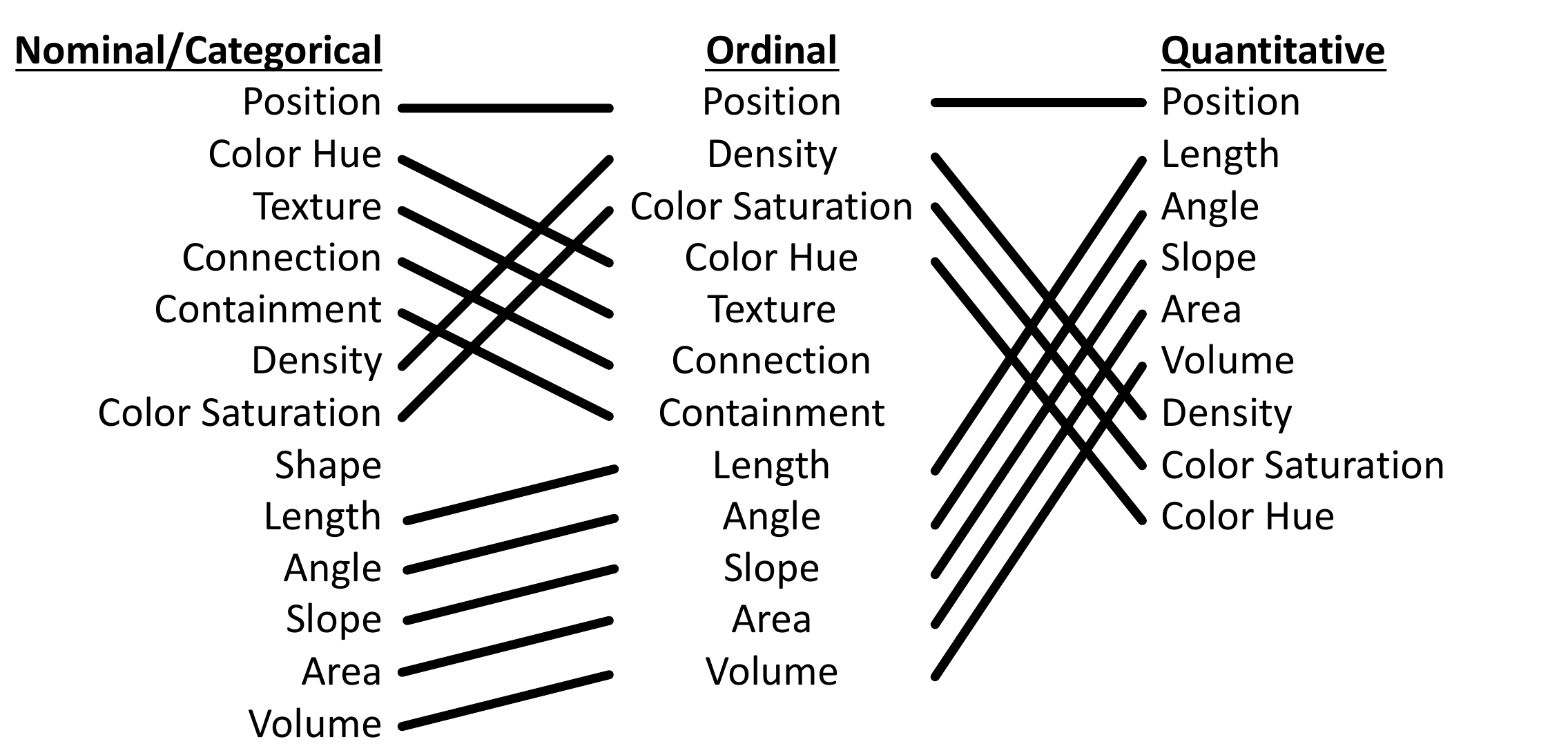}
    \caption{Reproduction of the Mackinlay visual encoding rankings~\cite{mackinlay1986automating}.}
    \label{fig:mackinlay_ranking}
\end{figure}

\para{Effectiveness of Visual Encodings}
Understanding the role of perception in the choice of visual encodings is critical to visualization designers.  Cleveland and McGill's study evaluated the efficacy of visual encodings by measuring the perceptual magnitude of judgments to determine their accuracy. Mackinlay produced the first comprehensive ranking of visual encodings by data type, as shown in \autoref{fig:mackinlay_ranking}~\cite{mackinlay1986automating}. The ranking has been further validated and elucidated through numerous follow-up studies, e.g.,~\cite{demiralp2014learning, gramazio2014relation, heer2009sizing, heer2010crowdsourcing, neumann1998perception, reda2018graphical, saket2018evaluating, sedlmair2012taxonomy, szafir2018modeling, talbot2014four}, many of which are further discussed throughout our taxonomy.

\para{Taxonomy on Visual Encodings}
Early works in visual encoding defined numerous individual visual channels, e.g., 10 in~\cite{cleveland1984graphical} and 13 in~\cite{mackinlay1986automating}. Given our taxonomy already defines 11 task categories, enumerating all visual encoding types would have resulted in far too many categories. Instead, we combined visual encodings into roughly two main categories: spatial encodings and color encodings, each with two subcategories. Spatial encodings encompass visual encodings having to do with position, size (i.e., length, area, and volume), direction, and shape. Mackinlay, as well as others, have shown that spatial encodings are particularly effective for quantitative data~\cite{mackinlay1986automating}. Color is another important visual encoding that includes properties of color such as hue, saturation, luminance, and opacity. Stimulus-based psychophysics experiments have demonstrated that color can be used to represent ordering~\cite{chung2016ordered}, category~\cite{setlur2016linguistic}, quantity~\cite{lin2013selecting}, and perceived-difference judgment~\cite{szafir2018modeling} in visualization. The categories are further divided as follows:

    \vspace{4pt}
    \noindent
    \begin{minipage}[t]{\linewidth}
    \begin{wrapfigure}[3]{L}{0.09\linewidth}         \vspace{-10pt}         \begin{minipage}[t]{1.2\linewidth}
        \includegraphics[width=28pt]{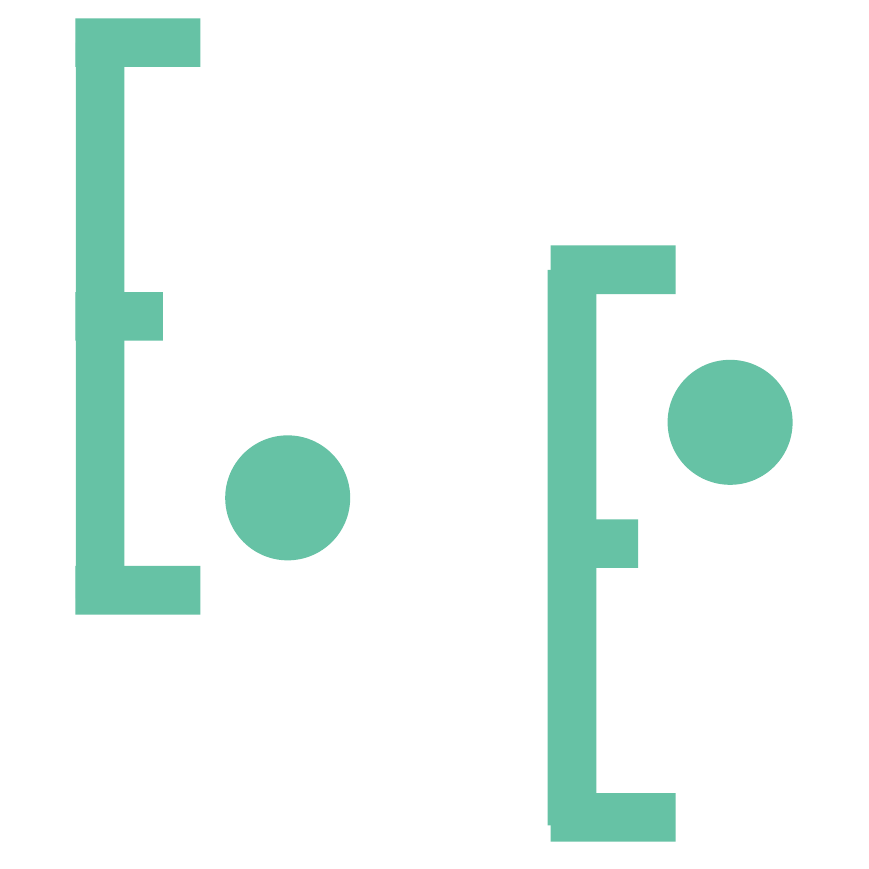}
        \end{minipage}
    \end{wrapfigure} 
    \noindent
    \textbf{Spatial Position and Shape} encodings are combined because of their geometric relationship. The category encompasses any encoding concerning the layout, e.g., position, or shape, e.g., direction/angle or curvature.
    \end{minipage} 
    
    \vspace{4pt}
    \noindent
    \begin{minipage}[t]{\linewidth}
    \begin{wrapfigure}[3]{L}{0.09\linewidth}         \vspace{-10pt}         \begin{minipage}[t]{1.2\linewidth}
        \includegraphics[width=28pt]{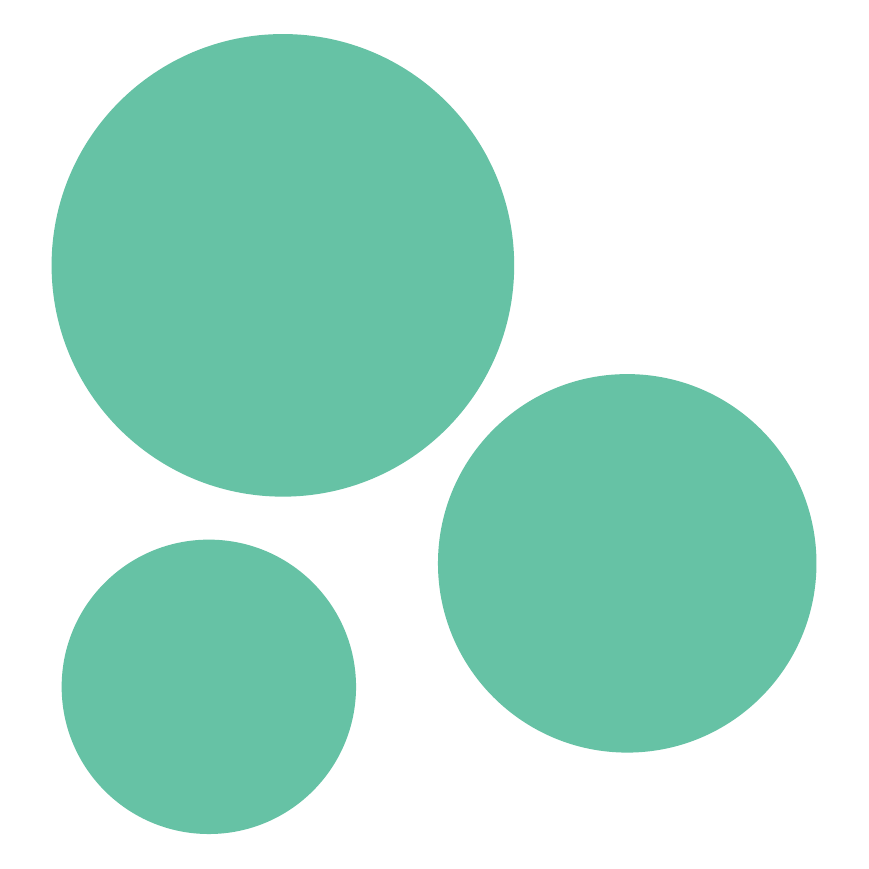}
        \end{minipage}
    \end{wrapfigure} 
    \noindent
    \textbf{Spatial Size} encodings are those where the size of objects, i.e., length, area, or volume, encode the relevant data in the visualization.
    \end{minipage}     
    
        \vspace{4pt}
    \noindent
    \begin{minipage}[t]{\linewidth}
    \begin{wrapfigure}[3]{L}{0.09\linewidth}         \vspace{-10pt}         \begin{minipage}[t]{1.2\linewidth}
        \includegraphics[width=28pt]{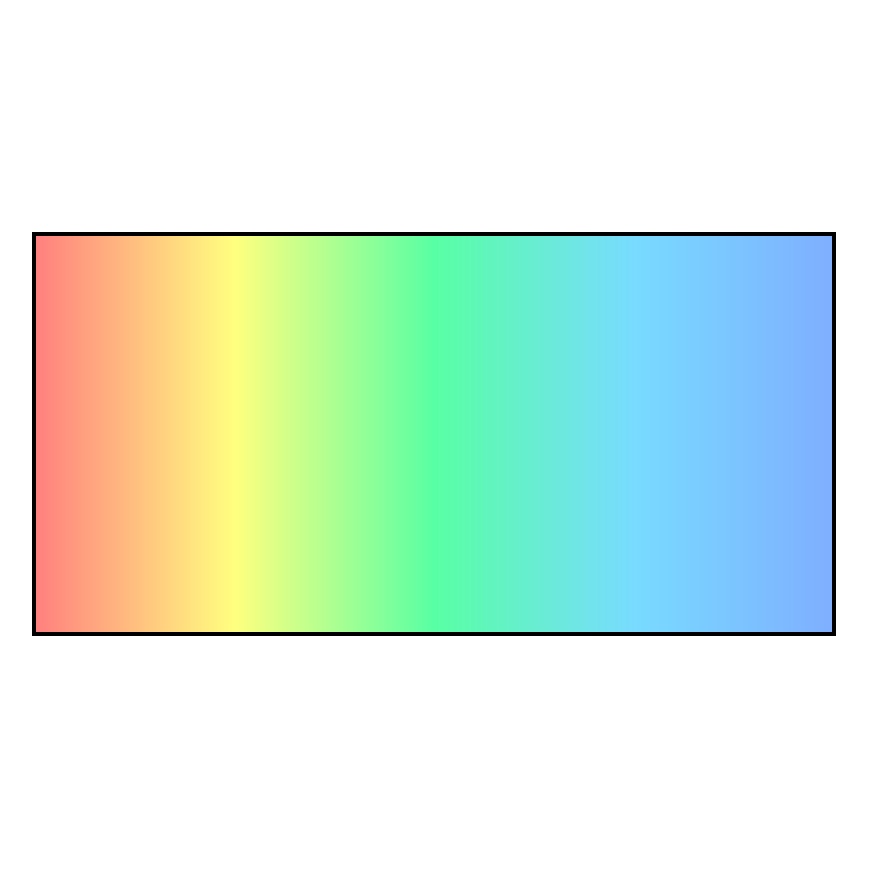}
        \end{minipage}
    \end{wrapfigure} 
    \noindent
    \textbf{Color Hue} deals primarily with variations in color (in the colloquial sense), usually dealing with categorical colors or colormaps. 
    \end{minipage} 
    
        \vspace{4pt}
    \noindent
    \begin{minipage}[t]{\linewidth}
    \begin{wrapfigure}[3]{L}{0.09\linewidth}         \vspace{-10pt}         \begin{minipage}[t]{1.2\linewidth}
        \includegraphics[width=28pt]{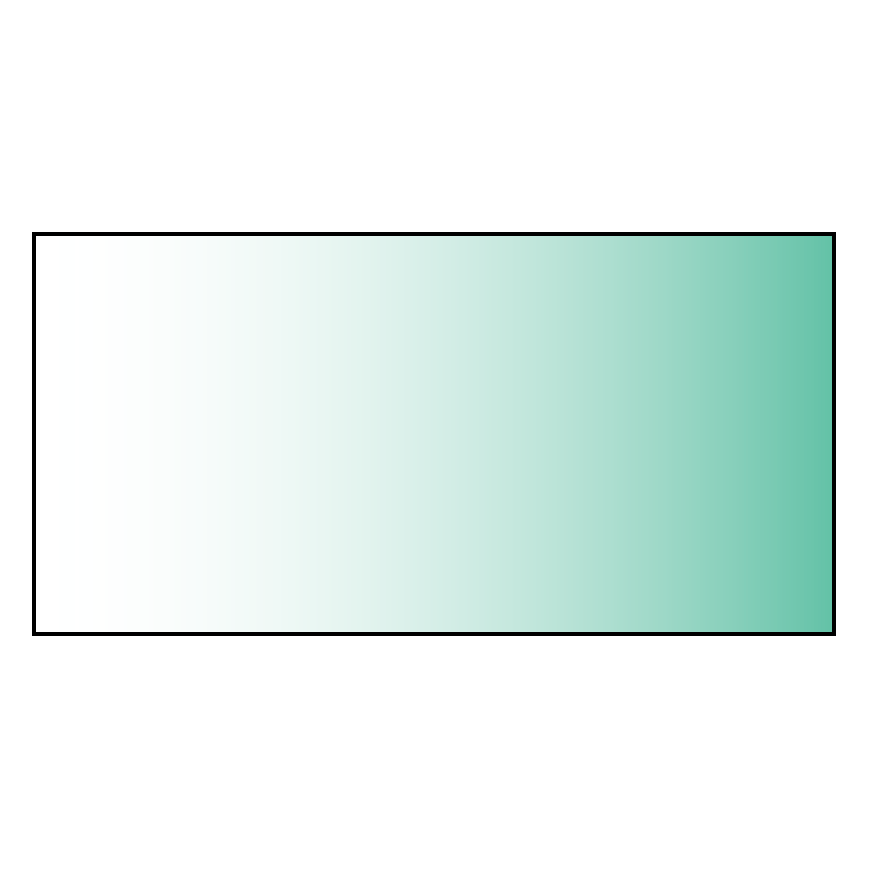}
        \end{minipage}
    \end{wrapfigure} 
    \noindent
    \textbf{Color Intensity} encodings capture data by the intensity of the color representation, e.g., in color saturation, luminance, shading, or opacity, which have combined due to their close inter-relatedness, e.g., changes in opacity, luminance, or saturation can all influence the intensity of a color.
    \end{minipage}

\subsection{Visualization}

The final category of our taxonomy is visualization type. Most papers studied standard visualization types, quite often with a variation in their design. We extracted all visualization types and combined related types into the following 10 categories and one ``other'' category:

\vspace{10pt}
\noindent
\begin{minipage}[t]{\linewidth}
\centering
\begin{minipage}[t]{0.24\linewidth}
    \centering
    \small
        \includegraphics[width=28pt]{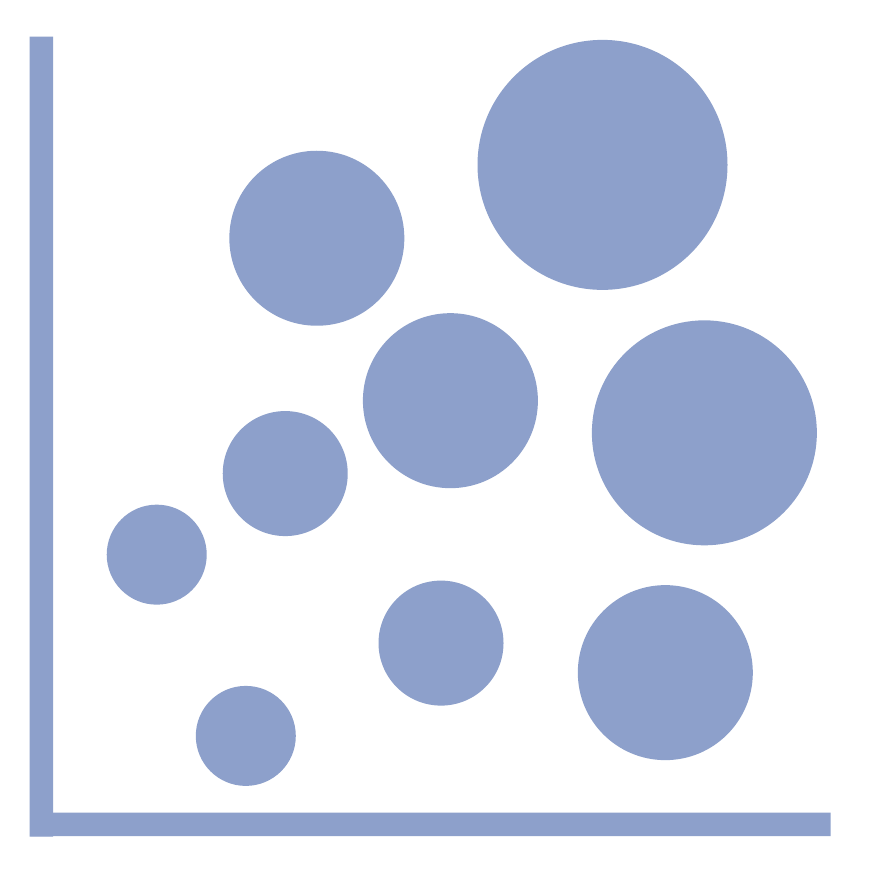}
        Scatterplot
\end{minipage}
\begin{minipage}[t]{0.24\linewidth}
    \centering
    \small
        \includegraphics[width=28pt]{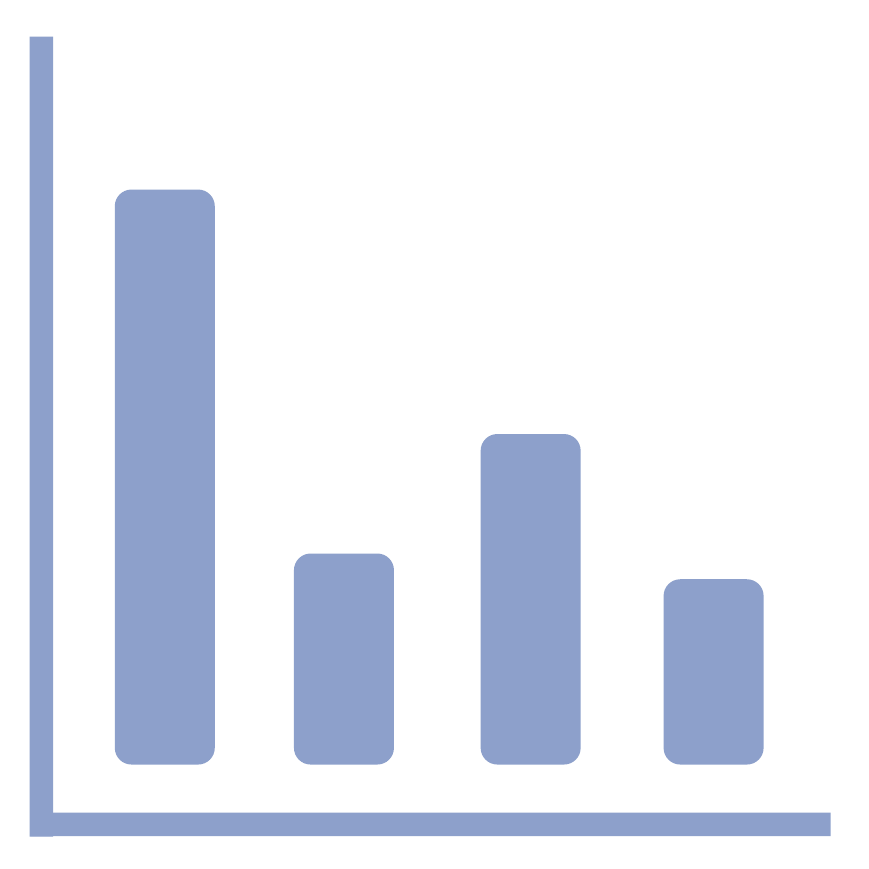}\\
        Bar Chart
\end{minipage}
\begin{minipage}[t]{0.24\linewidth}
    \centering
    \small
        \includegraphics[width=28pt]{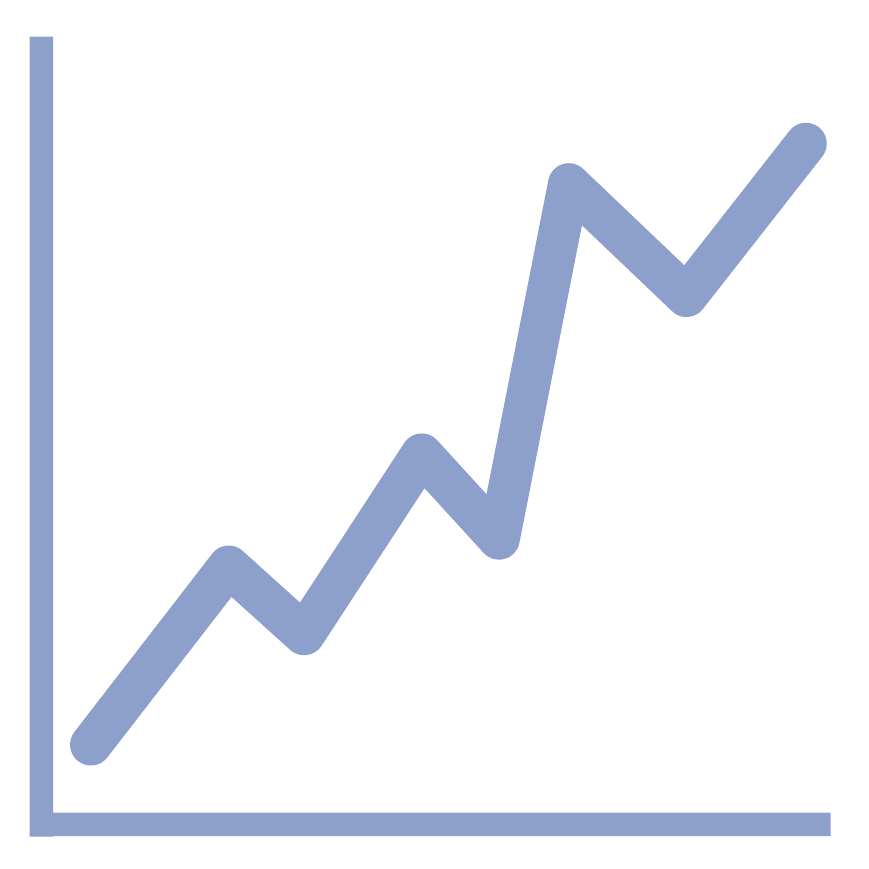}\\
        Line Chart
\end{minipage}

\vspace{10pt}
\begin{minipage}[t]{0.325\linewidth}
    \centering
    \small
        \includegraphics[width=28pt]{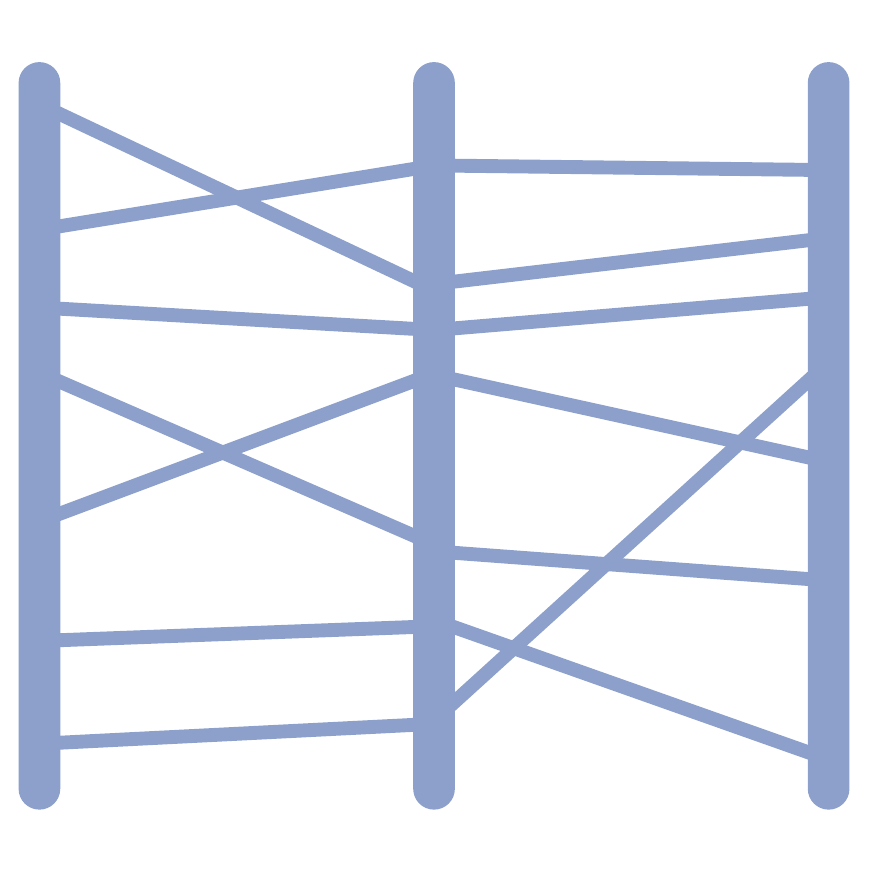}\\
        Parallel Coordinates
\end{minipage}
\begin{minipage}[t]{0.12\linewidth}
    \centering
    \small
        \includegraphics[width=28pt]{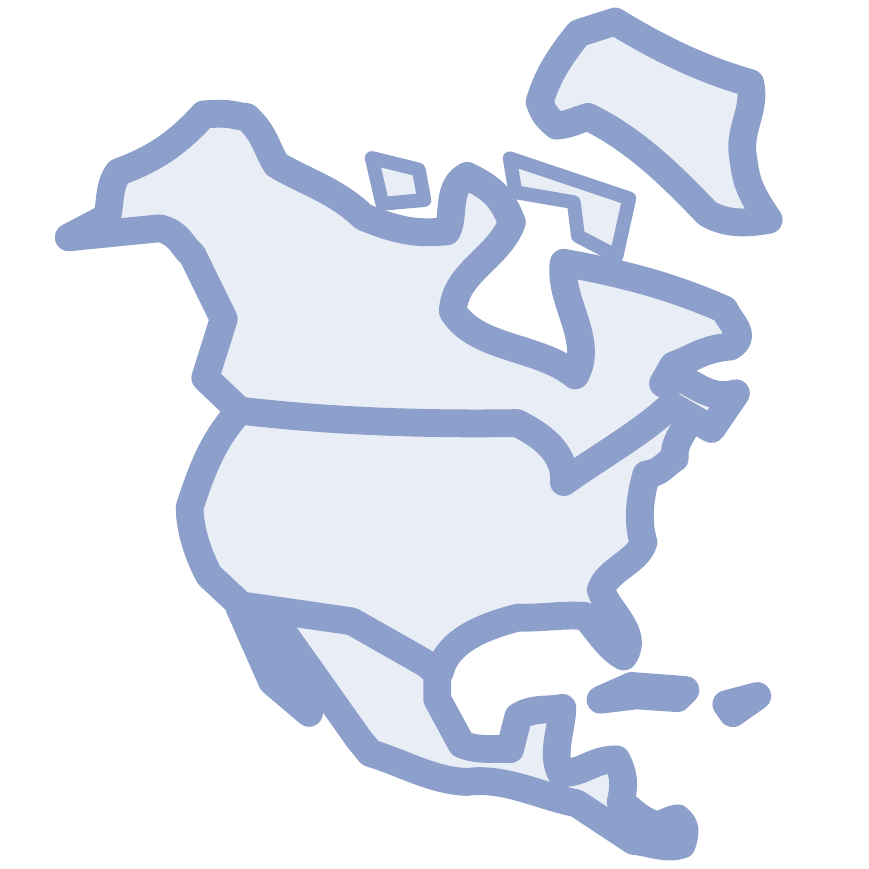}\\
        Map
\end{minipage}
\hfill
\begin{minipage}[t]{0.12\linewidth}
    \centering
    \small
        \includegraphics[width=28pt]{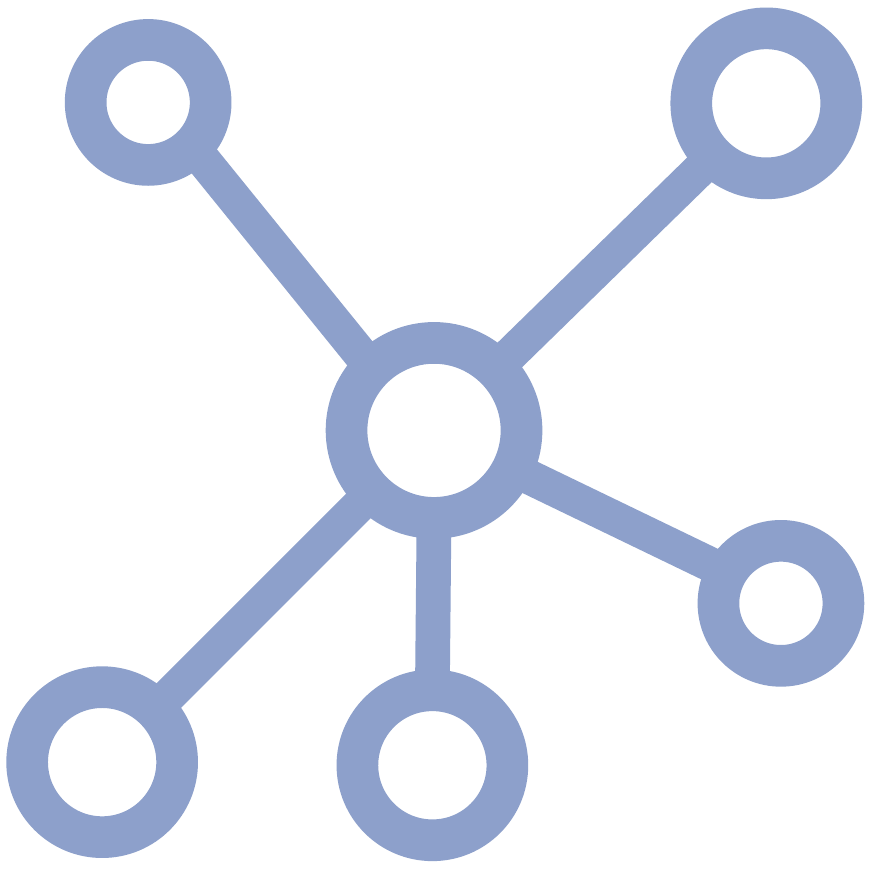}\\
        Network
\end{minipage}
\begin{minipage}[t]{0.325\linewidth}
    \centering
    \small
        \includegraphics[width=28pt]{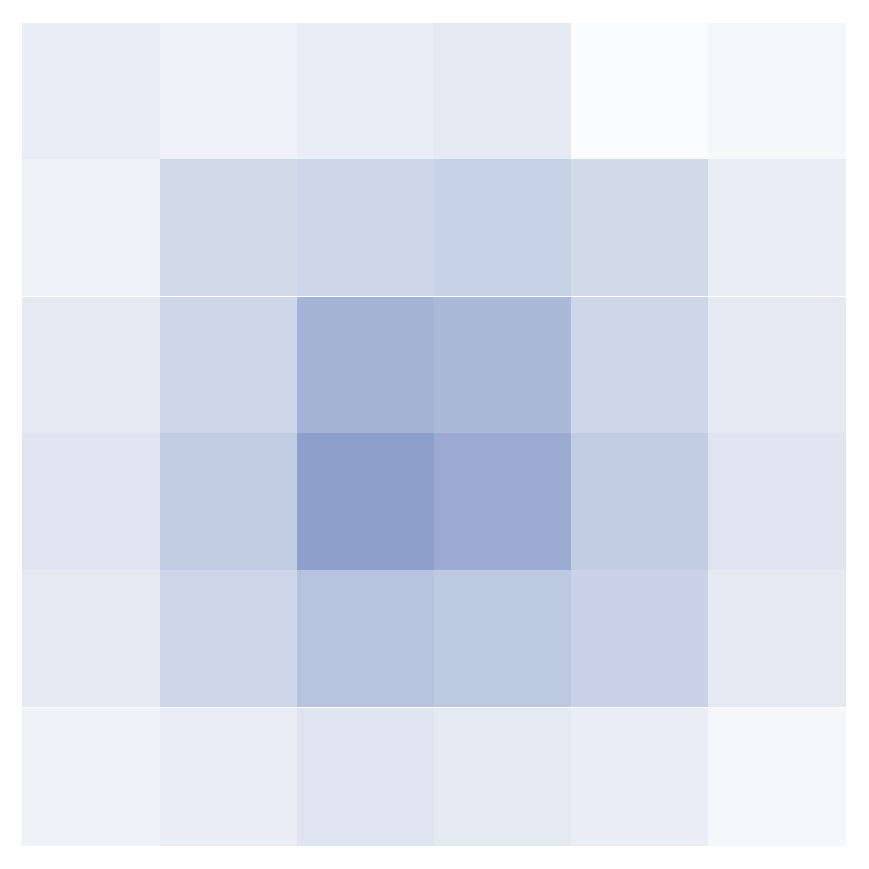}\\
        Heatmap    
\end{minipage}

\vspace{10pt}
\begin{minipage}[t]{0.24\linewidth}
    \centering
    \small
        \includegraphics[width=28pt]{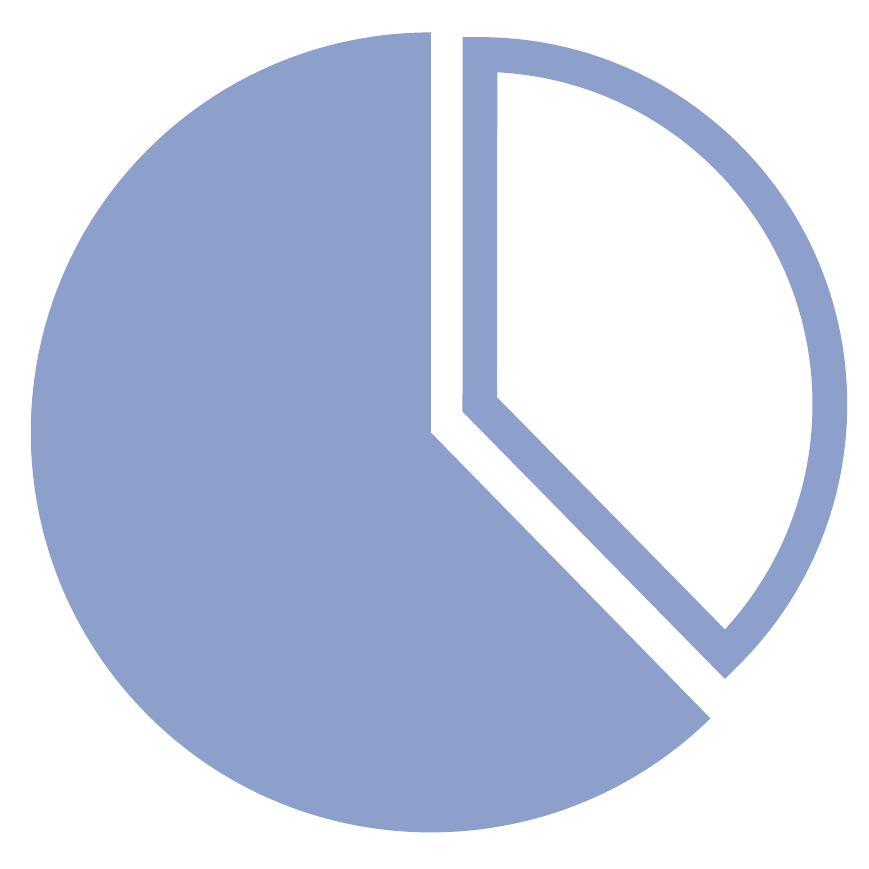}\\
        Pie Chart
\end{minipage}
\begin{minipage}[t]{0.24\linewidth}
    \centering
    \small
        \includegraphics[width=28pt]{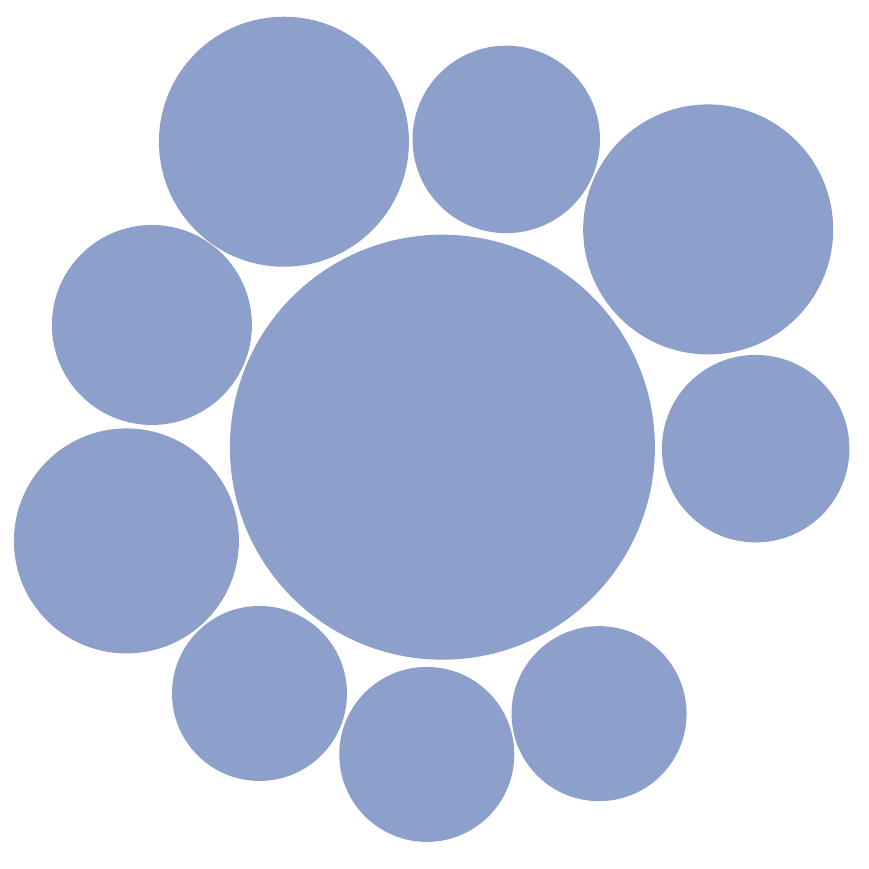}\\
        Bubble Chart
\end{minipage}
\begin{minipage}[t]{0.24\linewidth}
    \centering
    \small
        \includegraphics[width=28pt]{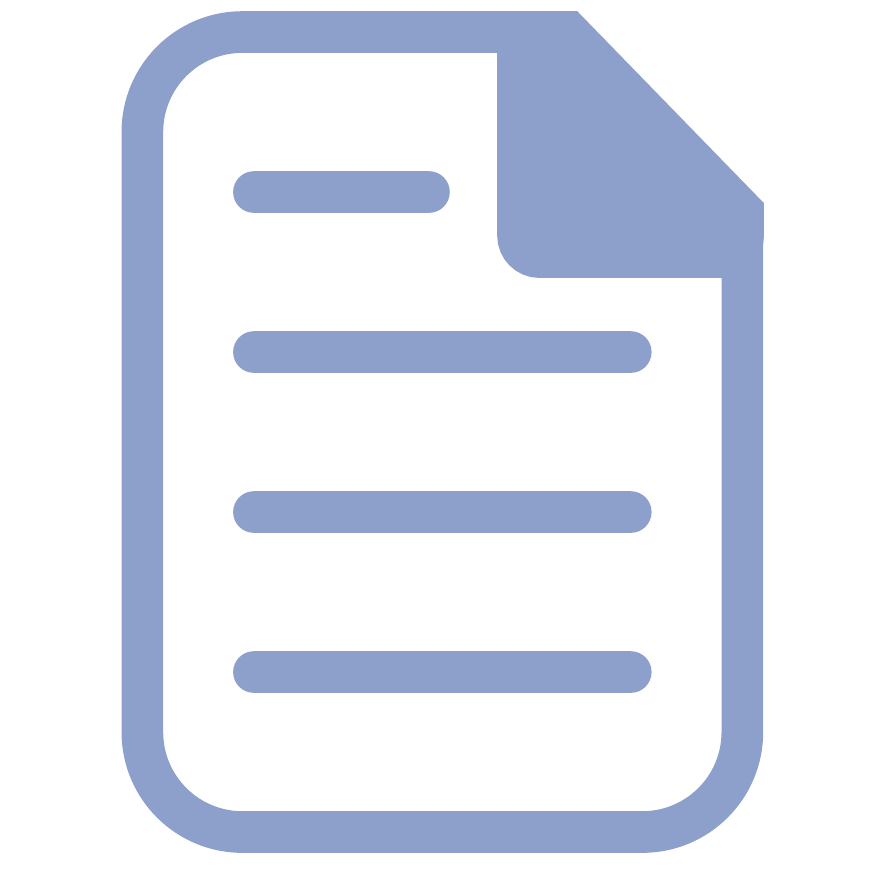}\\
        Text
\end{minipage}
\end{minipage}

\section{Survey}
\label{sec.task}

One of our survey's goals was collecting and framing studies in a way that one could generate effective visual designs considering graphical perception. However, our investigation led us to experiments with limited scope, e.g., focus on a small set of visual encodings, visualization types, limited tasks, or limited data set size and variety. Furthermore, most findings are never replicated or externally validated beyond peer review, with only a few exceptions.

These limitations have two effects. First, when the scope of work is limited, it puts transferability of findings on unsteady ground. E.g., if the real-world conditions associated with a data visualization do not match a contrived scenario, often necessary for studying graphical perception, will the findings be relevant? The second limitation is that finding common links between various studies is quite often difficult. E.g., if two studies using different sized data and slightly different tasks have contradictory results, relating the findings in an actionable way is difficult at best. As such, our summary of works in graphical perception tries to highlight these differences while leaving room for the reader to interpret real-world consequences for their particular situation.

\subsection{Retrieve Value}
\label{sec:task-retrieve}

    \vspace{-5pt}
    \noindent
    \begin{minipage}[t]{\linewidth}
    \begin{wrapfigure}[4]{L}{0.145\linewidth}
        \vspace{-10pt}
        \begin{minipage}[t]{1.2\linewidth}
            \includegraphics[width=45pt]{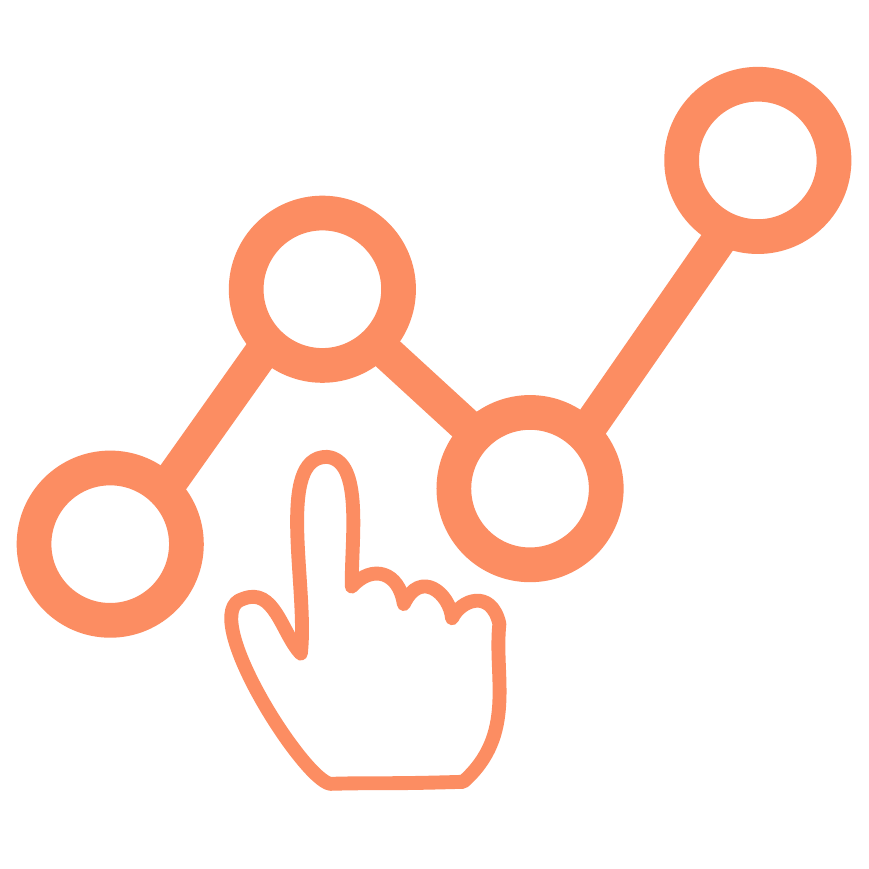}
        \end{minipage}
    \end{wrapfigure} 
    The retrieve value task requires identifying the data or attributes that satisfy a given set of specific criteria. In the literature, retrieve value tasks are often combined with other tasks, e.g., computing a derived value, comparison, sort, etc. We did identify several works that studied the retrieve value task individually with various visualization methods.
    \end{minipage}

\subsubsection{Visual Encoding} \ 

    \vspace{-10pt}
    \begin{wrapfigure}[6]{L}{0.09\linewidth}         \vspace{-5pt}         \begin{minipage}[t]{1.2\linewidth}
        \includegraphics[width=28pt]{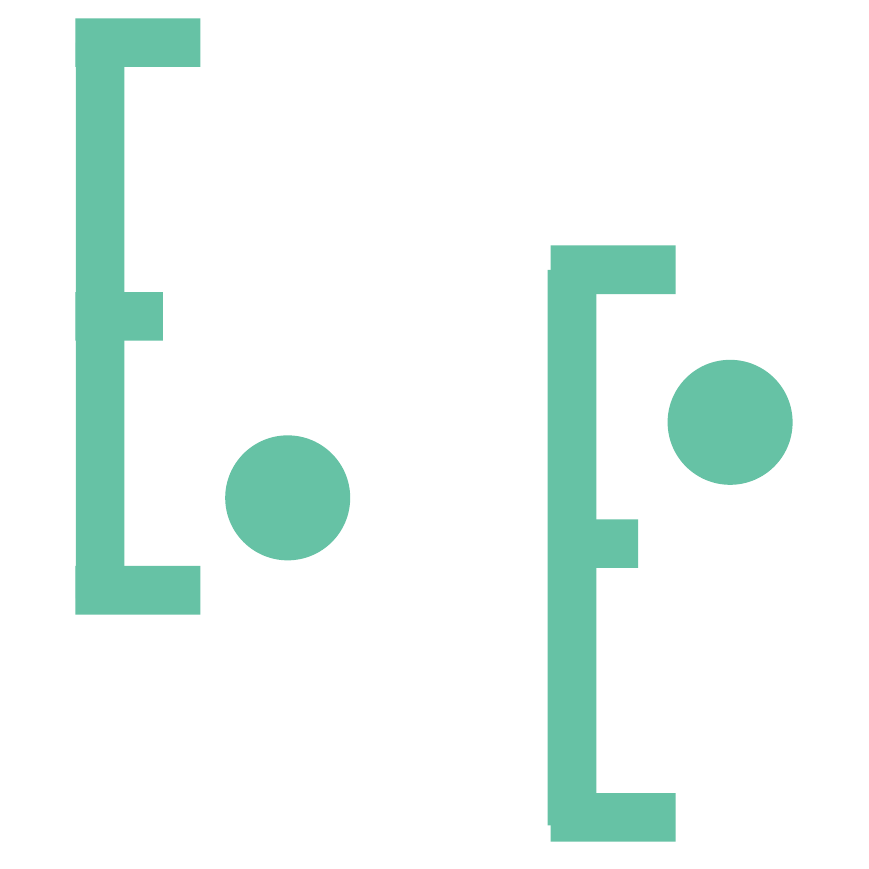}
        
        \vspace{3pt}
        \includegraphics[width=28pt]{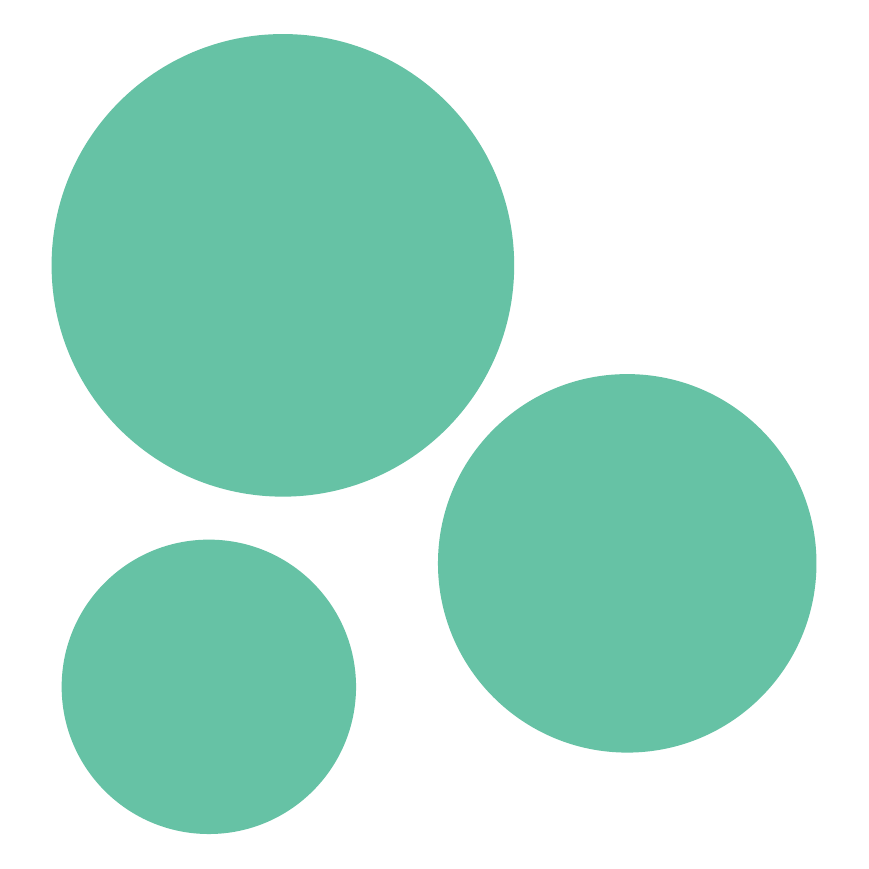}
        \end{minipage}
    \end{wrapfigure} 
    \noindent
    \para{Spatial Position, Shape, \& Size}  Visual factors, such as position along common and unaligned scales, have been observed to produce more accurate judgments than length, direction, and angle on simple tasks~\cite{cleveland1984graphical,mackinlay1986automating} (see \autoref{fig:mackinlay_ranking} for the ranking of other encodings). However, a recent study demonstrated in position-based visualizations, e.g., scatterplots, encoding additional attributes with size required more time for retrieving values than alternatively encoding with color~\cite{kim2018assessing}. On the other hand, subjects' accuracy improved with size. The findings further explored the symmetry between the {size} and {color} as the visual encoding for quantitative information, and they suggested that marks with varied sizes might interfere with decoding the quantity value in position channels. The results further indicated that chart orientation influenced performance by comparing x- and y-faceted charts, with y-faceted charts performing 0.9 times faster.
    Nevertheless, our understanding of how variations in the visualization design potentially influence user performance on this task is evolving.

    \begin{wrapfigure}[3]{L}{0.09\linewidth}         \vspace{-10pt}         \begin{minipage}[t]{1.2\linewidth}
        \includegraphics[width=28pt]{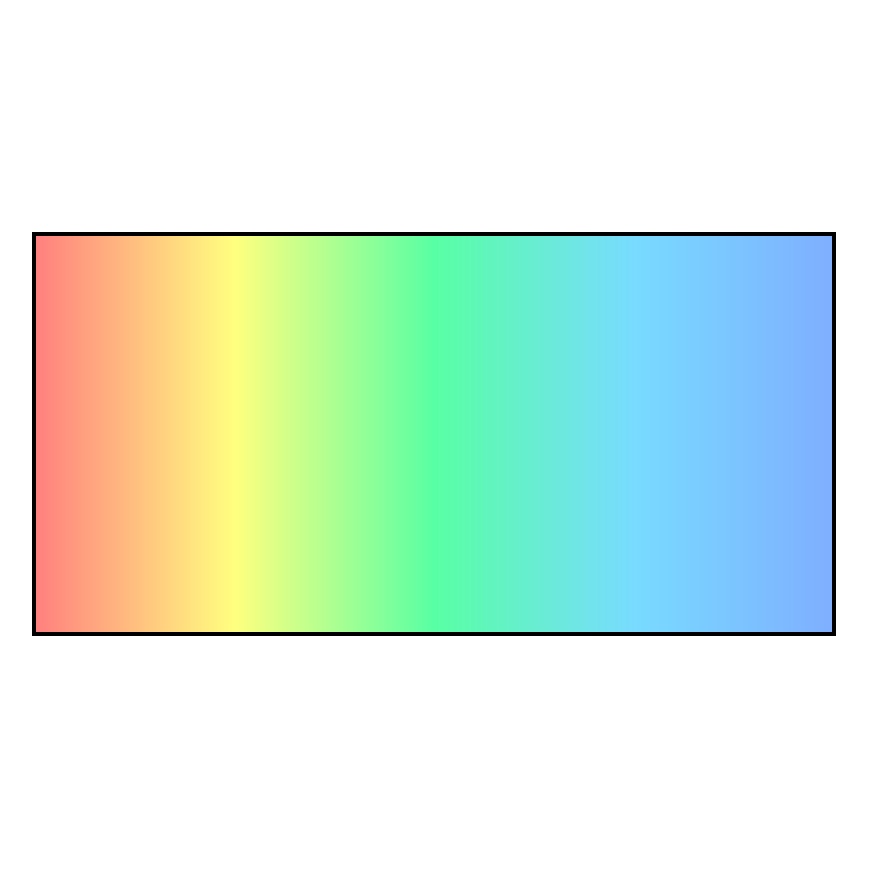}
        \end{minipage}
    \end{wrapfigure} 
    \noindent
    \para{Color Hue} 
    Several studies have evaluated people's ability to identify values using color-coded visualizations. In one instance, the performance in identifying or determining value in visualization was shown to be affected by color and other visual features, such as motion or layout~\cite{haroz2012capacity}. We have identified several studies focused on how color can be applied such that finding targets or reading a value becomes less strenuous. For example, the issue of color vision deficiency, i.e., colorblindness, was studied in automated colormap design~\cite{turton2017crowdsourced}. In the study, colormaps were created starting with a single seed color that was used to generate color ramps that mimic designer practices. Their experiments measured the viewer's potential to accurately identify or read a value in the given color-coded scatterplot, heatmap, or choropleth maps. Ultimately, the performance of their automated system was as good as human-designer-specified color ramps~\cite{smart2019color}. Reasonable precaution is necessary to improve accessibility, i.e., for colorblind, low vision, or other vision impairments, to make the task less challenging and provide an unbiased user response.

    \begin{wrapfigure}[3]{L}{0.09\linewidth}         \vspace{-10pt}         \begin{minipage}[t]{1.2\linewidth}
        \includegraphics[width=28pt]{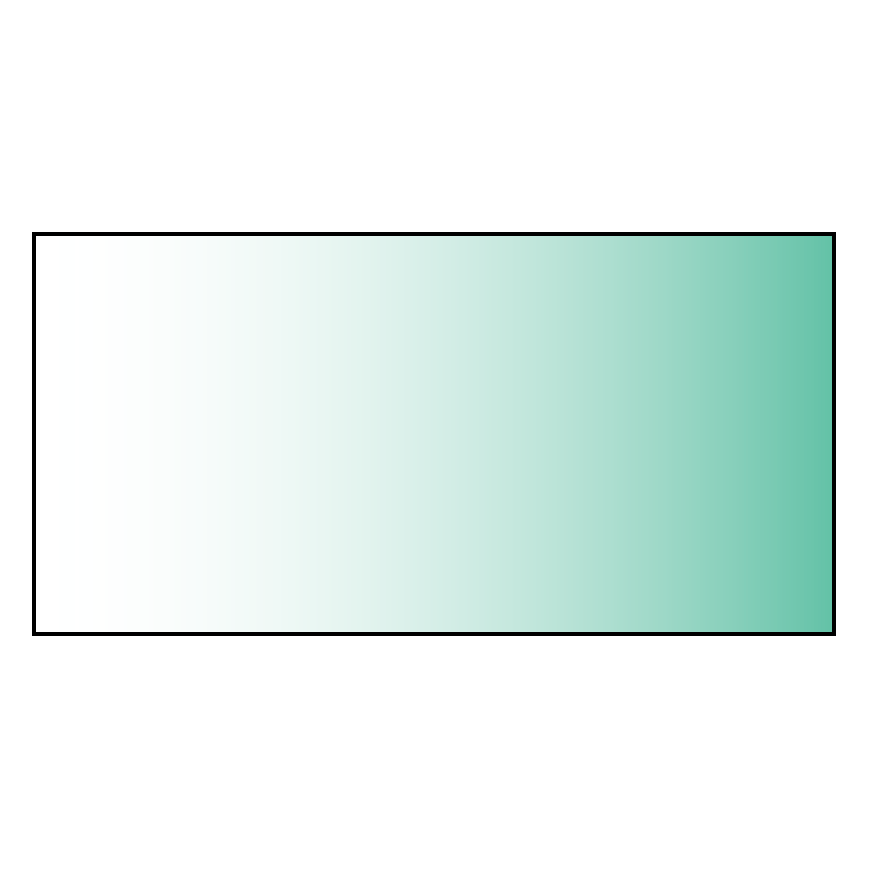}
        \end{minipage}
    \end{wrapfigure} 
    \noindent
   \para{Color Intensity}  One study showed that lightness in the data point symbol could be an impacting factor for visual tasks, such as locating and identifying values in sparse scatterplots on a white background~\cite{li2010lighntness}.  This experiment identified how lightness could be modeled as a combination of two opposite power functions and used to determine discriminability.
   Related to lightness, {opacity} has been used as a constructive factor in a scatterplot design to mitigate overdraw in tasks, such as to read and identify values for higher-level tasks~\cite{l2019toward}.

\begin{figure*}[!t]
    \centering
    
    \includegraphics[width=0.975\linewidth]{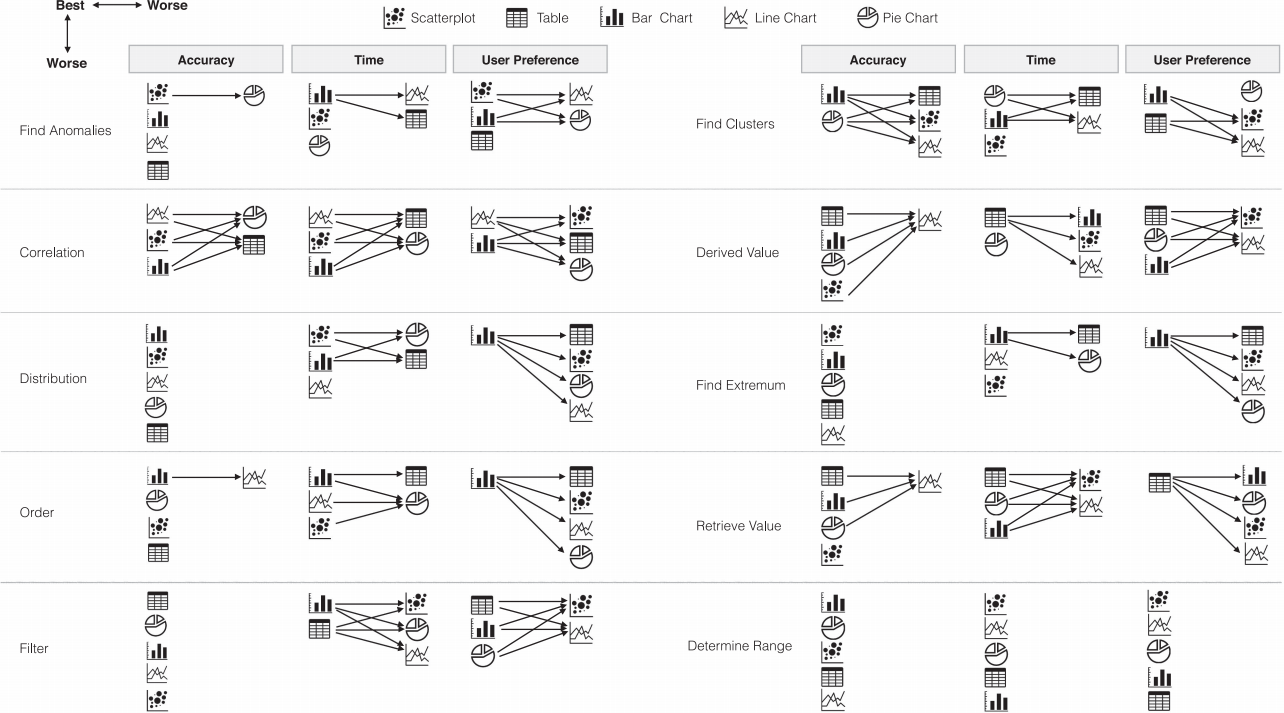}\hspace{15pt}
    
    \caption{A recent study evaluated a pairwise relation between visualization types across tasks and performance metrics on small datasets. Arrows show that the source is significantly better than the target. Image reproduced with permission~\cite{saket2018task}.}
    \label{fig:saket2018task}
    
\end{figure*}

\subsubsection{Visualizations} 

A recent study by Saket et al.~\cite{saket2018evaluating} comparing five visualization types---tabular visualization, scatterplots, bar charts, line charts, and pie chart---evaluated the efficacy of these visualizations with \textit{small data} for all ten of Amar et al.'s tasks~\cite{amar2005low}, in terms of accuracy, time, and user preference (see \autoref{fig:saket2018task}). For the retrieve value task, they found tabular visualization outperformed all others for accuracy, time, and preference, bar and pie charts performed well in terms of accuracy and time, and scatterplots performed well only in terms of accuracy. A similar comparative study of three multidimensional visualizations---parallel coordinates, scatterplot matrix, and tabular visualization---studied their user performance on analytical and decision-making tasks. For the retrieve value task, parallel coordinates had the highest accuracy. The evaluation also demonstrated that tabular visualization was familiar, accurate, and time-efficient for the retrieve value task~\cite{dimara2018conceptual}. 
In both studies, tabular visualizations allowed subjects to reach decisions faster with better accuracy levels than the other visualizations because of their familiarity with tables.

    \begin{wrapfigure}[3]{L}{0.09\linewidth}         \vspace{-10pt}         \begin{minipage}[t]{1.2\linewidth}
        \includegraphics[width=28pt]{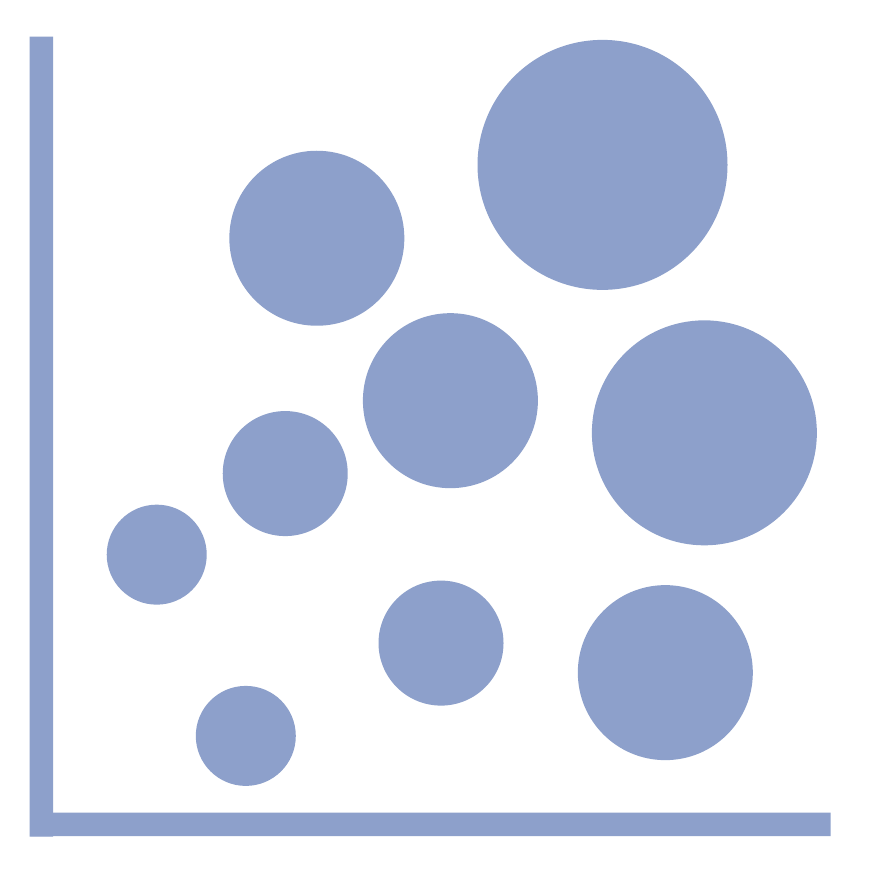}
        \end{minipage}
    \end{wrapfigure} 
    \noindent
    \para{Scatterplot} A scatterplot is one of the most effective forms of visualization, which allows users to identify, read, or retrieve a value based on different visual encodings. For example, scatterplots colored with additional categorical data were found to be highly effective for comparing individual data and were preferred over dot plots~\cite{kim2018assessing}.

    \begin{wrapfigure}[3]{L}{0.09\linewidth}         \vspace{-10pt}         \begin{minipage}[t]{1.2\linewidth}
        \includegraphics[width=28pt]{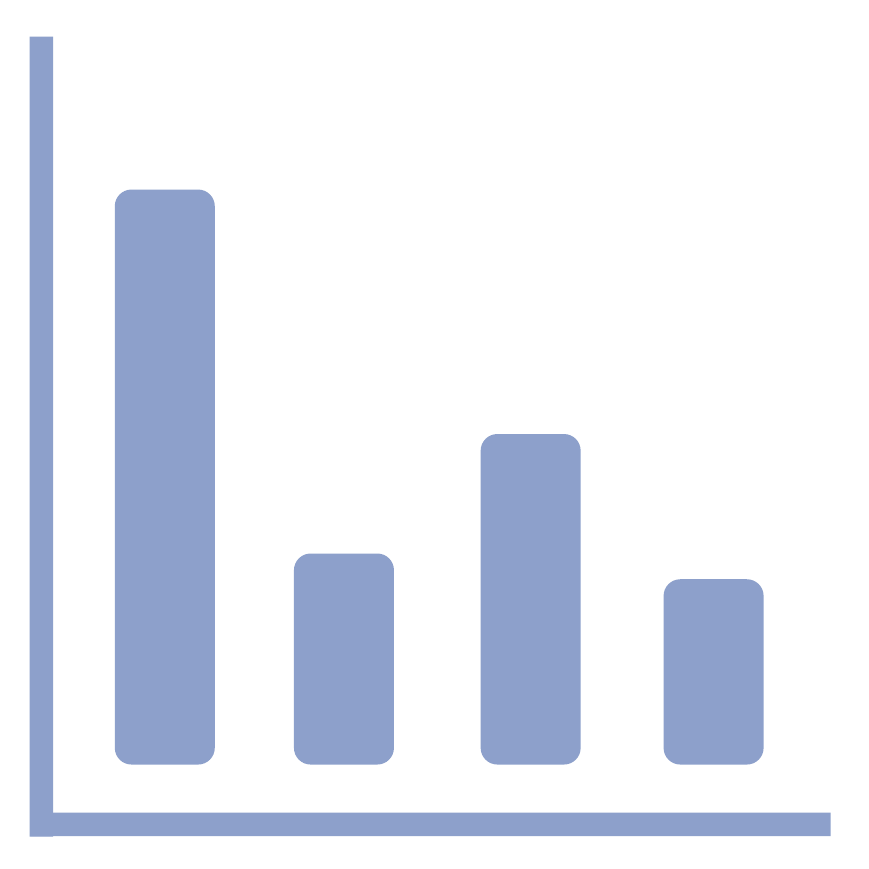}
        \end{minipage}
    \end{wrapfigure} 
    \noindent
    \para{Bar chart} While bar charts represent information in a way that helps the user identify and read quantitative data easily, the design decisions used play an essential role in user performance. In one study, pictorial bar charts were found to reduce the user performance on retrieving a value task but not beyond the effect already observed based upon their shape~\cite{skau2017readability}. A recent study focusing on a ``reading a value'' task concluded that there is a need for direct encoding of absolute values and their relationship in bar charts or dot plots~\cite{nothelfer2019measures}. Ultimately, while embellishments are used to improve aesthetics and memorability, they can affect user performance on the task.

    \begin{wrapfigure}[3]{L}{0.09\linewidth}         \vspace{-10pt}         \begin{minipage}[t]{1.2\linewidth}
        \includegraphics[width=28pt]{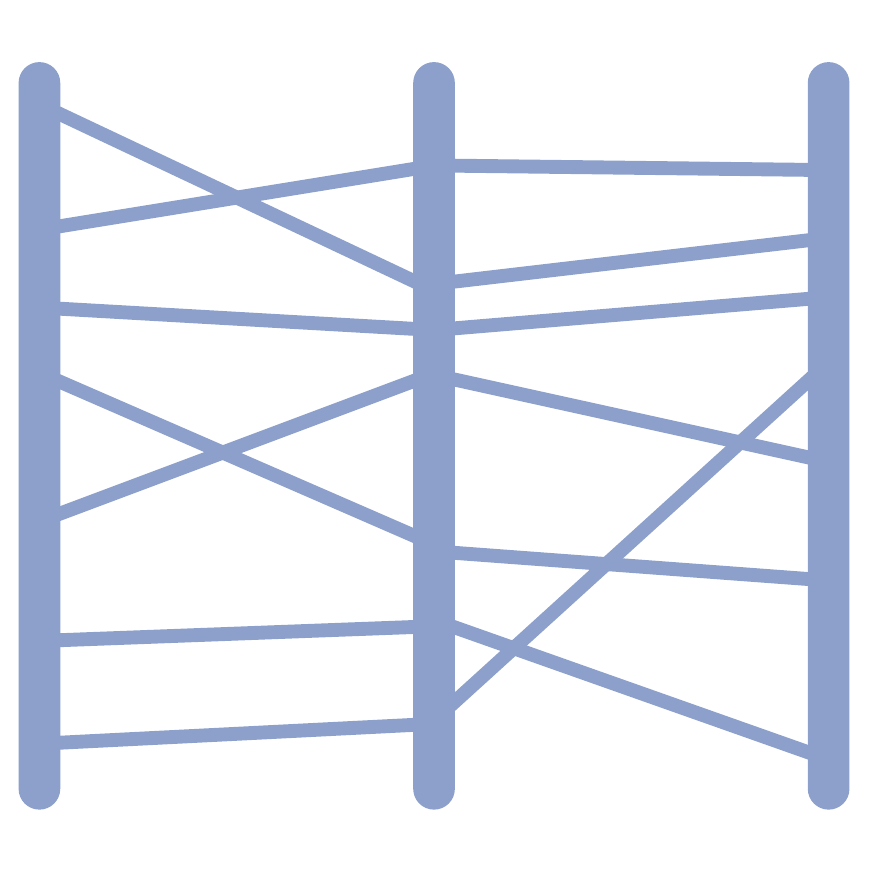}
        \end{minipage}
    \end{wrapfigure} 
    \noindent
    \para{Parallel Coordinates} Parallel coordinates are often disparaged as being challenging to use and understand for first-time users. However, one study used the retrieving a value task to show otherwise---the experiment assessed the task using eye-tracking, and it indicated that first-time users quickly learned to use them~\cite{siirtola2009visual}. The study plotted eight vehicle attributes as axes for 406 cars and used eye-tracking to measure the user's performance on identifying the values in the data.

    \begin{wrapfigure}[3]{L}{0.09\linewidth}         
    \vspace{-10pt}         
    \begin{minipage}[t]{1.2\linewidth}
        \includegraphics[width=28pt]{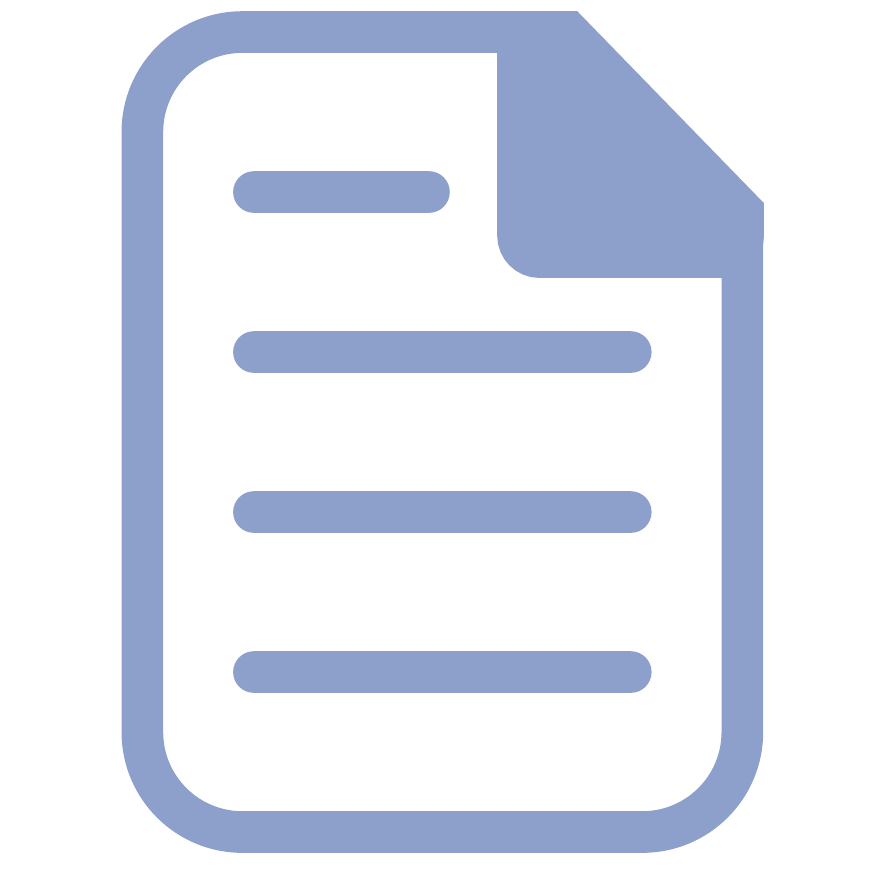}
        \end{minipage}
    \end{wrapfigure} 
    \noindent
    \para{Text} In visualizations where text is involved, visual complexity affects a user's ability to grasp aspects of the overall structure of the visual display.  Optimizing the use of typography in visualizations, e.g., the location of the labeling and parameters including typeface, font size, font weight, color, orientation, intensity (boldness), spacing, case, border, background, underline, and shadow, determines the legibility of text and, thus, inﬂuences the understandability of the visualization~\cite{strobelt2015guidelines}. One experiment indicated the biases caused by word length, height, and width could impact user accuracy, and the findings provide practical guidelines for improving the user experience~\cite{alexander2018perceptual}. Therefore, the design of text in a visualization should consider the word typography for any potential biases.

    \begin{wrapfigure}[3]{L}{0.09\linewidth}         \vspace{-10pt}         \begin{minipage}[t]{1.2\linewidth}
        \includegraphics[width=28pt]{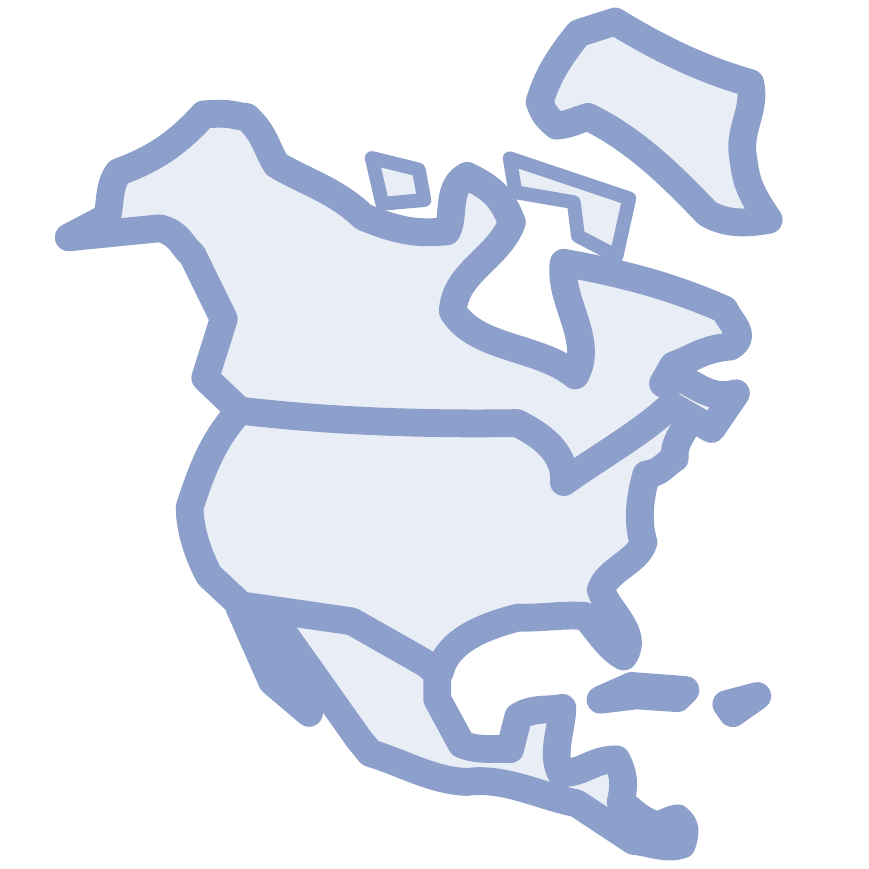}
        \end{minipage}
    \end{wrapfigure} 
    \noindent
    \para{Map} 
    We use larger and higher resolution displays to increase the scalability of visualizations, particularly when data is small, such as in maps. By sheer size, such displays could make it more difficult to attend to the visualization. A perceptual scalability study on maps investigated user performance focusing on retrieving or reading a value task in a larger display~\cite{yost2006perceptual}. They found that the larger display did \textit{not} result in a time increase or accuracy decrease.

\subsubsection{Summary} In summary, we identified several studies using the retrieve value task scattered over various visualization types. As one would expect, user performance varies with the choice of visualization and the visualization design. Prior work indicates that bar charts should be the first choice when accuracy is important. Thoughtful use of embellishments can improve aesthetics and memorability, but they can also affect user performance on the task.

\subsection{Filter}

\label{sec:task-filter}

    \vspace{-5pt}
    \noindent
    \begin{minipage}[t]{\linewidth}
    \begin{wrapfigure}[4]{L}{0.145\linewidth}
        \vspace{-10pt}
        \begin{minipage}[t]{1.2\linewidth}
            \includegraphics[width=45pt]{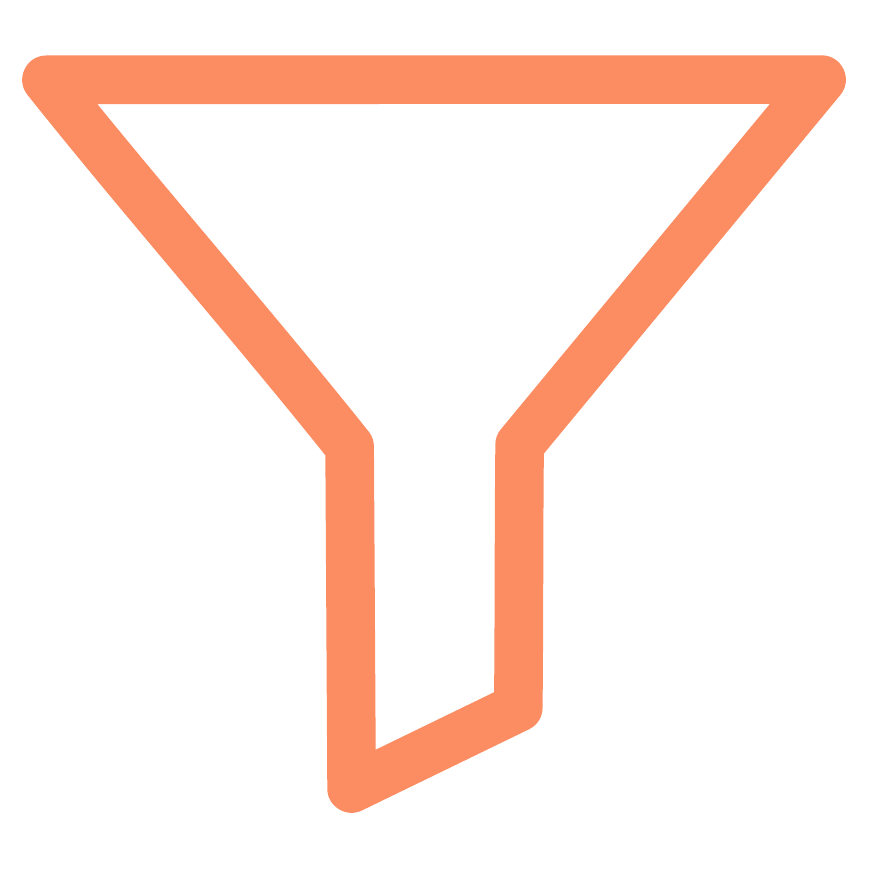}
        \end{minipage}
    \end{wrapfigure} 
    In our search of the literature, filtering essentially came down to two types of tasks. The first, what we would typically call filtering, encompassed eliminating subsets of data used in the visualization. The second was a search task, which focused on finding a target in the visualization.  
    \end{minipage}

    \subsubsection{Visual Encoding} \ 

    \vspace{-10pt}
    \begin{wrapfigure}[6]{L}{0.09\linewidth}         \vspace{-6pt}         \begin{minipage}[t]{1.2\linewidth}
        \includegraphics[width=28pt]{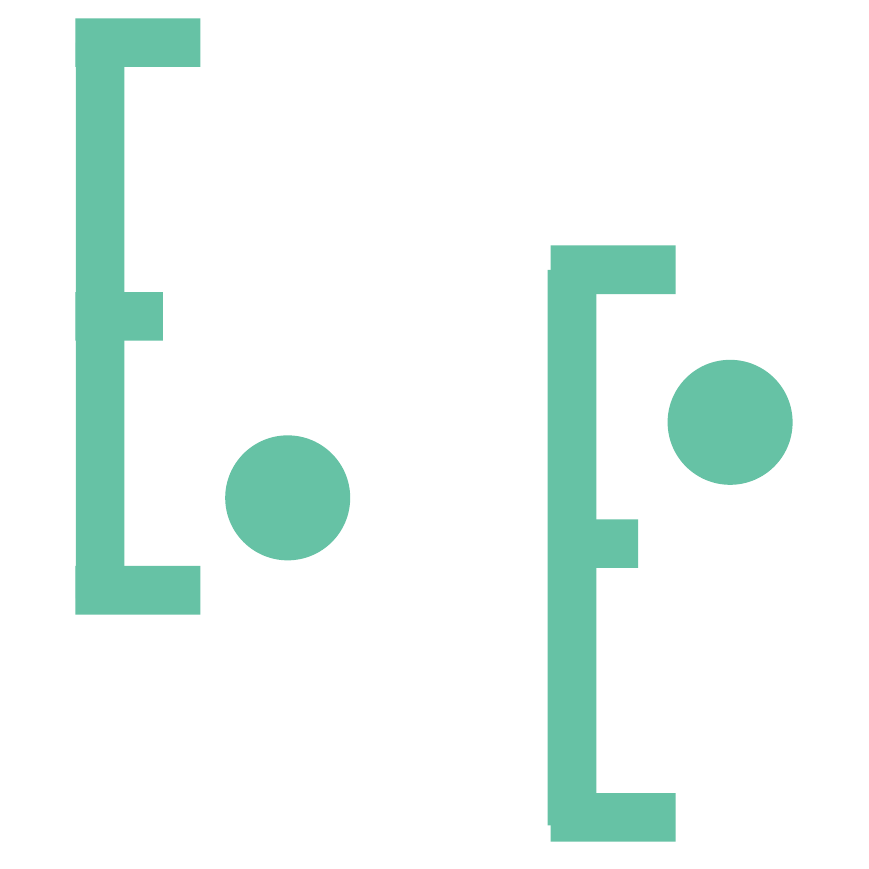}
        
        \vspace{3pt}
        \includegraphics[width=28pt]{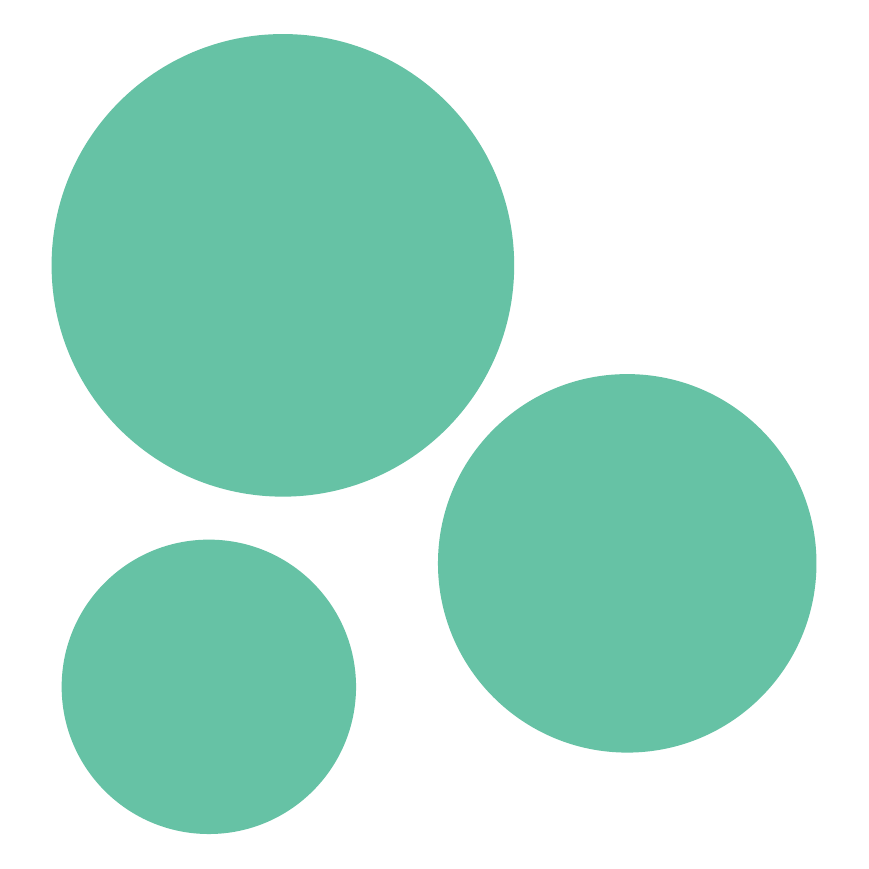}
        \end{minipage}
    \end{wrapfigure} 
    \noindent
    \para{Spatial Position, Shape, \& Size} Popout, a pre-attentive effect caused by variations in certain encodings, makes it easy to draw the user's attention to the critical elements of the visualization.  A study examined and demonstrated the efficacy of popout in target identification in a group of symbols in a scatterplot using different visual channels, including color, shape, luminance, flashing, motion, and size~\cite{gutwin2017peripheral}. On three metrics of performance---perceived success, visibility, and accuracy---shape required high effort with a low performance, whereas motion showed little effort and a high level of performance. The remaining four channels showed a definite increase in perceived visibility and accuracy across intensity. 
    
    A similar experiment investigated searching for a target in a grid matrix using three visual channels---mark size, set size, and color---suggested design guidelines based on grouping, quantity set, and size of visual symbols~\cite{gramazio2014relation}. While searching for an item was faster when colors were spatially grouped in the grid, the number of symbols had little effect on search time. At the same time, if the number of symbols increased, the performance slowed for random data displays.

    \begin{wrapfigure}[3]{L}{0.09\linewidth}         \vspace{-10pt}         \begin{minipage}[t]{1.2\linewidth}
        \includegraphics[width=28pt]{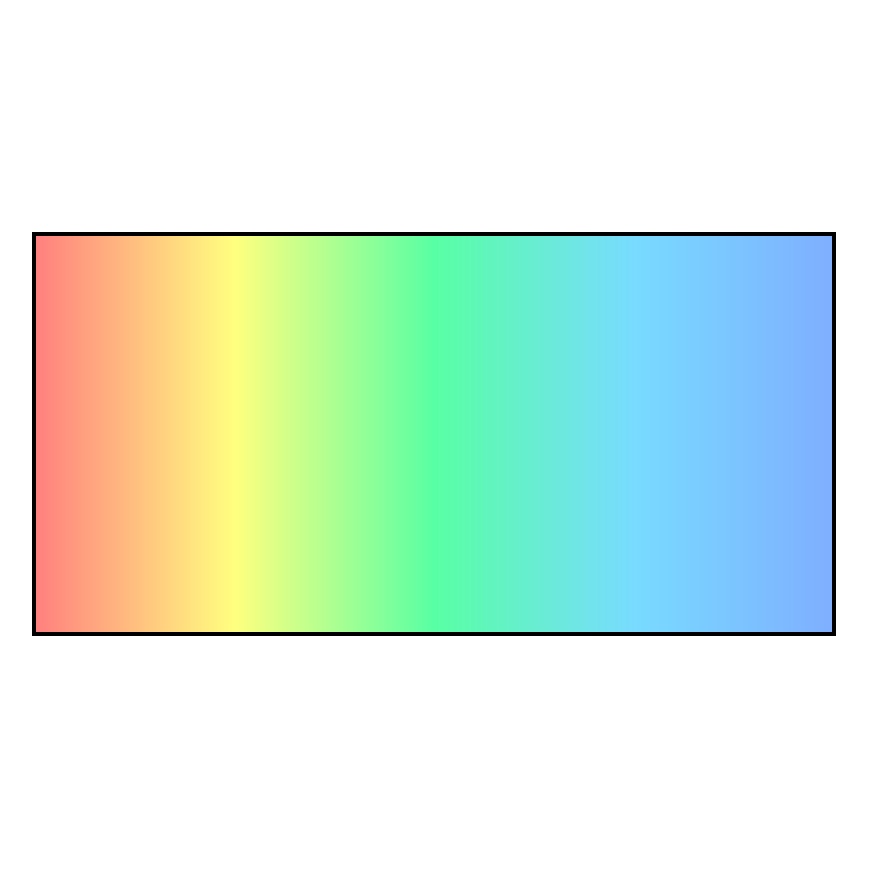}
        \end{minipage}
    \end{wrapfigure} 
    \noindent
    \para{Color Hue} Color encoding optimization can increase the effectiveness of categorical data visualization. While categorical colors are easily distinguishable in large color bars or individual plots, when they are inserted into maps or other plots with varying sizes, the variation reduces the visibility of categorical differences. {Class visibility} is an important measure of how color and spatial distribution of each class affect its perceptual intensity to the human visual system~\cite{lee2013perceptually}.

    \subsubsection{Visualizations} 
    
    Filtering task performance highlights the need to explicitly and directly encode numeric differences between data values~\cite{nothelfer2019measures}. In an experiment with image visualization, finding an image within a set of images showed that both latency and task complexity play a significant role in search behavior~\cite{battle2019role}. In Saket et al.'s study on small data~\cite{saket2018task}, they found tabular visualizations and bar charts performed well in terms of accuracy, time, and preference on filtering tasks (see \autoref{fig:saket2018task}).

    \begin{wrapfigure}[3]{L}{0.09\linewidth}         \vspace{-10pt}         \begin{minipage}[t]{1.2\linewidth}
        \includegraphics[width=28pt]{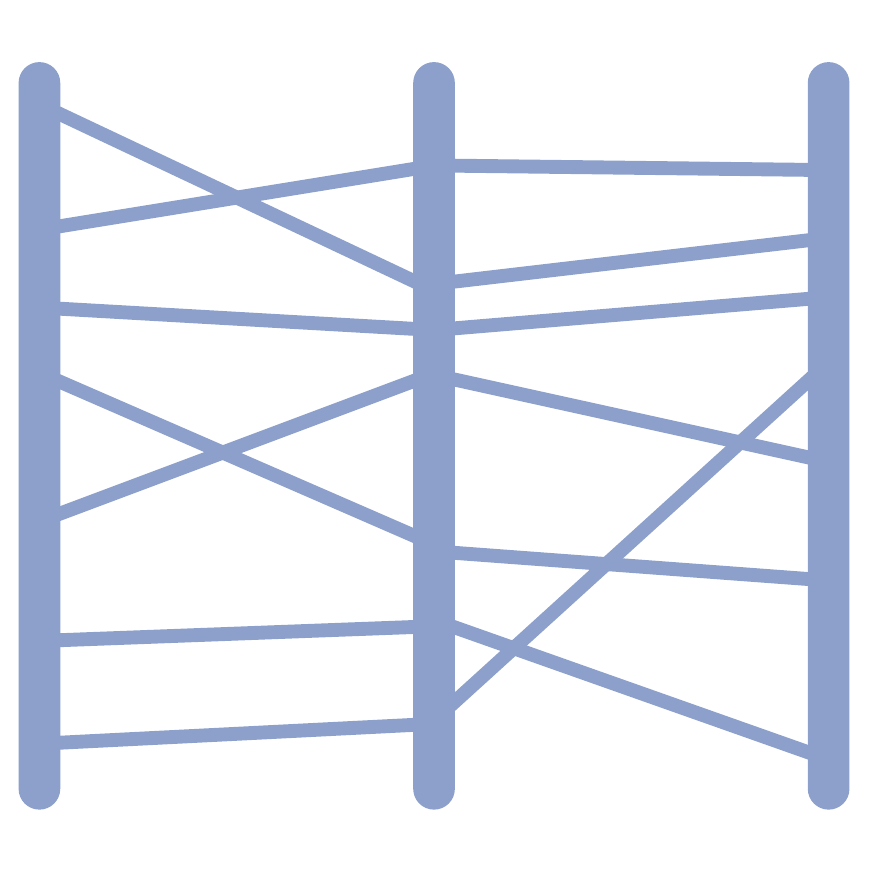}
        \end{minipage}
    \end{wrapfigure} 
    \noindent
   \para{Parallel Coordinates} In a decade-old experiment, a filtering task was used to assess the usability of parallel coordinates. The performance of the filtering task using eye-tracking showed that users had high fixation time in order to confirm their interpretation of the results~\cite{siirtola2009visual}. The study showed that the effect of latency in an interactive visual system is more gradual than binary.

   \begin{wrapfigure}[3]{L}{0.09\linewidth}        \vspace{-10pt}        
   \begin{minipage}[t]{1.2\linewidth}
        \includegraphics[width=28pt]{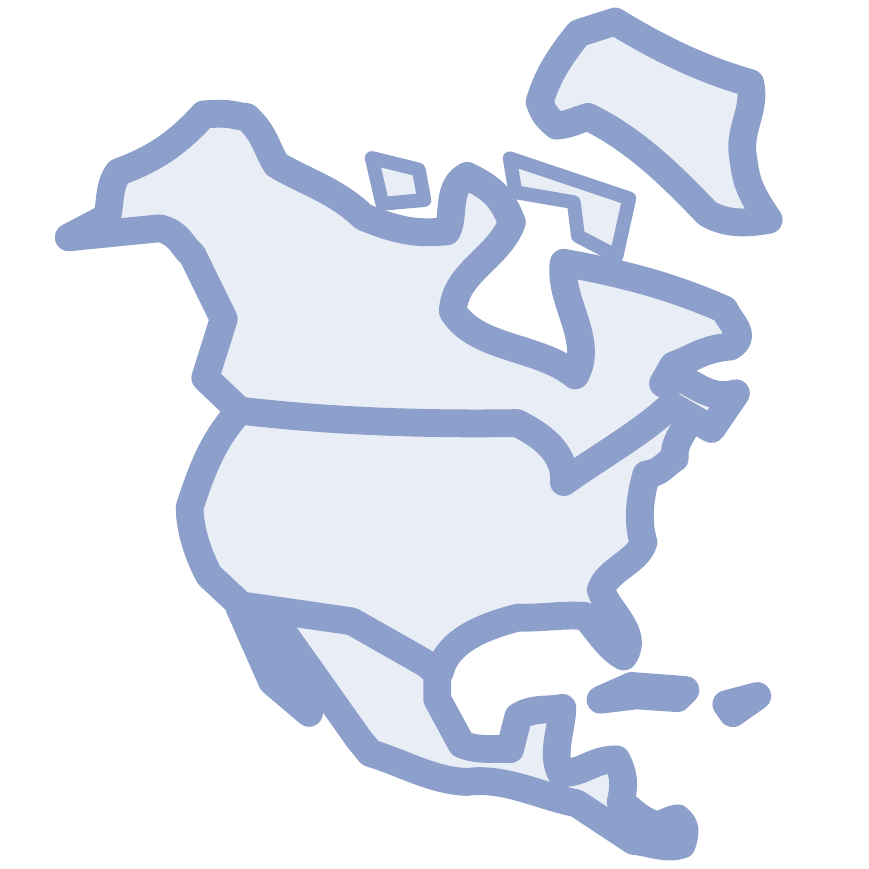}
        \end{minipage}
    \end{wrapfigure} 
    \noindent
   \para{Map} Maps are used to represent structures, patterns, and relations in spatial data. Researchers have worked on a variety of map types, e.g., choropleth~\cite{beecham2017map}, cartogram~\cite{nusrat2018evaluating}, geographical map~\cite{lee2013perceptually}, typographic maps~\cite{afzal2012spatial}, etc., to evaluate the human perception in identifying the pattern and structures, and optimize the design process based on the encoding method. As pointed out earlier, {class visibility} has been studied on maps and shown how color and the spatial distribution can influence perceptual intensity~\cite{lee2013perceptually}.
   Filter and search tasks on maps are affected by visual encoding choices, as well as the physical construction of maps.
   Cartograms were evaluated on the four classes---contiguous, non-contiguous, rectangular, and Dorling---using a qualitative performance analysis~\cite{nusrat2018evaluating}. Cartograms that preserve the relative position of the regions in the geographical maps, facilitating faster search, while Dorling and rectangular maps had better accuracy.

    \begin{figure}[!t]
        \centering
        \includegraphics[width=0.975\linewidth]{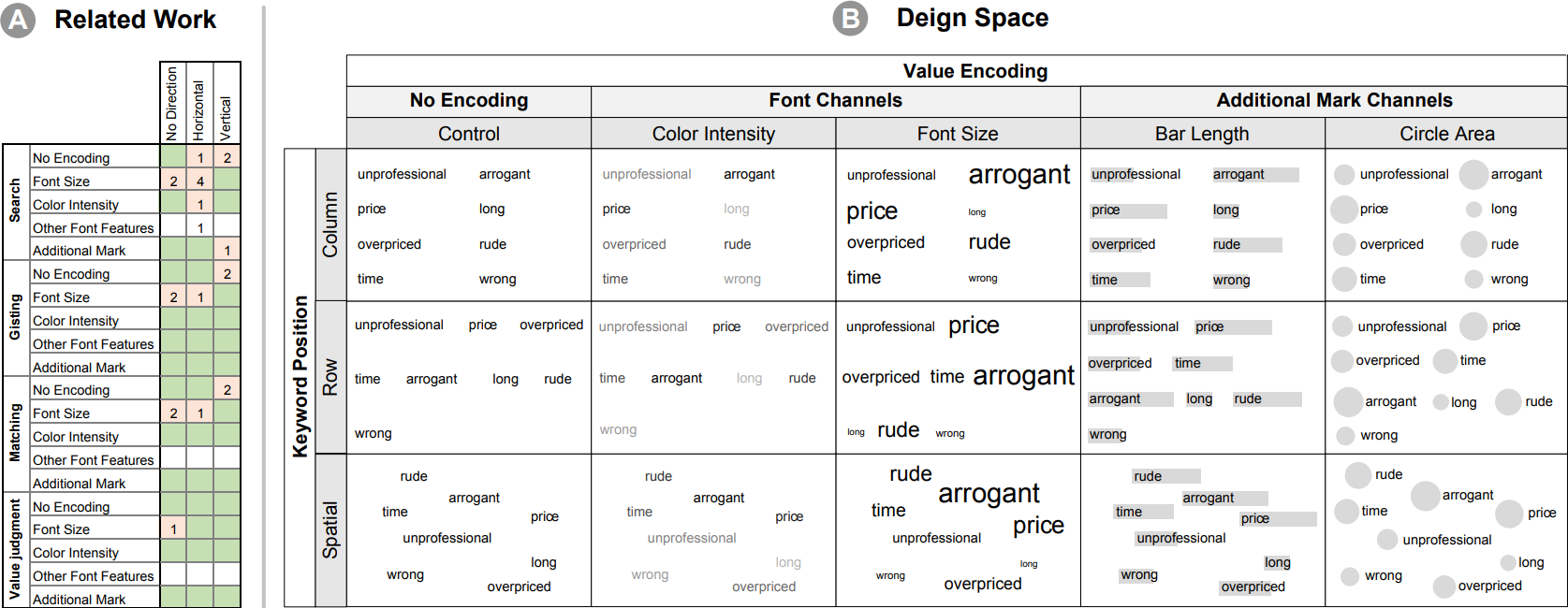}
        \caption{Different representations affect people’s performance in extracting information from text visualizations, such as {word clouds}. This design space summary shows examples of visualizations generated by different visual encodings and layouts. Image reproduced with permission~\cite{felix2018taking}.}
        \label{fig:felix2018taking}
        
    \end{figure}

 \begin{wrapfigure}[3]{L}{0.09\linewidth}         \vspace{-10pt}         \begin{minipage}[t]{1.2\linewidth}
        \includegraphics[width=28pt]{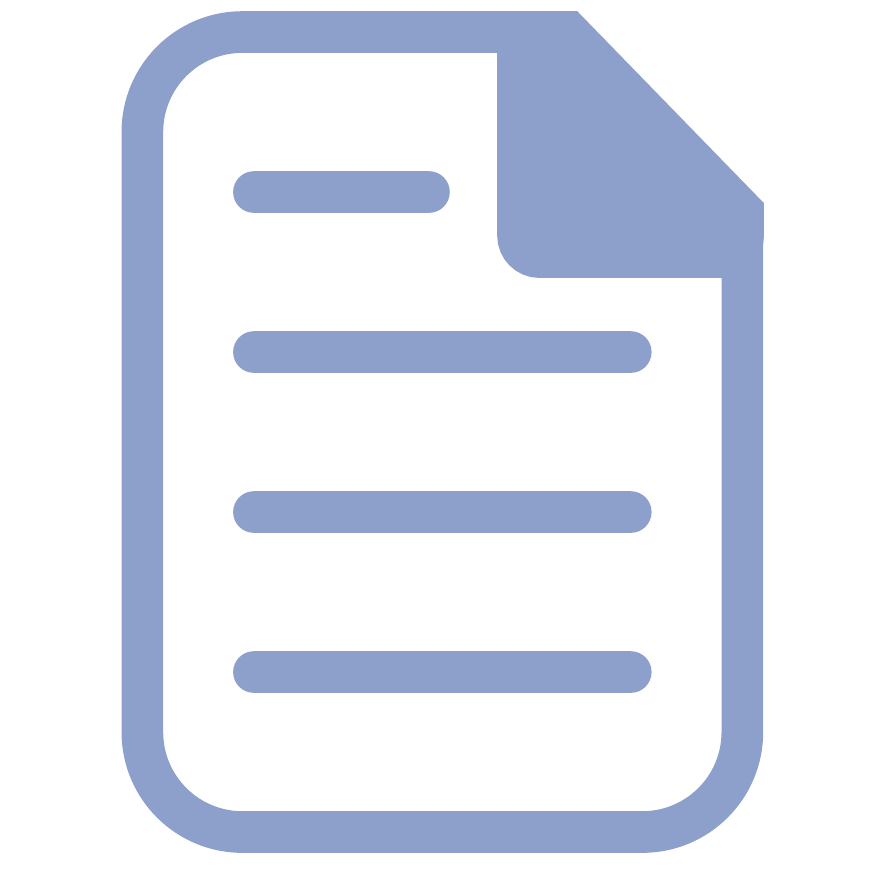}
        \end{minipage}
    \end{wrapfigure} 
    \noindent
   \para{Text} 
    The optimal setting of typography parameters determines the legibility of text, thus, inﬂuences the understandability of visualizations. In a word cloud study focused on different layout and encodings effects, font size and color performed well compared to encodings using additional marks, such as bars and circles on keywords, in a task of searching for a keyword~\cite{felix2018taking} (see \autoref{fig:felix2018taking}). Furthermore, spatial and column layout of the keywords similarly outperformed the row version of keyword arrangement on keyword searches. A similar study on text confirmed the prior results by showing how searching for target words could be influenced by font sizes~\cite{alexander2018perceptual}.

    Another study evaluated the effects of four types of layouts, and the results showed that perhaps unsurprisingly, alphabetic ordering was faster for searching for a keyword, order and font size were found to have no impact on searching for a tag belonging to an assigned topic, and words with bigger fonts were more likely to be recalled~\cite{schrammel2009semantically}. This confirmed a prior study that investigated word recall based on classic word clouds with unordered layouts and font size encoding frequency. In a study, participants recalled the words with larger font sizes more often. Their guidelines considered optimizations based on typography (i.e., font weight, size, and color) and word placement (i.e., sorting, clustering, and spatial layout)~\cite{rivadeneira2007getting}. 
    In addition to layout and font size, good highlighting mechanisms facilitate search and comparison in textual views or labels. Another study empirically investigated the effective use of highlighting techniques for visualization applications for text data and suggested design guidelines for the effectiveness of nine web-friendly text highlighting techniques~\cite{strobelt2015guidelines}. In summary, search tasks on the text and word clouds benefit from the thoughtful selection of size, layout, ordering, and highlight.

    \subsubsection{Summary} In summary, the filter task is one in which user performance varies based on the design of visualization, choice of visualization, and related set of tasks being performed. Tabular visualizations and bar charts perform well in terms of accuracy, time, and preference. Searching for a keyword in text or a value in a scatterplot can benefit from utilizing encodings font/symbol size and color, as well as alphabetic ordering for text.

\subsection{Compute Derived Value}
\label{sec:task-compute}

    \vspace{-5pt}
    \noindent
    \begin{minipage}[t]{\linewidth}
    \begin{wrapfigure}[4]{L}{0.145\linewidth}
        \vspace{-10pt}
        \begin{minipage}[t]{1.2\linewidth}
            \includegraphics[width=45pt]{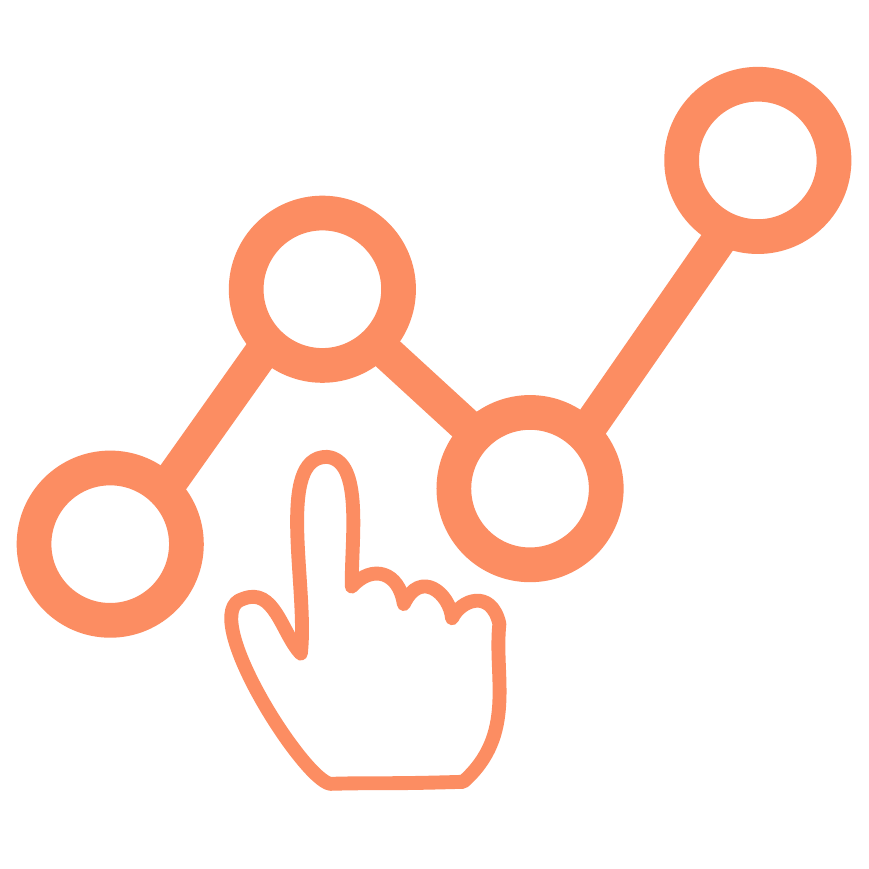}
        \end{minipage}
    \end{wrapfigure} 
    Given a set of data, computing derived values is similar to calculating an aggregate of that data. Many visualization tasks require users to create an aggregated abstraction or statistical summary, which are often referred to as visual aggregations.
    \end{minipage}

\subsubsection{Visual Encoding} \ 
   
   \vspace{-10pt}
    \begin{wrapfigure}[6]{L}{0.09\linewidth}         \vspace{-5pt}         \begin{minipage}[t]{1.2\linewidth}
        \includegraphics[width=28pt]{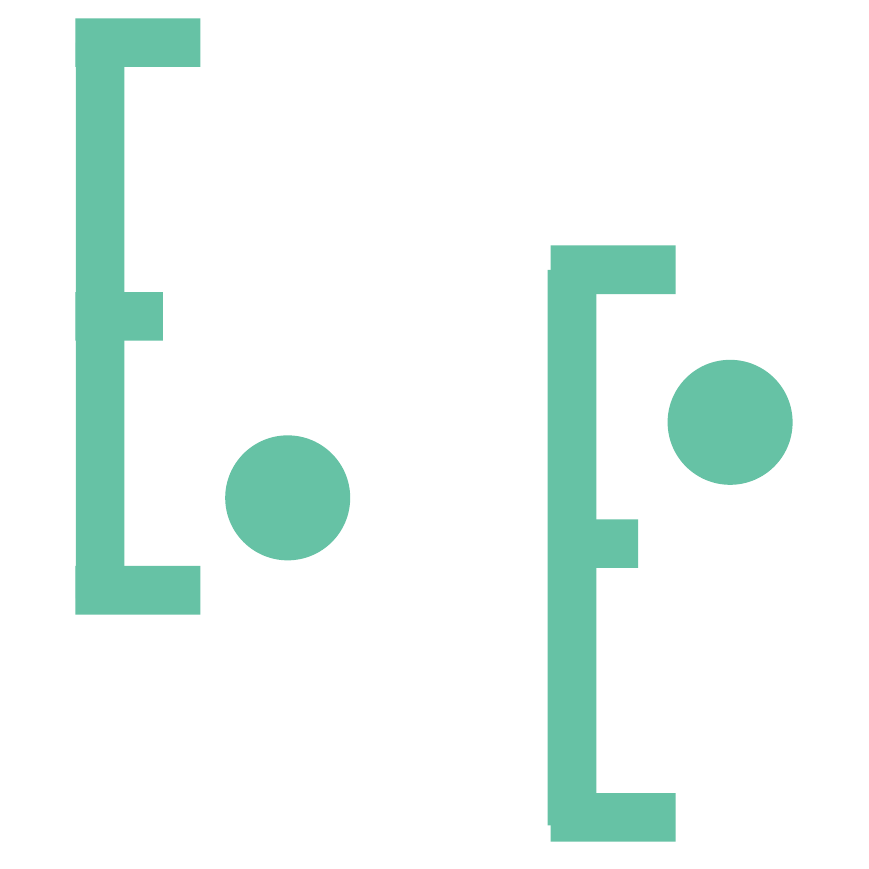}
        
        \vspace{3pt}
        \includegraphics[width=28pt]{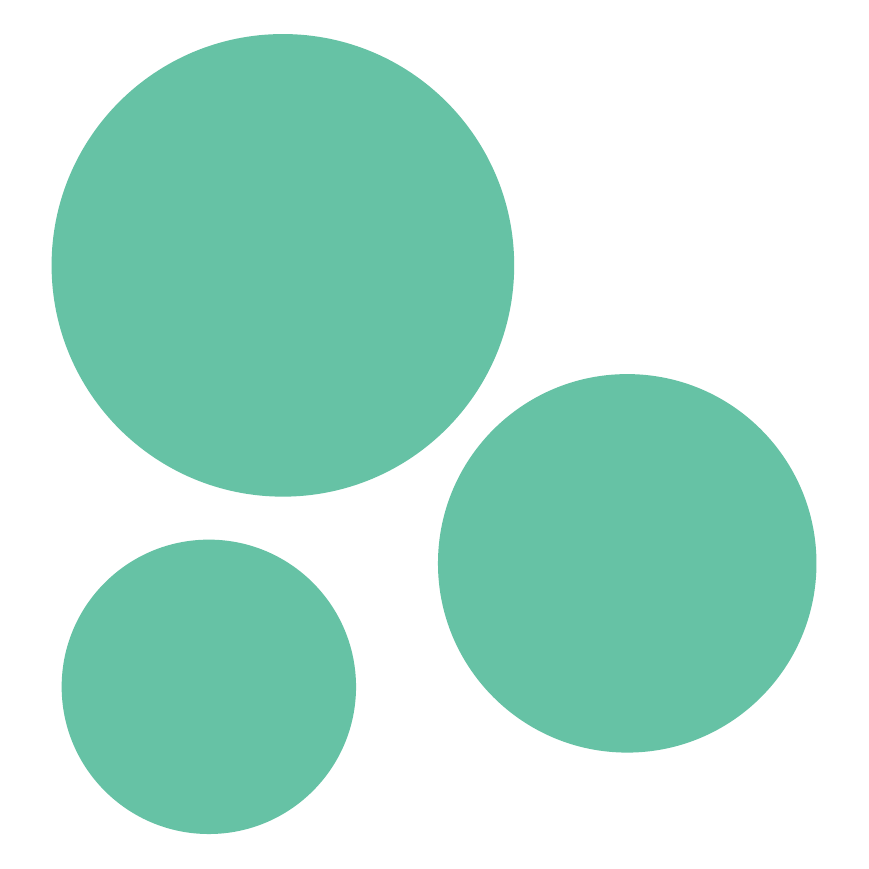}        
        \end{minipage}
    \end{wrapfigure} 
    \noindent
    \para{Spatial Position, Shape, \& Size} 
    Cleveland and McGill's seminal work on visual perception utilized the task of computing derived values from the visual representation~\cite{cleveland1984graphical}. 
    In their results, position and length were among the visual encodings with the highest accuracy in quantitative judgment. That study was replicated on a crowdsourced platform using proportional judgment tasks to confirm the effectiveness rankings of visual encoding~\cite{heer2010crowdsourcing}. 
    A recent study, concentrating on various channels in scatterplots, suggested point size encoding performs well for summary tasks on quantitative encoding~\cite{kim2018assessing}.
    
    Another experimental study investigated the estimation of average position in line and bar charts~\cite{xiong2019biased}. Even though positions are considered a precise form of data encoding, reports of average positions were biased because of underestimation or overestimation of bar positions due to the introduction of a bias called \textit{perceptual pull}---position estimates for a target data were ``pulled’' toward the irrelevant data in the series.

    \begin{wrapfigure}[5]{L}{0.09\linewidth}         
    \begin{minipage}[t]{1.2\linewidth}
        \includegraphics[width=28pt]{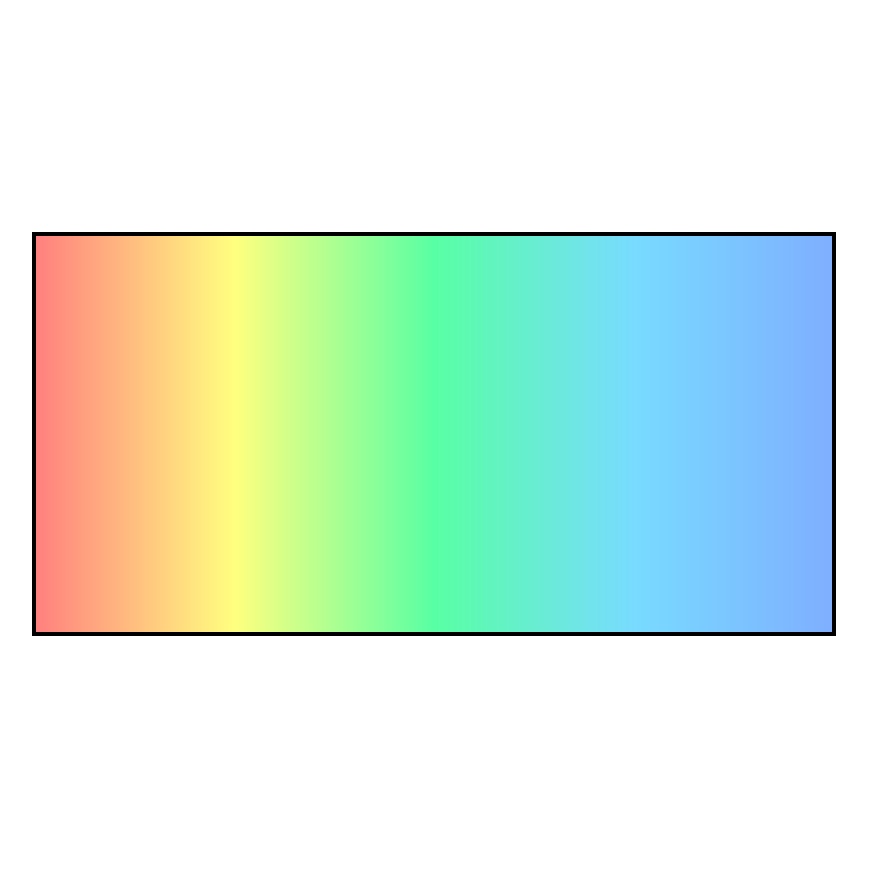}
        
        \vspace{-3pt}
        \includegraphics[width=28pt]{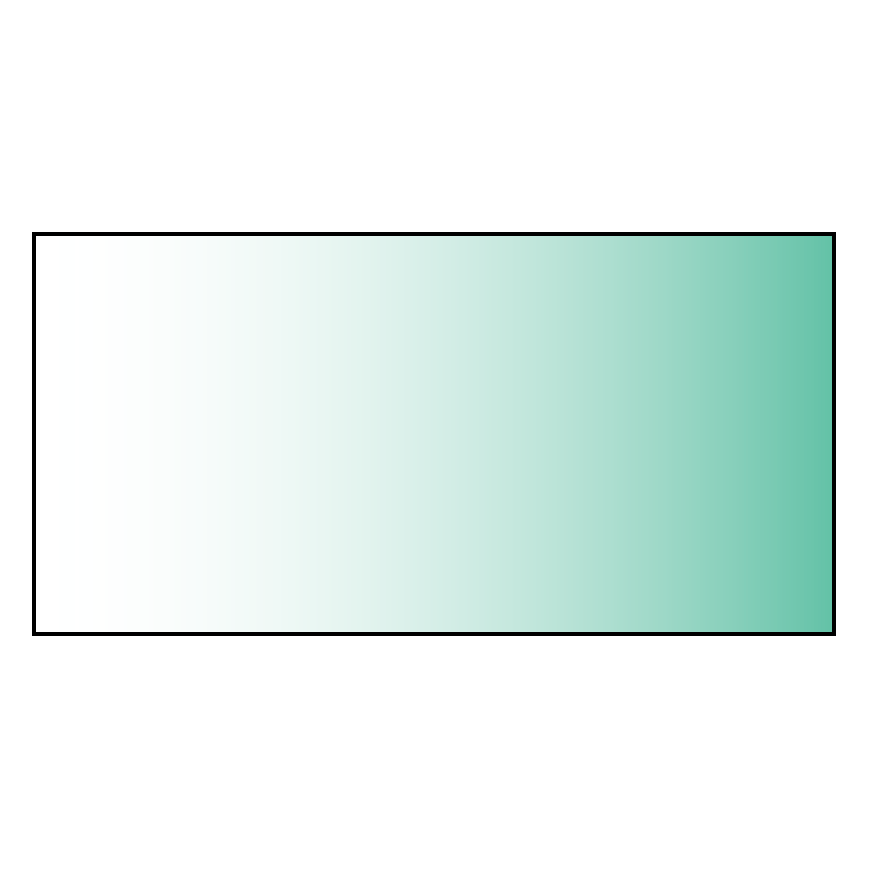}
        \end{minipage}
    \end{wrapfigure} 
    \noindent
    \para{Color Hue \& Intensity} Quite often, colormaps are applied to data, and aggregation tasks are performed. Studies have demonstrated the perception of continuous colormaps is affected by colormap characteristics and spatial frequency. Estimating the values based on colormaps in a continuous quantitative colormap showed no relation between colormaps and spatial frequency, but with increased spatial complexity, estimation error increases~\cite{reda2018graphical}. The important point is that spatial frequency impacts the effectiveness of color encoding, but the true impact is task-dependent. 
    
    Similarly, another study assessed the efficacy of colormaps for encoding scalar information using binary-choice experiments~\cite{liu2018somewhere}. Relative distance judgments investigated accessing color similarity between the source and target using reference color. A combination of perceptually uniform color space and color naming more accurately predicted user performance than either alone, but the accuracy was low in both cases.

    A study, which used color as an encoding method on time-series data was performed to identify the average in data, and it demonstrated that color-coding techniques showed better average judgments~\cite{correll2012comparing}.
    Symbols are used in a scatterplot to encode values, and lightness is one of the dominant factors of color encoding. A model-based study on computing derived values compared two different types of scatterplots---with a circle as symbols and with spots as a symbol---with varying levels of lightness~\cite{li2010model}. The scatterplot with a circle as a symbol performed better in terms of error rate. 
    Gleicher et al.'s large-scale crowdsourced study on identifying aggregate relative judgment to read and the average value in the multi-class scatterplot was another form of computing derived values on various groups of objects~\cite{gleicher2013perception}.

    \para{Other} Factors, such as affective priming, social information, or individual user characteristics, directly impact the task performance. Accurate visual judgment is essential to performing the summarization, estimation, or other related computations of a derived value. Harrison et al.\ performed a series of experiments that showed users articles from the New York Times, followed by a judgment task~\cite{harrison2013influencing}. The evidence from the experiment showed that affective priming influences the user's visual judgment accuracy. In similar experiments, social information was shown to influence the users on summarization tasks, such as proportion judgment and linear estimation~\cite{hullman2016evaluating}. Other studies have evaluated how individual user characteristics, in particular, working memory (WM) (low, average, or medium), affected the user's performance on visual tasks, such as counting values~\cite{conati2014evaluating}. The experiment evaluated the task by comparing charts with vertical and horizontal layouts. Users with low visual WM performed faster with a horizontal layout.

\subsubsection{Visualization}

    When comparing the overall efficacy of different visualization methods for computing derived values, Saket et al.\ showed that for small data sizes: in terms of accuracy table, bar chart, pie chart, and scatterplots were best, in that order; in terms of completion time table and pie chart were best; and finally in terms of user preference, table, pie chart, and bar chart were best~\cite{saket2018task} (see \autoref{fig:saket2018task}).

    \begin{wrapfigure}[3]{L}{0.09\linewidth}         
    \begin{minipage}[t]{1.2\linewidth}
        \includegraphics[width=28pt]{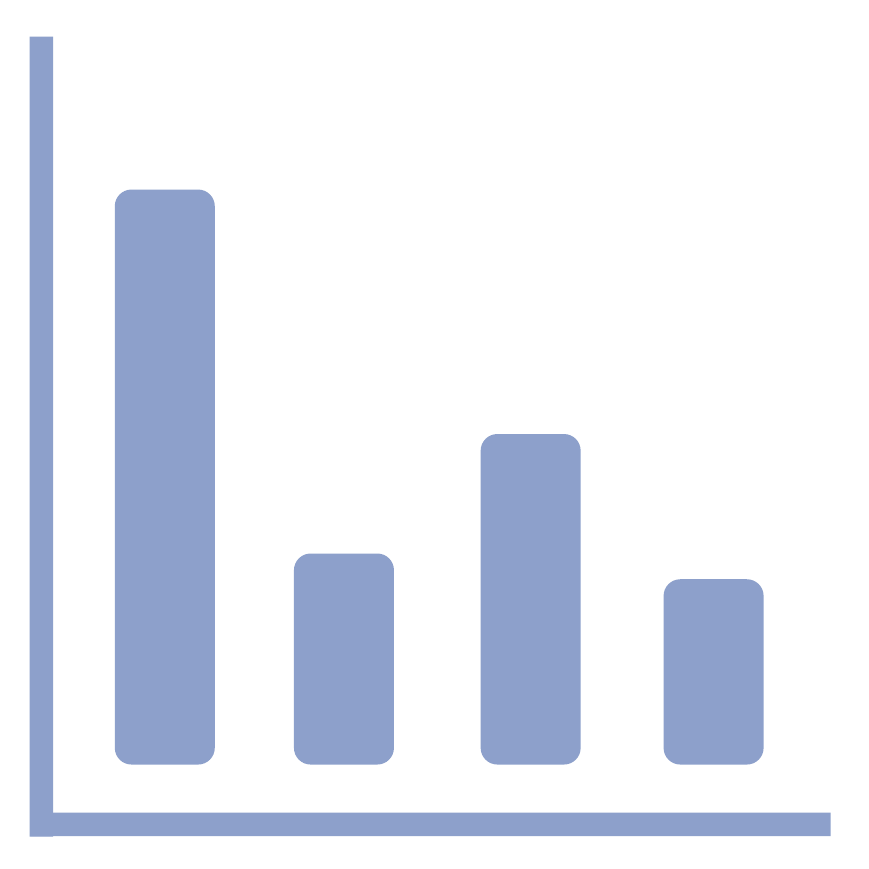}
        \end{minipage}
    \end{wrapfigure} 
    \noindent
    \para{Bar Chart} 
    People prefer linear bar charts because of their familiarity.
    A comparison-based study between using radial and linear bar charts to visualize daily patterns suggested that 24-hour linear bar charts are more accurate and efficient for summarization tasks~\cite{waldner2019comparison}. Along with the layout of bars in a bar chart, the design of the bars themselves also affects user performance. In a recent study on comparison tasks on various types of bar charts, two types of compute a derived value task were studied to identify the effects of bar variants on tasks~\cite{srinivasan2018s}. They found chart design affects the task completion time, and data conditions also influenced the completion time and accuracy.

    3D bar graphs are considered to be difficult to understand and generally bad for performing visual analysis tasks. In an experiment where the performance of relative magnitude estimation on pie charts and bar charts with and without 3D was evaluated, the task performance was shown to be better in 2D than 3D~\cite{siegrist1996use}.

    As better design improves performance, visual embellishments are often used to improve aesthetics. Embellishments were evaluated in bar charts on relative magnitude estimations, which confirmed that common embellishments significantly impact the task~\cite{skau2015evaluation}. In another follow-up study, pictorial charts reduced the user's performance on relative judgments or computing a retrieved value task, but not beyond the effect already observed for their shape~\cite{skau2017readability}.

    \begin{wrapfigure}[3]{L}{0.09\linewidth}         \vspace{-10pt}         \begin{minipage}[t]{1.2\linewidth}
        \includegraphics[width=28pt]{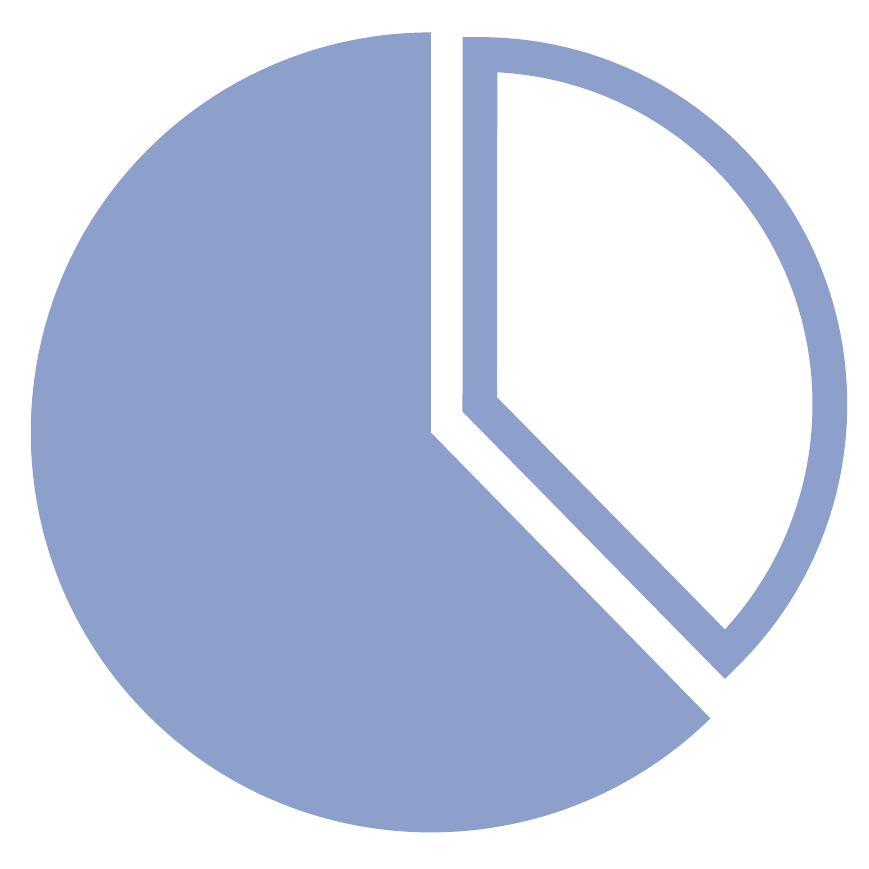}
        \end{minipage}
    \end{wrapfigure} 
    \noindent
    \para{Pie Chart} In an evaluation of pie and donut charts on relative magnitude estimation performance, the baseline donut chart was observed to be as good as the baseline pie chart. Furthermore, user performance on the arc length chart was similar to an area chart, angle pie chart, and angle donut chart~\cite{skau2016arcs}. In a similar study, where judgment error in pie chart variations was evaluated, variants of the pie charts, such as a chart with a larger slice, exploded pie, elliptical pie, and square pie caused more significant judgment errors than regular pie charts~\cite{kosara2016judgment}.

    \begin{wrapfigure}[3]{L}{0.09\linewidth}         \vspace{-10pt}         \begin{minipage}[t]{1.2\linewidth}
        \includegraphics[width=28pt]{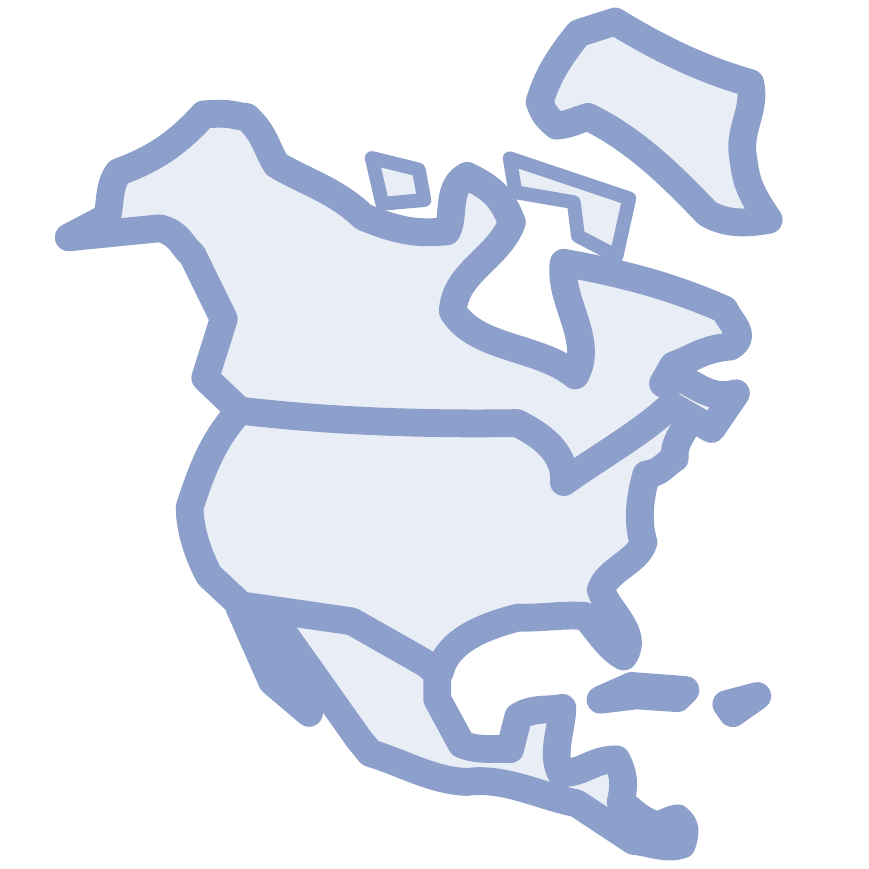}
        \end{minipage}
    \end{wrapfigure} 
    \noindent
    \para{Maps} A study compared different types of cartograms on tasks, such as {detecting change} and {summarizing}, found that contiguous cartograms performed better at {detecting change}, with the lowest error rate and completion time. For the summarize task, contiguous and Dorling cartograms had lower completion times, while the error rate in rectangular cartograms was highest among all four types of map visualization~\cite{nusrat2018evaluating}.

\subsubsection{Summary} In summary, computing derived values is one of the more basic low-level tasks, generally studied as a standalone task. User performance for this task varies greatly based upon visualization and encoding type. Though, the majority of the studies considered only bar charts and line charts. Overall, prior studies have indicated that bar charts, line charts, pie charts, and scatterplots can all be effectively used for summary-based task visualizations.

\subsection{Find Extremum}
\label{sec:task-extremum}

    \vspace{-5pt}
    \noindent
    \begin{minipage}[t]{\linewidth}
    \begin{wrapfigure}[4]{L}{0.145\linewidth}
        \vspace{-10pt}
        \begin{minipage}[t]{1.2\linewidth}
            \includegraphics[width=45pt]{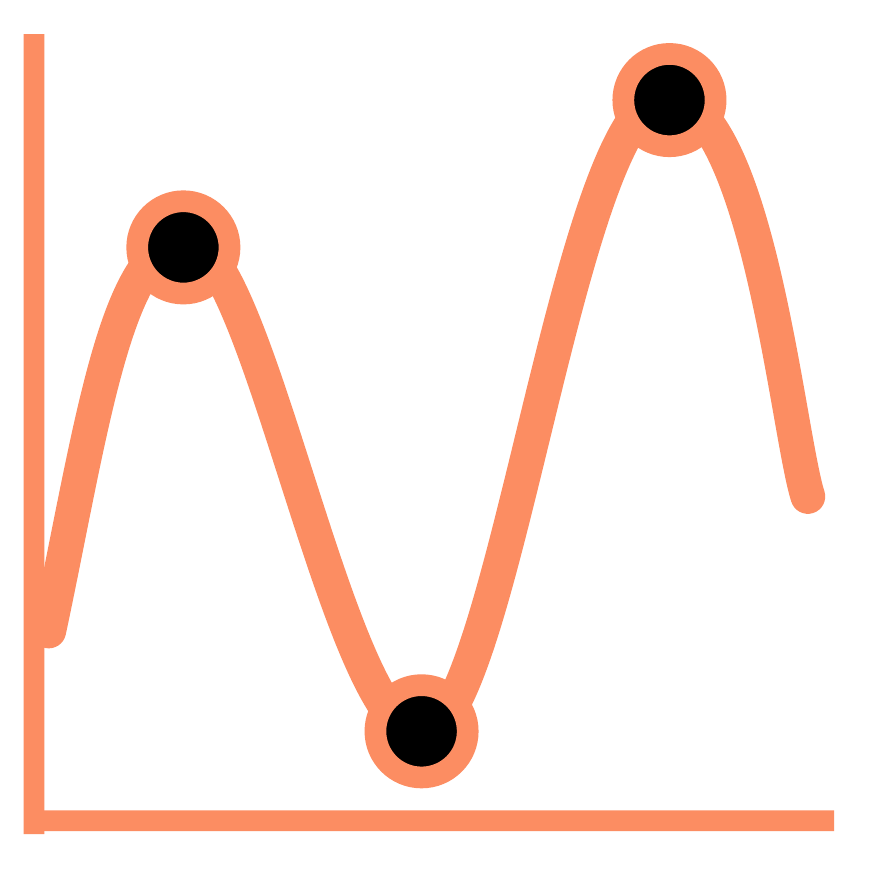}
        \end{minipage}
    \end{wrapfigure} 
     This task focuses on identifying data cases that possess an extreme value of an attribute over its range within the dataset. In this context, the find extremum values task can encompass finding both global and local maxima or minima in data.
     \end{minipage}

\subsubsection{Visual Encoding}

     Position encoded plots, e.g., scatterplots and dot plots, are often used to identify quantitative values~\cite{cleveland1984graphical}. Further, in a study of the effectiveness of visual encodings, the size of marks was more effective than color for quantitative values involving finding or identifying the extremum values~\cite{kim2018assessing}.

     \subsubsection{Visualization}
     
     For the find extremum task, Saket et al.\ found all five methods---tabular visualization, scatterplots, bar charts, pie charts, and line charts---performed well in terms of accuracy on small data. Bar charts, line charts, and scatterplots additionally performed well in terms of time. Finally, bar charts performed best in terms of user preference~\cite{saket2018task} (see \autoref{fig:saket2018task}). 
         Another recent work compared the performance of design variation in three visualizations---bar charts, line charts, and pie charts---on a task of finding the maximum~\cite{ondov2018face}. The evaluation confirmed the prior hypothesis of better performance in overlaid design versus small multiples and suggested additional new performance improvements.

     \begin{wrapfigure}[3]{L}{0.09\linewidth}         \vspace{-10pt}         \begin{minipage}[t]{1.2\linewidth}
        \includegraphics[width=28pt]{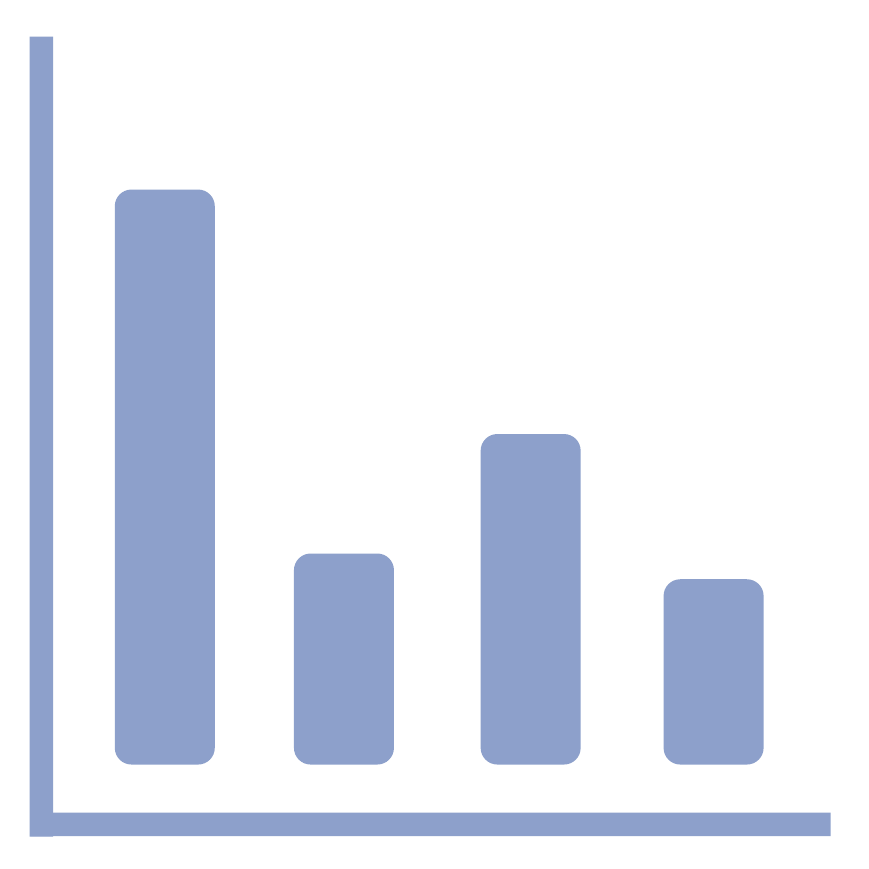}
        \end{minipage}
    \end{wrapfigure} 
    \noindent
    \para{Bar Chart} Bar charts are often used to read the minimum/maximum quantitative values in data, and recent studies have suggested have indicated that bar design in bar charts influences the user's performance on identifying those values~\cite{srinivasan2018s}. 
    In another example, a comparison between radial charts and linear bar charts investigated the visualization of 24-hours time. Users had higher accuracy for identifying the maximum value in linear bar charts compared to radial charts~\cite{waldner2019comparison}.
    A series of studies were performed for determining the maximum value of delta and mean in bar charts, slope, and donut chart for three variations of their design---small multiples, overlaid, and mirrored~\cite{ondov2018face, jardine2019perceptual}.
    Animation in the design made identifying maximum delta value particularly salient, but the effect did not carry over when the change was large. The identification of maximum mean value held high accuracy for the bar charts with mirrored and stacked bar arrangements.

     \begin{wrapfigure}[3]{L}{0.09\linewidth}         \vspace{-10pt}         \begin{minipage}[t]{1.2\linewidth}
        \includegraphics[width=28pt]{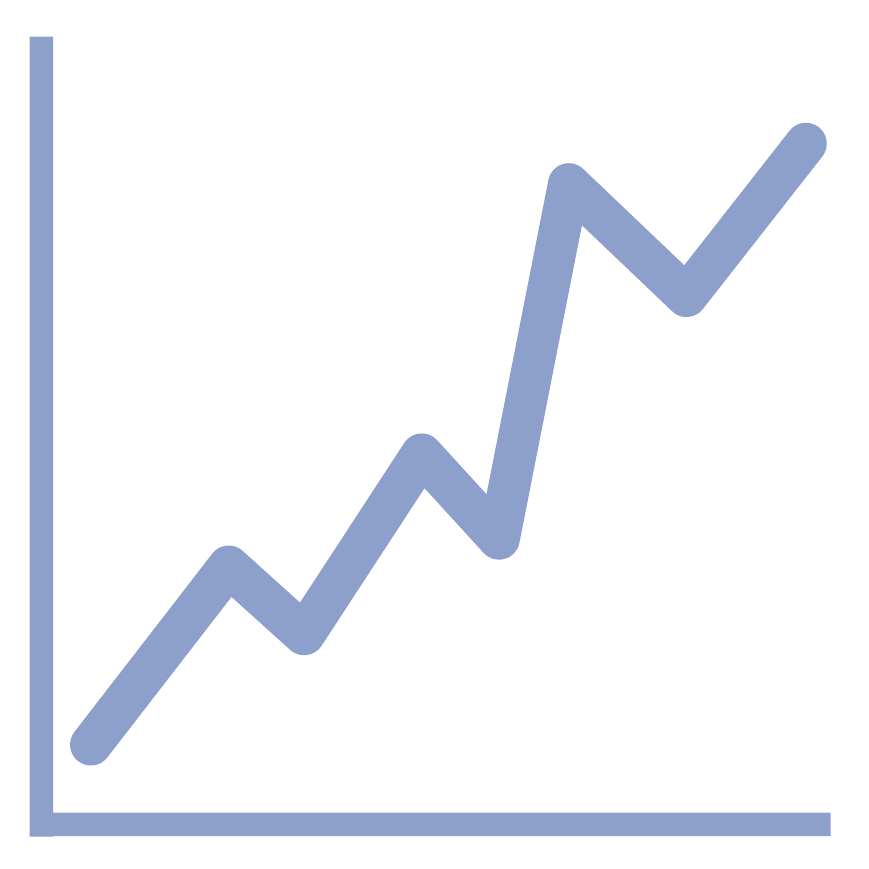}
        \end{minipage}
    \end{wrapfigure} 
    \noindent
    \para{Line Chart} Line charts are used to visualize time-series data because of the inherent potential of showing the trend. Evaluation of variation in line graph design demonstrates user performance of the finding minimum or maximum values also depends on the design of the visual representation~\cite{albers2014task}. A modified form of line charts using position encoding, called a ``stock chart,''  \footnote{The modified stock chart is a line chart with a layering of moving average over the original series to supplement summary judgments~\cite{albers2014task}.} performed significantly better on finding a minimum, whereas the composite line chart\footnote{A composite line chart is a line chart layered over a bar chart representing averages of discrete subregions~\cite{albers2014task}.} performed well on finding a maximum. However, line charts using a colorfield had the highest accuracy on this task.

     \begin{wrapfigure}[6]{L}{0.09\linewidth}     
     \vspace{-10pt}
     \begin{minipage}[t]{1.2\linewidth}
        \includegraphics[width=28pt]{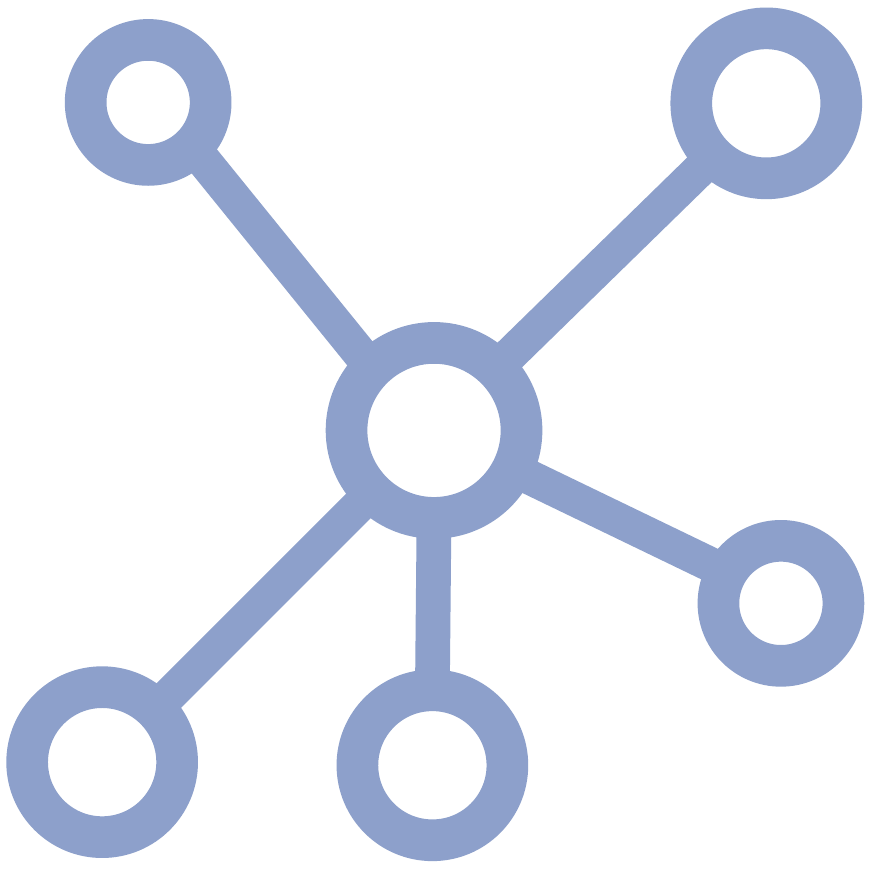}
        
        \vspace{3pt}
        \includegraphics[width=28pt]{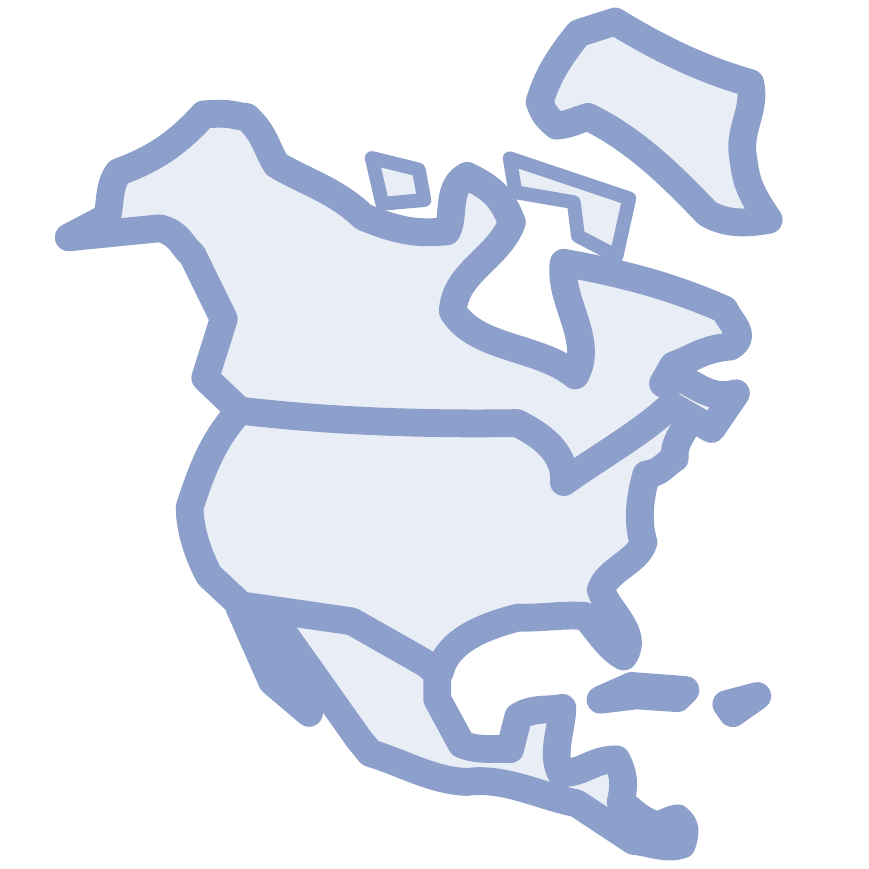}
        \end{minipage}
    \end{wrapfigure} 
    \noindent
     \para{Network and Map} A comparison study between node-link graphs and matrix-based representation reported that the task of ``finding extremum,'' in the form of finding the most connected node, had the highest accuracy and lower completion time for a matrix representation~\cite{ghoniem2004comparison}. The results were particularly strong as the size of the graph became large, or the link density increased. 
     
     The task of identifying the highest value of an attribute in maps, in the form of top-k identification, a standard cartogram has the highest accuracy when compared to other types of cartograms, i.e., rectangular, non-continuous, or Dorling cartograms~\cite{nusrat2018evaluating}.

\subsubsection{Summary} 
In summary, finding extremum values is a low-level task mostly studied on bar charts and line charts. 
Generally speaking, line charts can be used to represent the time-series data when finding an extremum value, while scatterplots and bar charts can be used for quantitative values encoded by position or length.

\subsection{Determine Range}
\label{sec:task-range}

    \vspace{-5pt}
    \noindent
    \begin{minipage}[t]{\linewidth}
    \begin{wrapfigure}[4]{L}{0.145\linewidth}
        \vspace{-10pt}
        \begin{minipage}[t]{1.2\linewidth}
            \includegraphics[width=45pt]{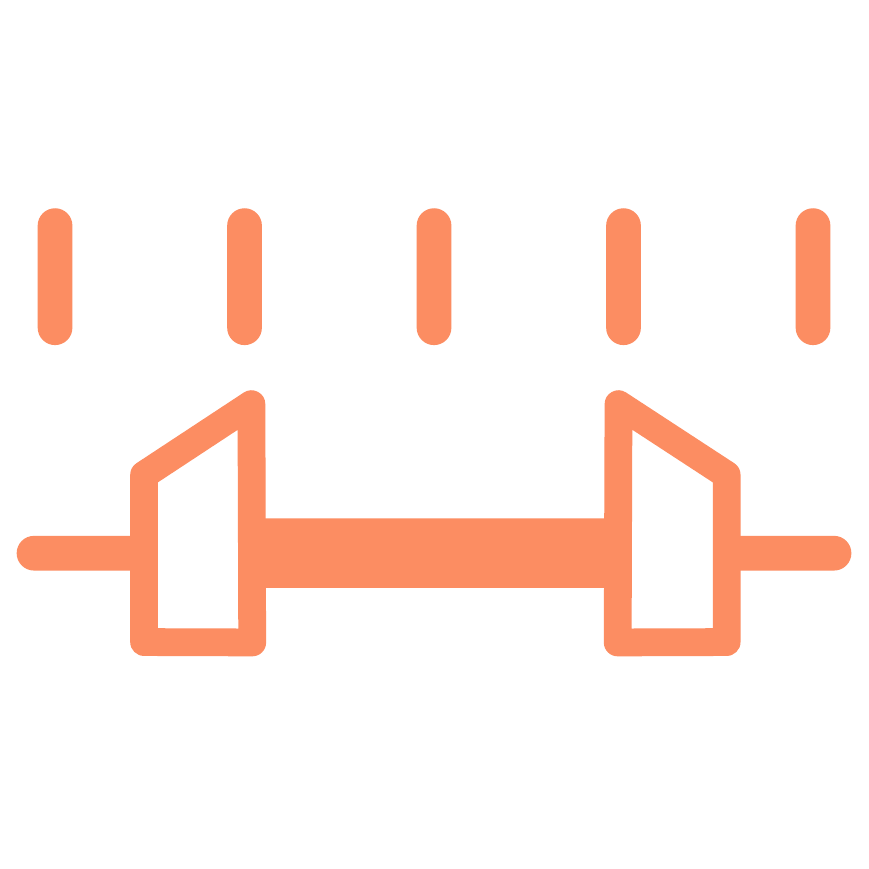}
        \end{minipage}
    \end{wrapfigure} 
    A {determine range} analysis task has users finding the span of values in a given data for an attribute of interest. The determine range task is another task that has received limited attention. 
    \end{minipage}

   \subsubsection{Visualization}
   
   Saket et al.'s evaluation on small data found that for the task of determining a range of values, bar charts had high accuracy, whereas scatterplot performed better on completion time and user preference~\cite{saket2018task} (see \autoref{fig:saket2018task}). 
   In an evaluation of multidimensional visualizations, where the task of finding range is evaluated on three user performance metrics of accuracy, completion time, and user satisfaction, parallel coordinates and tabular visualization were found to be effective in terms of accuracy and completion time, but parallel coordinates were least preferred by users~\cite{dimara2018conceptual}. However, this result is only applicable to this limited context.

   \begin{wrapfigure}[3]{L}{0.09\linewidth}         \vspace{-10pt}         \begin{minipage}[t]{1.2\linewidth}
        \includegraphics[width=28pt]{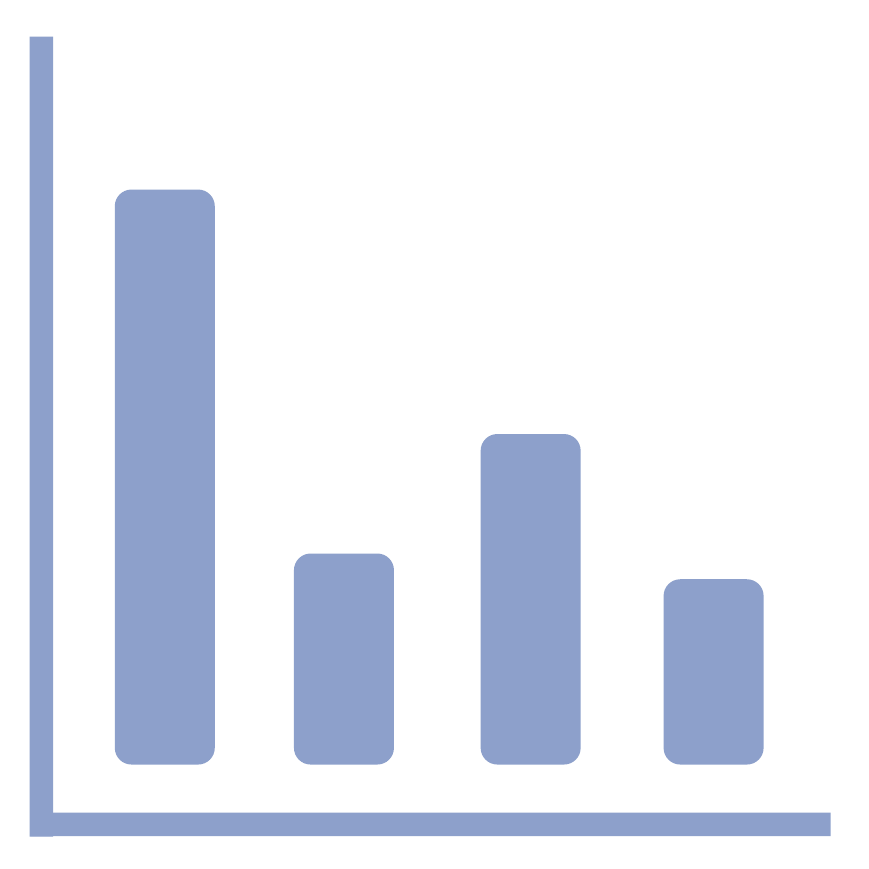}
        \end{minipage}
    \end{wrapfigure} 
    \noindent
   \para{Bar Chart} 
   A variety of bar charts arrangements designs were investigated on a maximum range task~\cite{jardine2019perceptual}. A stacked arrangement of bar charts gave the highest visual comparison accuracy, while superimposed charts gave the lowest.
    
         \begin{wrapfigure}[3]{L}{0.09\linewidth}         \vspace{-10pt}         \begin{minipage}[t]{1.2\linewidth}
        \includegraphics[width=28pt]{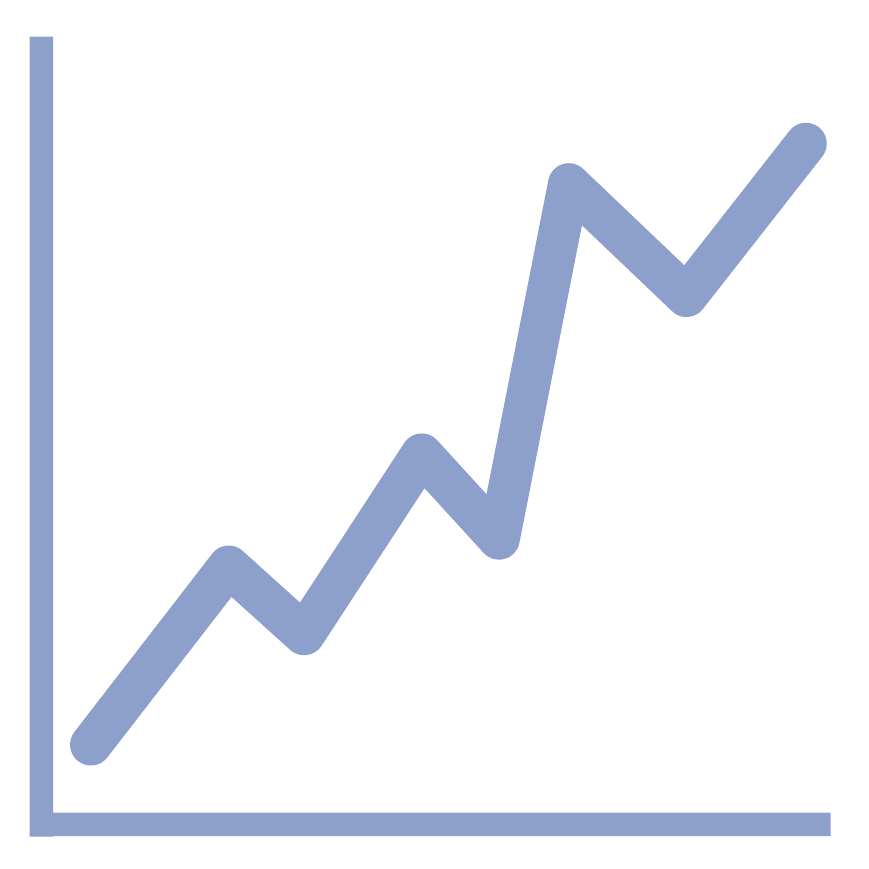}
        \end{minipage}
    \end{wrapfigure} 
    \noindent
    \para{Line Chart} For identifying the range of values on time-series data visualizations, line charts based on position encoding had higher accuracy over a modified stock chart, box plot, and a composite line graph, respectively~\cite{albers2014task}. Though these are variations of the same visualization type, i.e., a line chart, their design variations significantly impact their efficacy.

\subsubsection{Summary} In summary, determining a range of values has many unstudied aspects. For example, we did not find any studies which directly investigate the performance on visual encodings.  That said, the limited depth and breadth still indicate that scatterplots are preferred when requiring faster performance, bar charts should be chosen when accuracy is needed, and parallel coordinates work well for multidimensional data.

\subsection{Sort}
\label{sec:task-sort}

    \vspace{-5pt}
    \noindent
    \begin{minipage}[t]{\linewidth}
    \begin{wrapfigure}[4]{L}{0.145\linewidth}
        \vspace{-10pt}
        \begin{minipage}[t]{1.2\linewidth}
            \includegraphics[width=45pt]{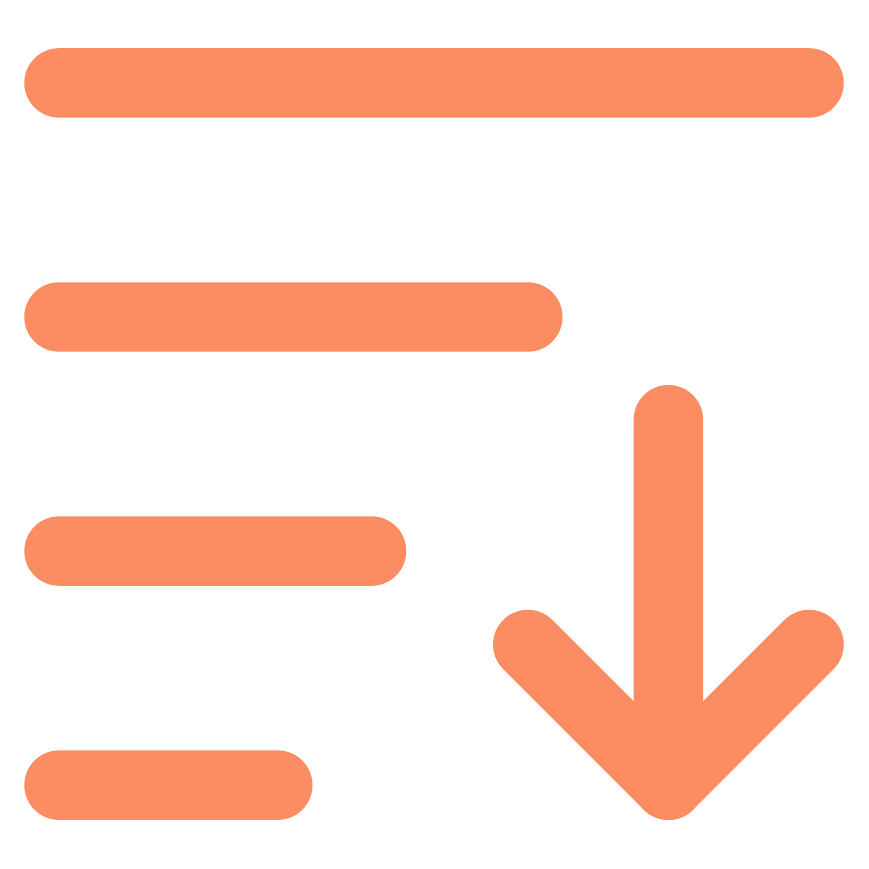}
        \end{minipage}
    \end{wrapfigure} 
    Sorting generally implies ranking the given set of data according to some ordinal attribute. Sorting, as a low-level task, has not received much attention. 
    \end{minipage}

    \subsubsection{Visualization}
    
    Saket et al.\ also studied ordering tasks, which are synonymous with sorting~\cite{saket2018task} (see \autoref{fig:saket2018task}). The bar chart stood out in terms of accuracy, timely completion, and user preference.  Pie chart, scatterplot, and tabular visualization were next best in terms of accuracy, whereas the line chart and scatterplot were next best for the time completion metric.

    \begin{wrapfigure}[3]{L}{0.09\linewidth}         \vspace{-10pt}         \begin{minipage}[t]{1.2\linewidth}
        \includegraphics[width=28pt]{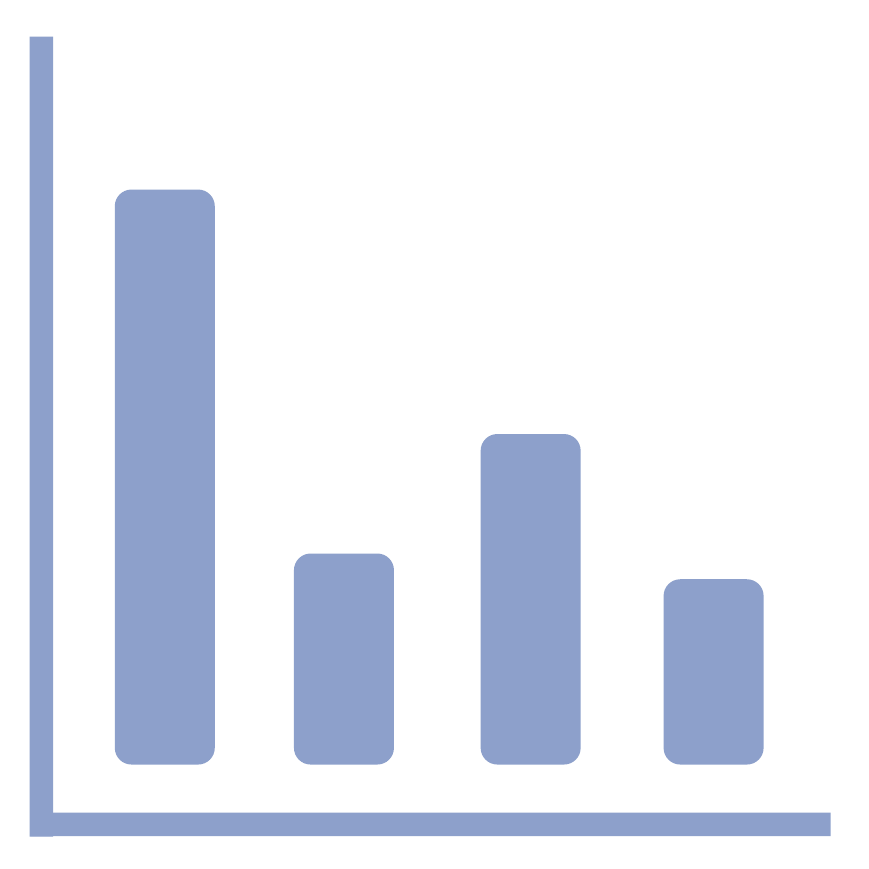}
        \end{minipage}
    \end{wrapfigure} 
    \noindent
    \para{Bar Chart} 
     Different types of ranked-list visualizations, such as scrolled bar charts, wrapped bars, piled bars, packed bars, treemaps, and Zvinca plots\footnote{A Zvinca plot is a layered plot where points replace bars, see~\cite{few_2017}.}, were evaluated in a study that suggested wrapped bars are best for visualizing ranked lists as they provide a simple, compact, and interaction friendly visualization, while treemaps reported the highest accuracy despite the use of area when length is generally a preferable encoding~\cite{mylavarapu2019ranked}.

        \begin{wrapfigure}[3]{L}{0.09\linewidth}         \vspace{-10pt}         \begin{minipage}[t]{1.2\linewidth}
        \includegraphics[width=28pt]{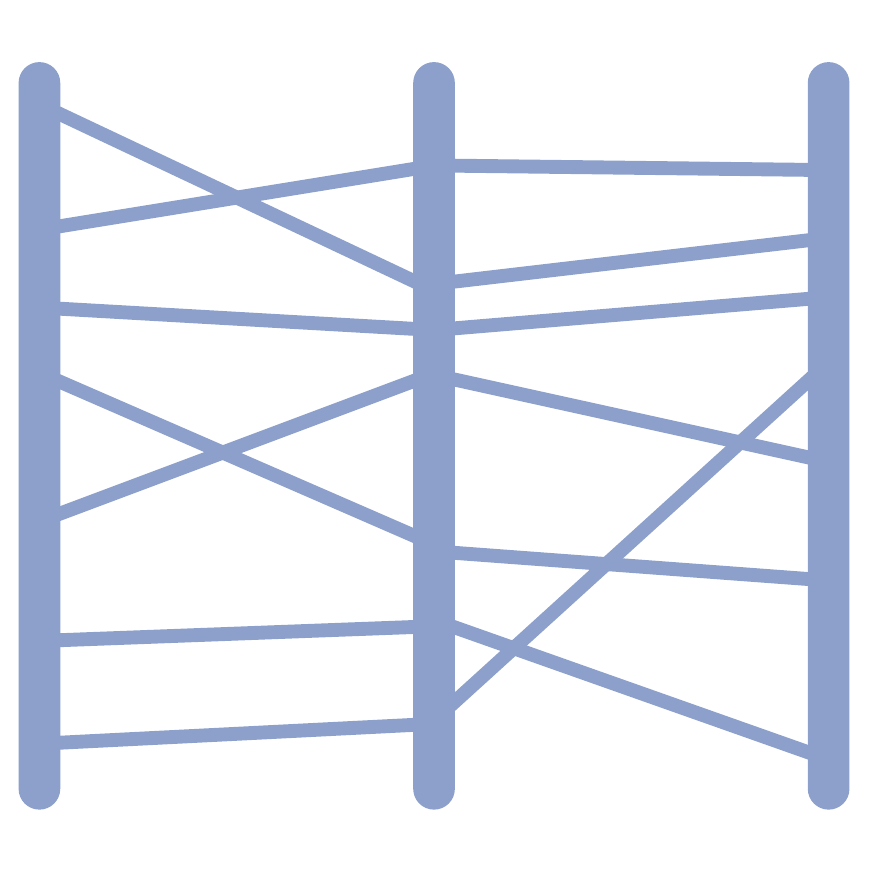}
        \end{minipage}
    \end{wrapfigure} 
    \noindent
    \para{Parallel Coordinates} An improved form of parallel coordinates, called the progressive parallel coordinates, was studied to understand how data order can mitigate scalability concerns~\cite{rosenbaum2012progressive}. The application of level-of-detail and randomly accessing individual values were based on an ordering activity.

    \begin{wrapfigure}[3]{L}{0.09\linewidth}         \vspace{-10pt}         \begin{minipage}[t]{1.2\linewidth}
        \includegraphics[width=28pt]{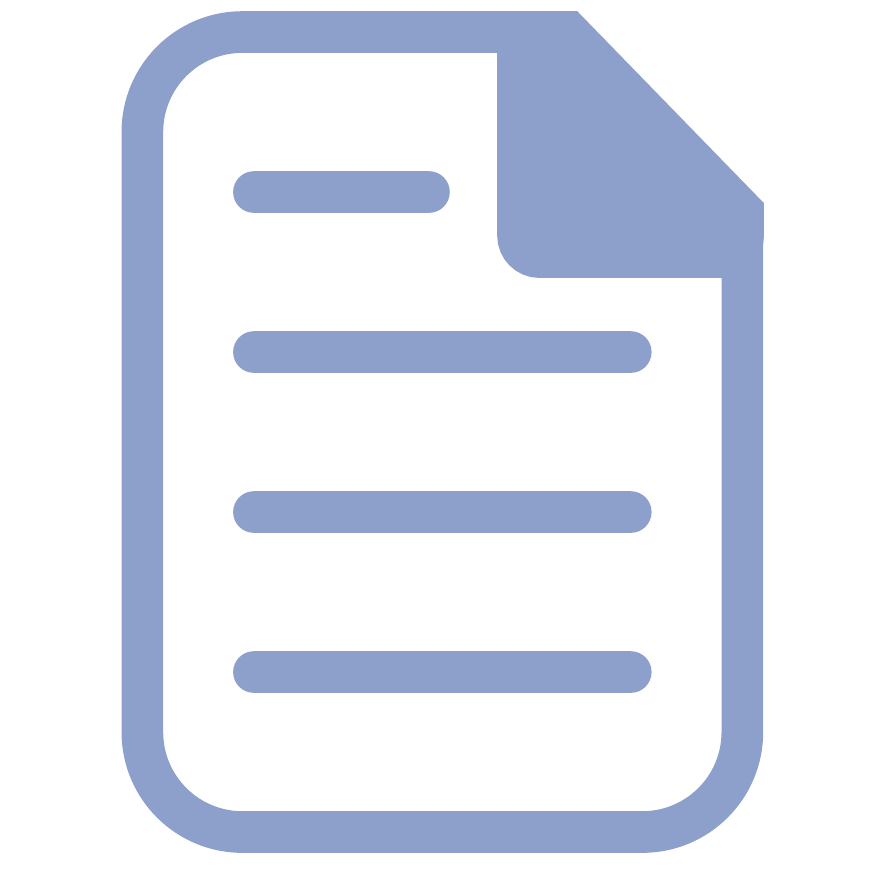}
        \end{minipage}
    \end{wrapfigure}
    \noindent
    \para{Text}  
   Sorting a keyword summary alphabetically reduces the time to find a word, and as such, it makes ordered layouts (e.g., column layouts) more effective than unordered layouts (e.g., spatial arrangements)~\cite{halvey2007assessment}. It was shown that a simple ordering of text data could be much more effective than font size encodings, most notably in searching and retention.
   In other cases, words can be sorted alphabetically, by frequency, or by a predetermined algorithm. In an experiment on word cloud effectiveness suggesting guidelines for construction, the authors evaluated impression formation and memory by varying font size and layout (e.g., alphabetical sorting, frequency sorting) of words~\cite{rivadeneira2007getting}. One important finding was that a list ordered by frequency might provide a more accurate impression as compared to other layouts.

    \para{Other} A study focused on user characteristics showed that user performance on sorting values depended upon participants' working memory (WM) (low, average, and medium)~\cite{conati2014evaluating}. 
    Due to individual differences in perceptual judgments, users with lower verbal WM were slower than others for the sorting task, and users with low visual WM required more time for the task than users with average visual WM. Users with higher cognitive processing speed were also more effective at deriving facts and insights from a visualization than others.
    
    In another study on \textit{perceptual kernels}---a matrix of aggregated pairwise subjective measures of judged similarity---participants were asked to rank the data categories from least to most similar to a target class~\cite{demiralp2014learning}. This experiment estimated perceptual kernels for visual encoding variables of shape, size, color, and combinations. Based on the judged similarities using Likert ratings among visual variables, findings can be applied to improve visualization design through automatic palette optimization.

\subsubsection{Summary} In summary, the sorting, which has many similarities to finding extremum and determining range, has not been studied extensively. Nevertheless, studies indicate that bar charts and text-based visualizations are the preferred techniques for quantitative and textual data, respectively.

\subsection{Find Anomalies}

\label{sec:task-anomalies}

    \vspace{-5pt}
    \noindent
    \begin{minipage}[t]{\linewidth}
    \begin{wrapfigure}[4]{L}{0.145\linewidth}
        \vspace{-10pt}
        \begin{minipage}[t]{1.2\linewidth}
            \includegraphics[width=45pt]{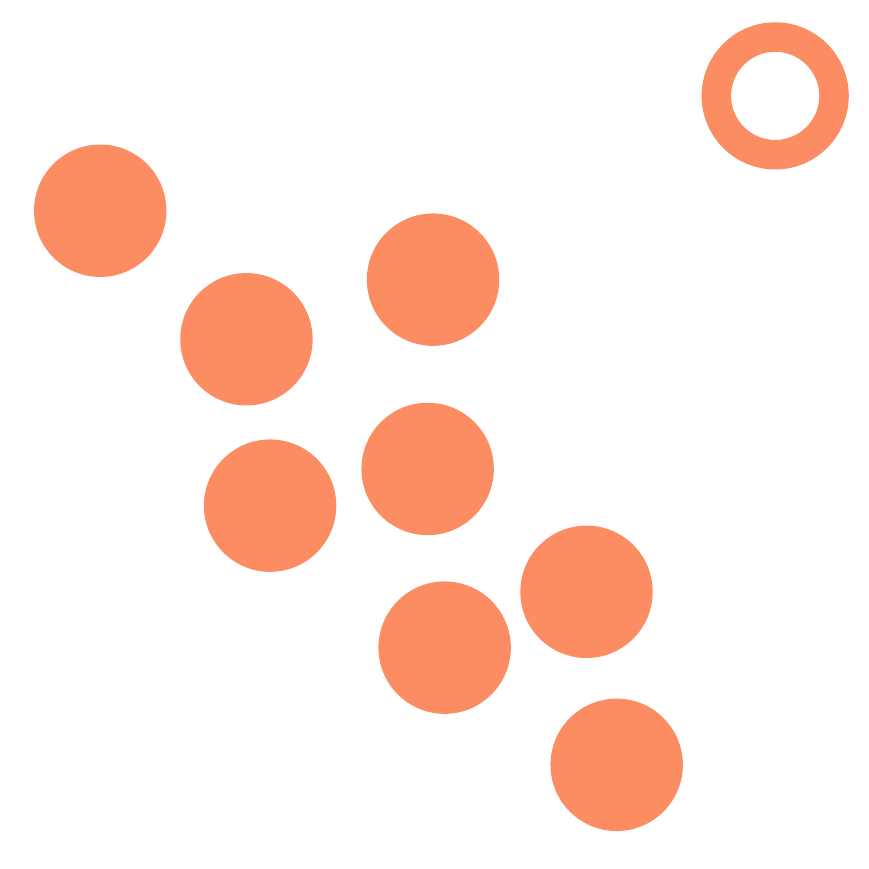}
        \end{minipage}
    \end{wrapfigure} 
    The find anomalies task is a form of visual aggregation task, generally involving identifying any outliers or unexpected cases within a given set of data.
    \end{minipage}

    \subsubsection{Visual Encoding} \ 
    
    \vspace{-10pt}
    \begin{wrapfigure}[3]{L}{0.09\linewidth}         \vspace{-10pt}         \begin{minipage}[t]{1.2\linewidth}
        \includegraphics[width=28pt]{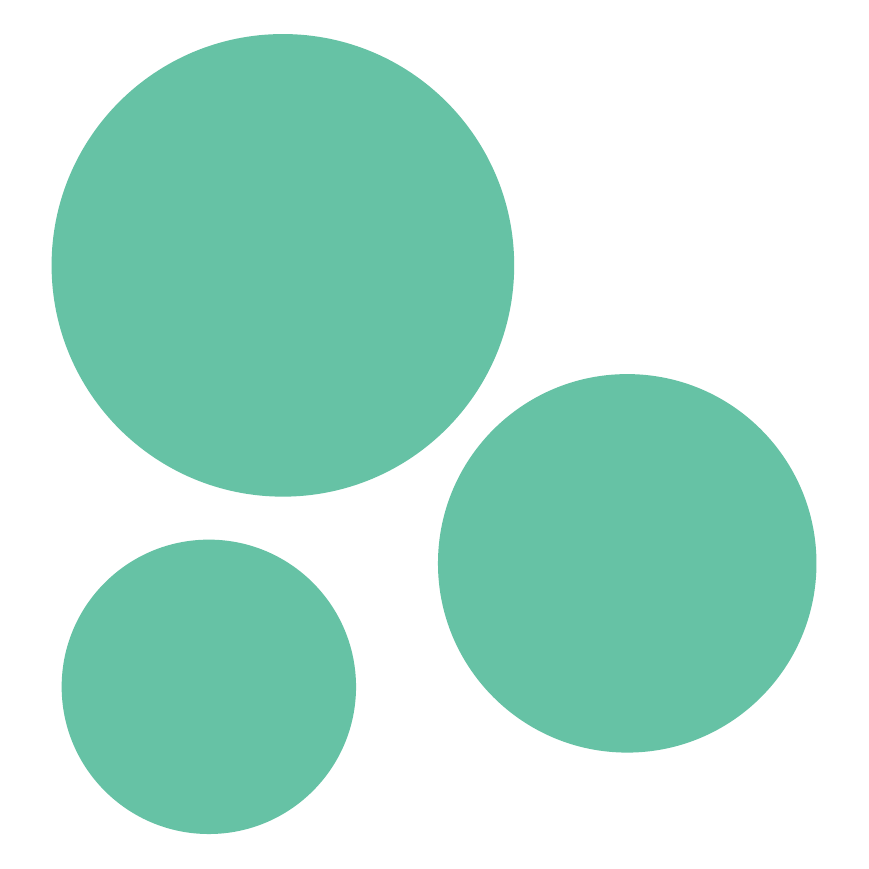}
        \end{minipage}
    \end{wrapfigure} 
    \noindent
    \para{Spatial Size} With the focus on investigating symbol size discrimination in scatterplots, one study developed a method to pick sizes that are effective for counting outliers and making the judgment as easy as possible~\cite{li2010size}.
    The results show that size perception 
    can be described by the Power law transformation to yield an optimal scale for symbol size discrimination.

    \begin{wrapfigure}[3]{L}{0.09\linewidth}         \vspace{-10pt}         \begin{minipage}[t]{1.2\linewidth}
        \includegraphics[width=28pt]{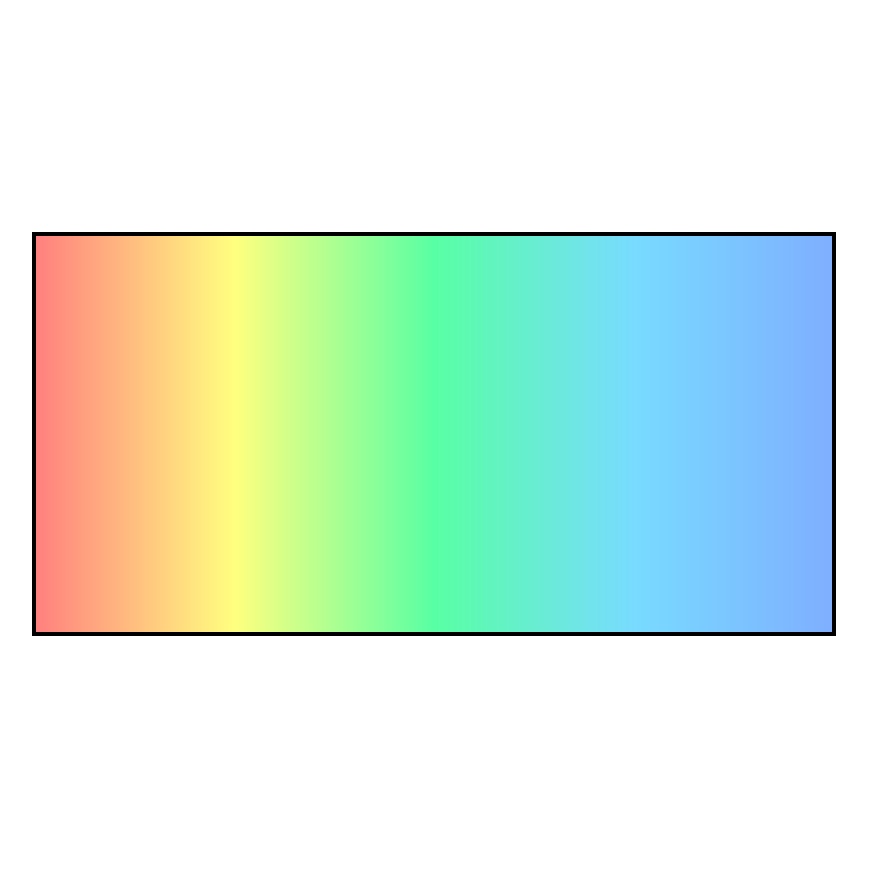}
        \end{minipage}
    \end{wrapfigure} 
    \noindent    
    \para{Color Hue} Color, motion, and layout (random or grouped) affect the users' attention capacity, and the same has been evaluated on identifying outliers~\cite{haroz2012capacity}. Search time for an outlier in a matrix form of symbols was shown to vary with color, motion, and layout. Further, color grouped and motion grouped had the best response time, whereas motion random cases were worst in response time. The color group also had the highest identification accuracy.
    %
  

    \begin{wrapfigure}[3]{L}{0.09\linewidth}         \vspace{-10pt}         \begin{minipage}[t]{1.2\linewidth}
        \includegraphics[width=28pt]{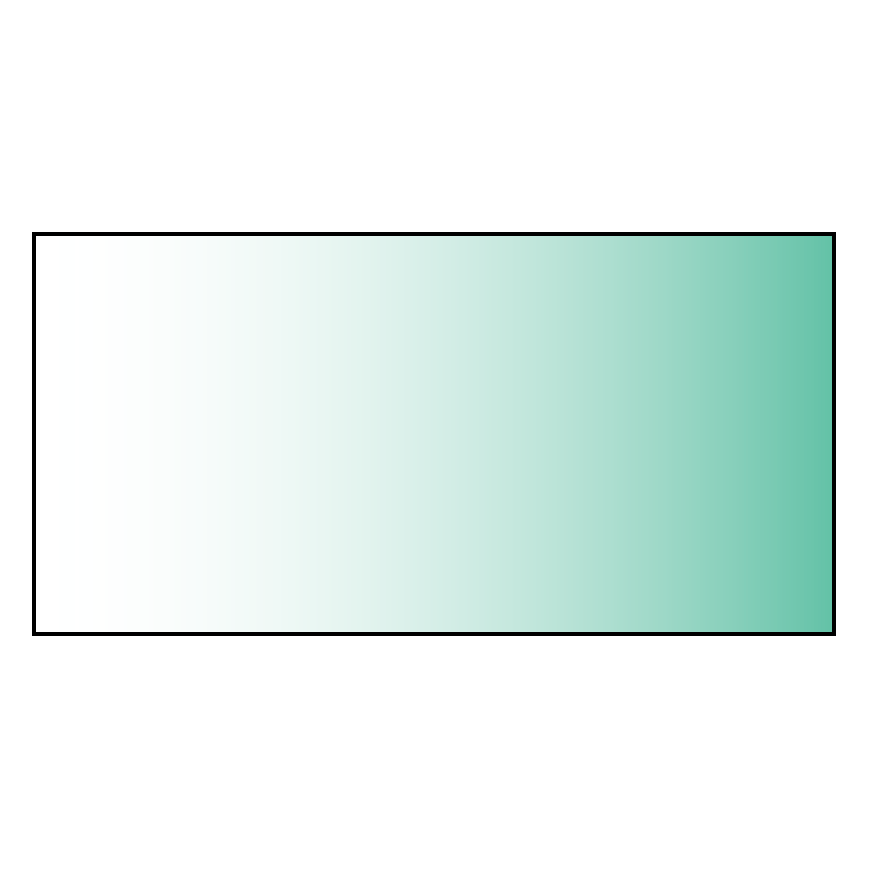}
        \end{minipage}
    \end{wrapfigure} 
    \noindent    
    \para{Color Intensity} Another discriminability study aimed to select luminance levels such that analytical tasks, including counting outliers, were as easy as possible~\cite{li2010lighntness}. The performance limit for the task was approached where high-perceived contrast was observed for the data symbols. Additional findings stated that with clustering/groupings of the same data symbols, the task might become easy for low contrast sets.

\subsubsection{Visualization}
   
    Saket et al.\ found that for finding anomalies, scatterplots and bar charts performed best in terms of accuracy, time, and user preference~\cite{saket2018task} (see \autoref{fig:saket2018task}).

    \begin{wrapfigure}[3]{L}{0.09\linewidth}         \vspace{-10pt}         \begin{minipage}[t]{1.2\linewidth}
        \includegraphics[width=28pt]{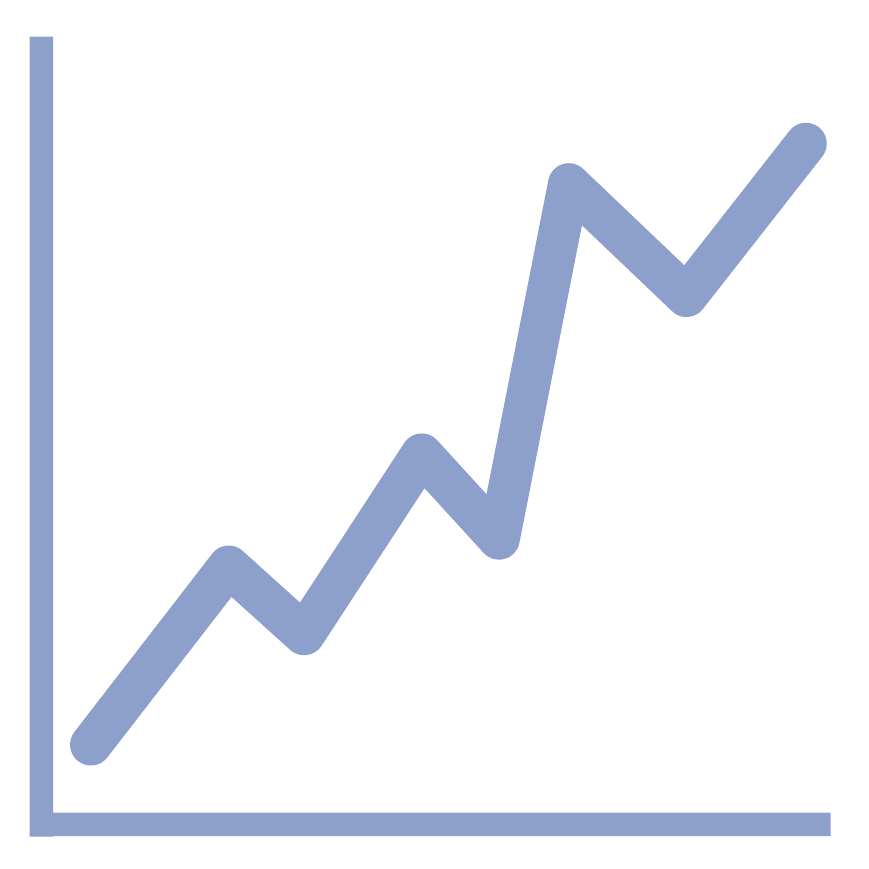}
        \end{minipage}
    \end{wrapfigure} 
    \noindent
    \para{Line chart} Identification of outliers in time-series data using line charts and their variants is complex and reported low user response accuracy~\cite{albers2014task}. 
    Basic line chart, modified stock chart, composite line chart, and color-field have the same level of accuracy, whereas event stripping outperformed all of these visualizations. The accuracy was low in this visualization for outlier tasks because data spread was confounding as an outlier and vice-versa. Essentially, when data values that would increase or decrease in the opposite direction were focused on for outlier behavior~\cite{yau2019bridging}.

    \para{Other} A recent study on three types of visualization---density plot, histogram, and dot plot---pointed out the outlier detection task as a part of data quality issues~\cite{correll2018looks}. Participants were better able to identify outliers as the flaws in data, but there was no single visualization suggested, which was significantly better among the three.

    In an algorithm-based outlier detection, the authors studied outliers and anomalies on box plots and letter-value-box plots~\cite{wilkinson2017visualizing}. These types of approaches can be paired with visualizations to help analysts explore anomalous data features, especially multidimensional outliers.

\subsubsection{Summary} In summary, finding anomalies is defined by the need to identify targets that are different from others in the given set, which varies based upon the visual features, e.g., position, size, orientation, color, and luminance~\cite{szafir2016four}. Scatterplots have been most heavily studied for this task. They are efficient in identifying outliers and detect anomalies in the data. Additionally, line charts easily represent any abnormal or outliers behavior in time-series data.

\subsection{Characterize Distribution}
\label{sec:task-distribution}

    \vspace{-5pt}
    \noindent
    \begin{minipage}[t]{\linewidth}
    \begin{wrapfigure}[4]{L}{0.145\linewidth}
        \vspace{-10pt}
        \begin{minipage}[t]{1.2\linewidth}
            \includegraphics[width=45pt]{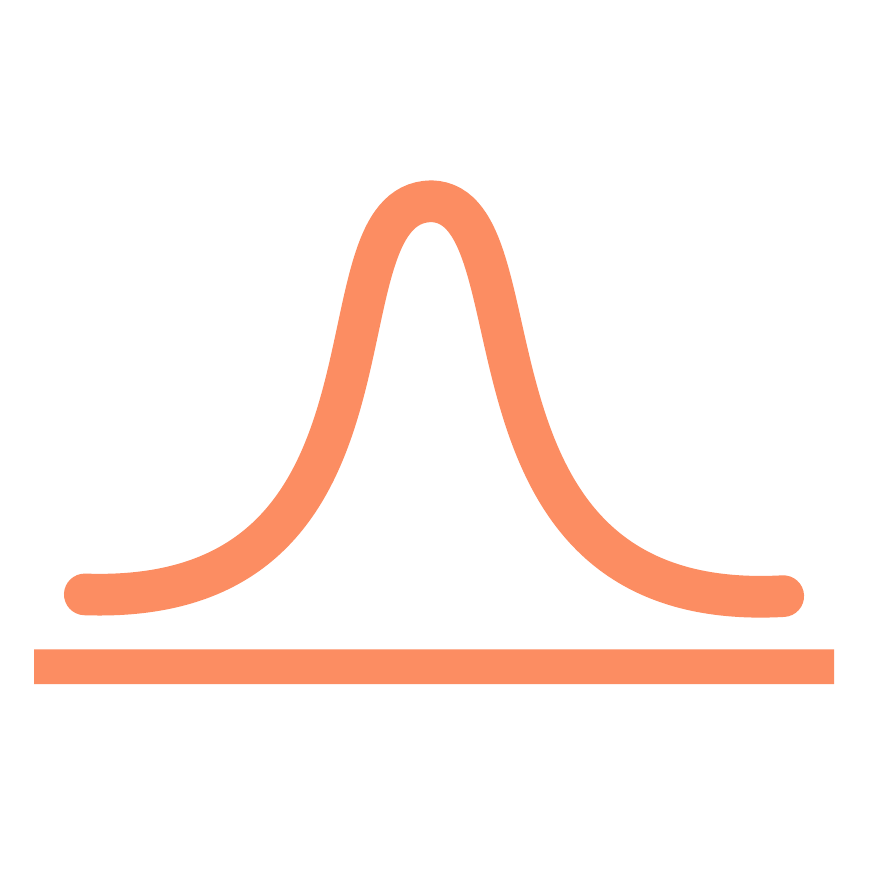}
        \end{minipage}
    \end{wrapfigure} 
    This task requires that for a given set of data and a quantitative attribute of interest, the distribution of that data should be characterized over that attribute's value.
    \end{minipage}

\subsubsection{Visual Encoding}
    
    Characterizing distribution is another visual abstraction task, with a specific focus on pattern or trend recognition. The data can be encoded using several visual features, e.g., position, size, orientation, color, and luminance, which can be crucial in identifying the distribution~\cite{szafir2016four}. For example, Gapminder uses size, position, and color to reveal patterns in global demographics~\cite{rosling2009gapminder}, and weather maps use color and orientation to visualize information about wind speed, temperature, and other meteorological data~\cite{ware2013designing}.

     \begin{wrapfigure}[3]{L}{0.09\linewidth}         \vspace{-10pt}         \begin{minipage}[t]{1.2\linewidth}
        \includegraphics[width=28pt]{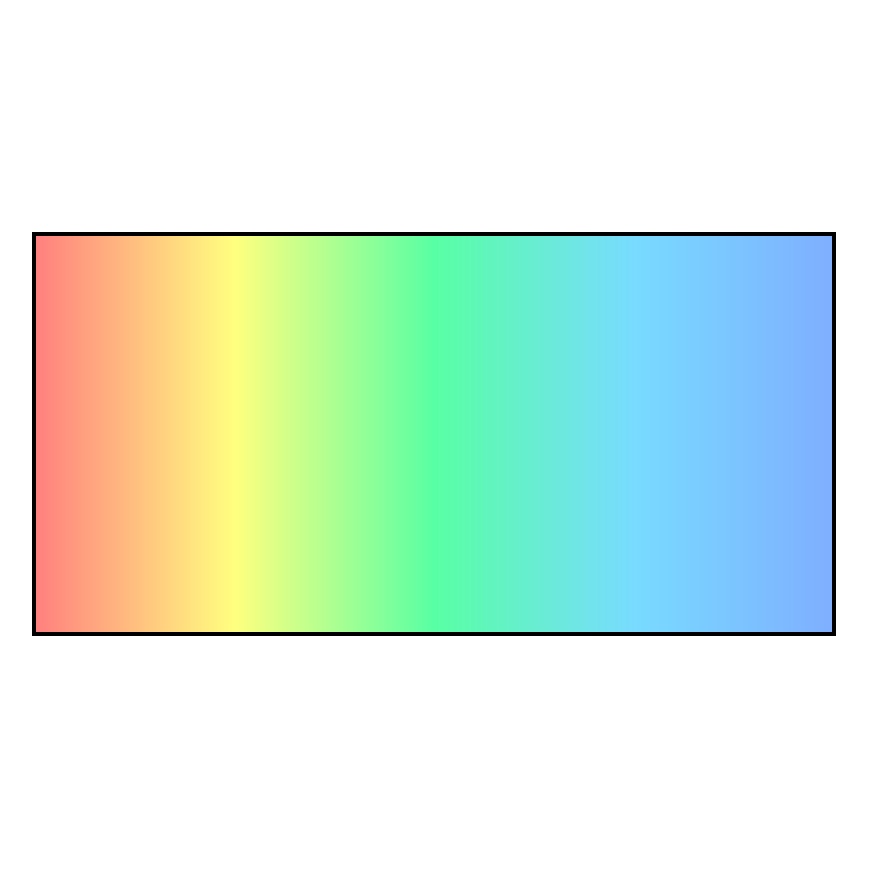}
        \end{minipage}
    \end{wrapfigure}     
    \noindent
    \para{Color Hue}
    Colormap design in continuous quantitative maps was used to evaluate the perception of patterns~\cite{reda2018graphical}. There was a negative main effect of spatial frequency on pattern perception. Further, the use of low color at low spatial frequency was unlikely to improve pattern perception as compared to a plain grayscale ramp.    

    \subsubsection{Visualization}
    
    In Saket et al.'s study, the bar chart was shown to have high accuracy and to be the user preferred method for characterizing distribution tasks, but the scatterplot had better completion time~\cite{saket2018task} (see \autoref{fig:saket2018task}). The line chart was next most accurate after the bar chart and scatterplot.

    \begin{wrapfigure}[3]{L}{0.09\linewidth}         \vspace{-10pt}         \begin{minipage}[t]{1.2\linewidth}
        \includegraphics[width=28pt]{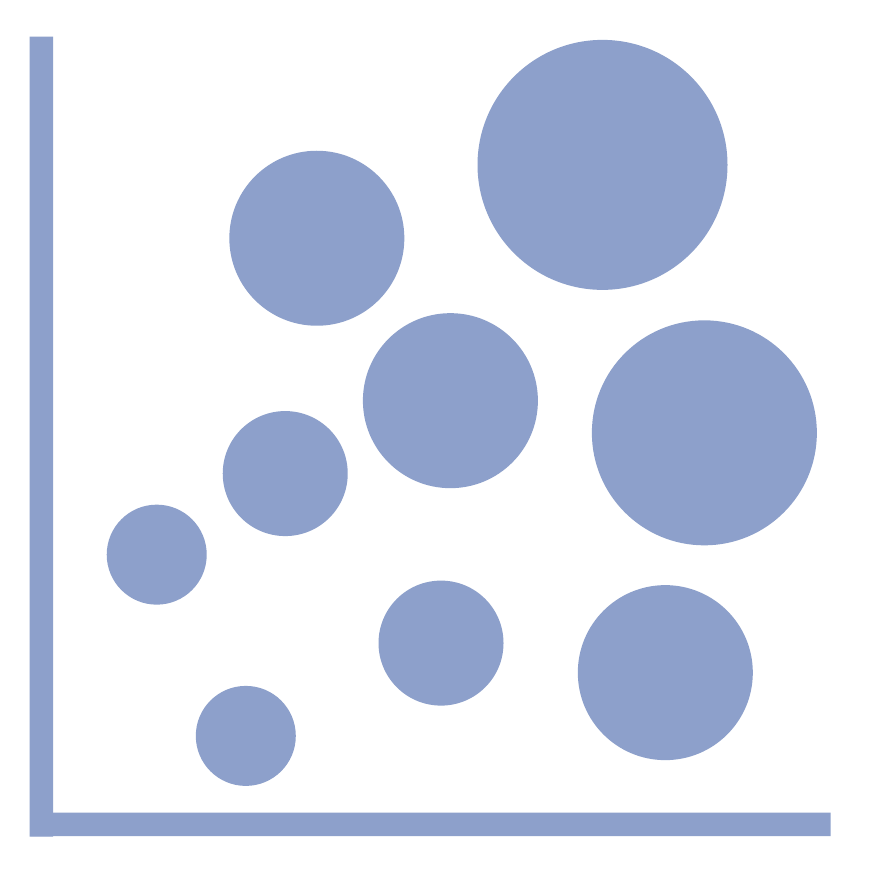}
        \end{minipage}
    \end{wrapfigure}     
    \noindent
    \para{Scatterplot}
    The scatterplot uses position encoding for the data, and users can identify or detect a data pattern or distribution easily. Varying densities and gaps between data points were shown to influence the user's perception of the distribution or pattern of data points~\cite{o1974human}. Further, multiple studies demonstrated the task of characterizing a distribution is influenced by encodings, specifically symbol size and lightness, in a scatterplot~\cite{li2010model, li2010size}.
    
    With scagnostics, density is assumed to be a property that shows the concentration of points, which is directly influenced by the distribution of points~\cite{wilkinson2005graph}. This observation led the way to investigating how people interpret trends in a scatterplot and was studied using a sensitivity model. Visual augmentation of scatterplots introduces sensitivity information, which was used to study how people interpret trends in scatterplots~\cite{chan2013generalized}. Orientation cues provided by the flow lines give an idea of the local data. Sensitivity or local trends helped in identifying the type of relationship between two variables.

     \begin{wrapfigure}[3]{L}{0.09\linewidth}         
     \begin{minipage}[t]{1.2\linewidth}
        \includegraphics[width=28pt]{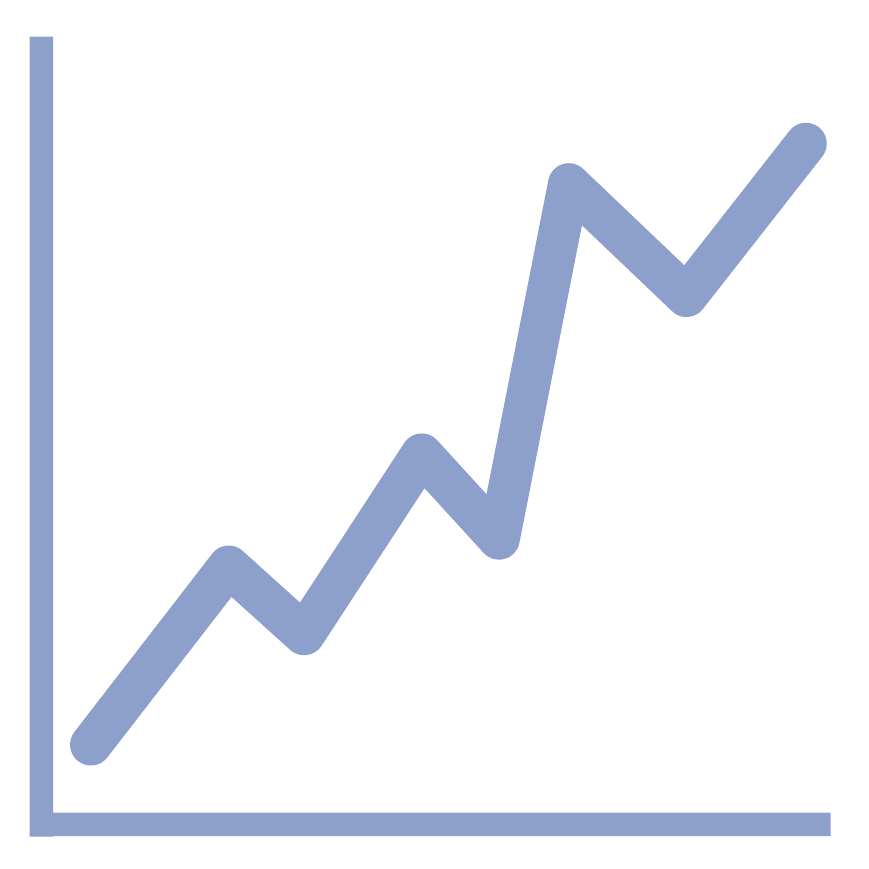}
        \end{minipage}
    \end{wrapfigure} 
    \noindent
    \para{Line chart} Variants of the line chart were evaluated on characterizing the distribution of data, also called the spread~\cite{albers2014task}. When encodings are position-based, the box plot performed with the highest accuracy, but in the case of color encoding, event stripping performed better. Conventional line charts stood third in the ordering after bar charts and scatterplots for pattern identification. While line charts are usually considered to be the best choice for time series visualization, scatterplots are more effective for showing trends. A study was conducted to merge the two, automatically selecting the right representation for trend exploration in time series data~\cite{wang2018line}. The choice was affected by the amount of noise, outliers in the data, and aspect ratio. When the noise was small, a line chart is preferred, whereas a scatterplot was preferred with the larger noise.

     \begin{wrapfigure}[3]{L}{0.09\linewidth}         \vspace{-10pt}         \begin{minipage}[t]{1.2\linewidth}
        \includegraphics[width=28pt]{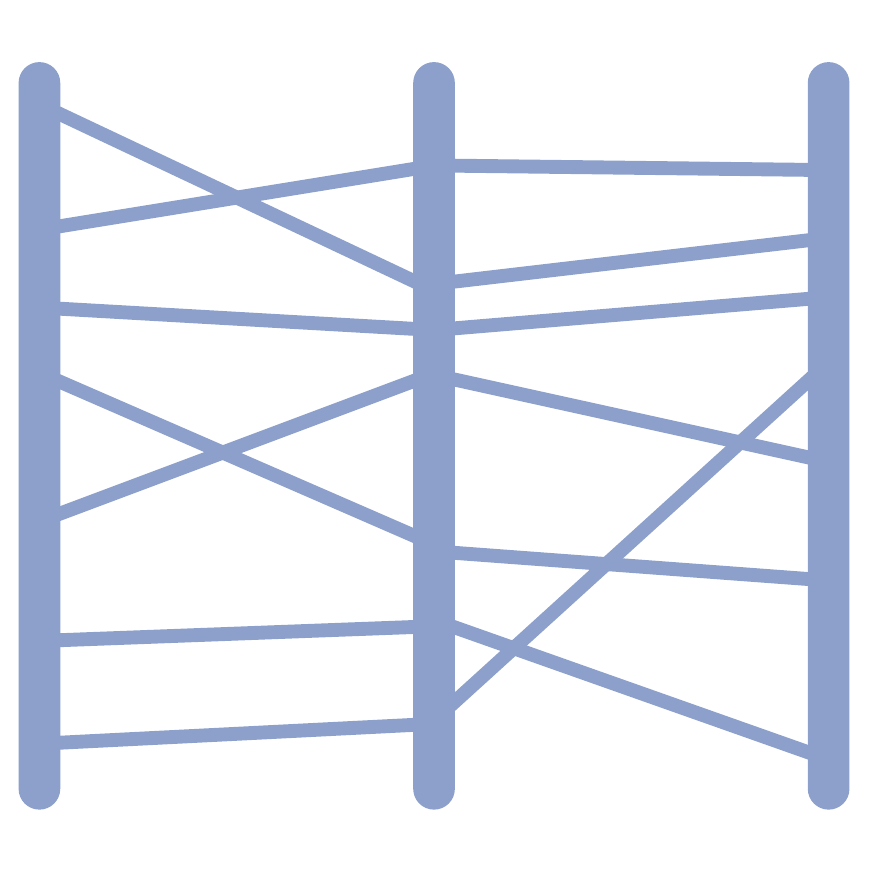}
        \end{minipage}
    \end{wrapfigure}     
    \noindent
    \para{Parallel Coordinates} A new technique was combined with parallel coordinate, called orientation-enhanced parallel coordinates, to overcome the clutter due to overplotting for characterizing the underlying data distribution~\cite{raidou2015orientation}. This method improves outlier discernibility by visually enhancing parts of each parallel coordinates polyline through its slope. Interactive evaluation verified the feature of the discernibility of information in complex data.

    \begin{wrapfigure}[3]{L}{0.09\linewidth}         \vspace{-10pt}         \begin{minipage}[t]{1.2\linewidth}
        \includegraphics[width=28pt]{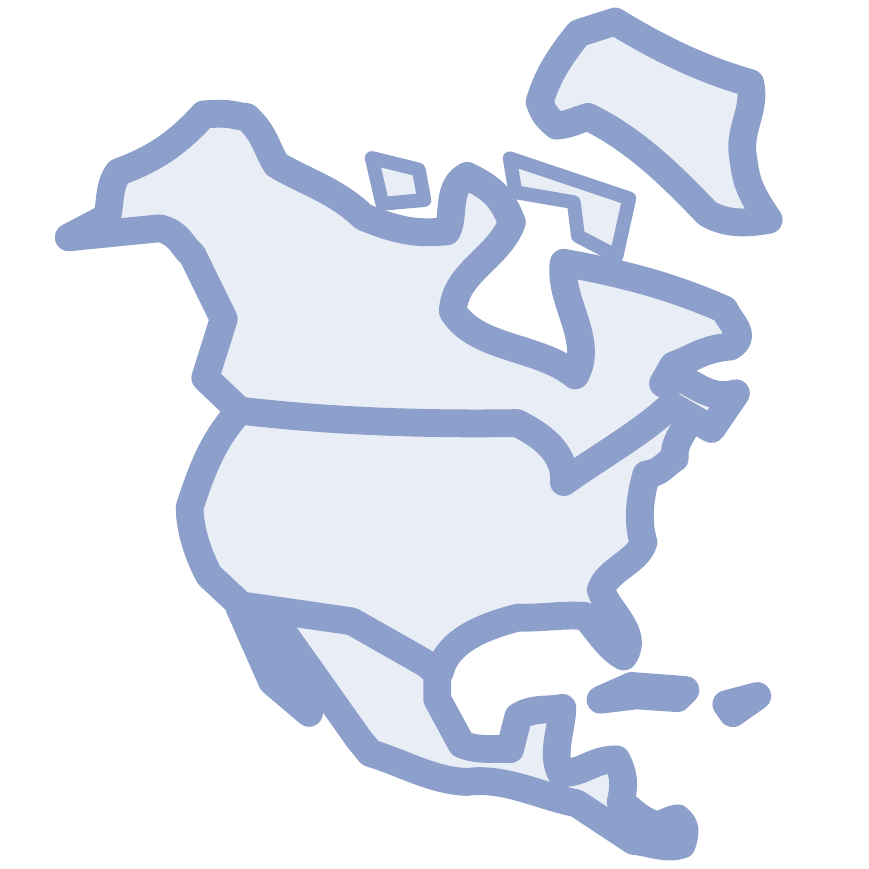}
        \end{minipage}
    \end{wrapfigure}     
    \noindent
    \para{Maps} Cartograms are popular for geo-referenced data visualizations used to illustrated patterns and trends in the map. Major types of cartograms (e.g., contiguous, non-contiguous, rectangular, Dorling, etc.) were evaluated for comparing trends and analyzed on quantitative performance analysis in terms of completion time and error and subjective preferences~\cite{nusrat2018evaluating}.  Dorling cartograms performed best on the task involving comparing trends, whereas rectangular performed worst.

    \para{Other} Viewers accurately estimate trends in many standard visualizations of bi-variate data. However, visual features, e.g., bias within a bar of visualization, and data features, e.g., outliers, can result in visual estimates that systematically diverge from standard least-squares regression models~\cite{correll2017regression}. Designers should be aware of the distinction between regression by eye and explicit statistical information.

\subsubsection{Summary} Scatterplots and line charts have received the majority of attention for the task of characterizing the data distribution. Scatterplots support faster and easier identification of distributions and patterns in data, followed by line charts. A line chart should still be used when the patterns or trends are from time-varying data. Pattern perception on colormaps or chloropleth maps should be sure not to use low color combined with lower spatial frequency. Otherwise, performance will suffer.

\subsection{Cluster}
\label{sec:task-cluster}

    \vspace{-5pt}
    \noindent
    \begin{minipage}[t]{\linewidth}
    \begin{wrapfigure}[4]{L}{0.145\linewidth}
        \vspace{-10pt}
        \begin{minipage}[t]{1.2\linewidth}
            \includegraphics[width=45pt]{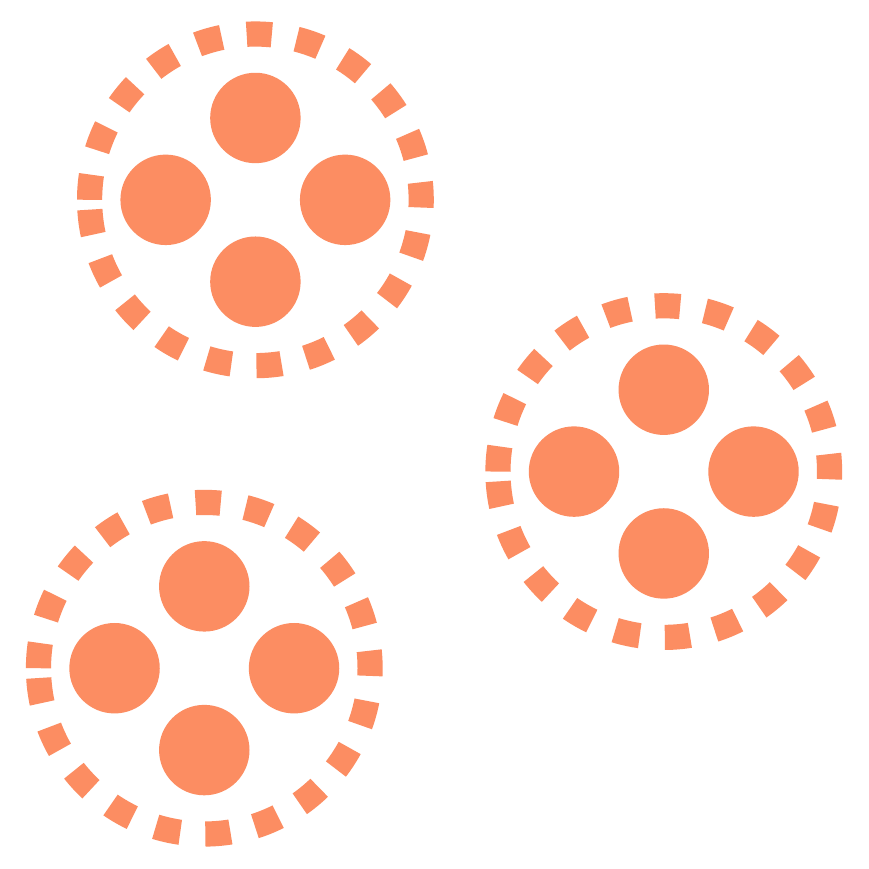}
        \end{minipage}
    \end{wrapfigure} 
    Clustering tasks are focused on identifying a similar attribute in a given set of data. A design factor survey study on information visualization defines clustering as a {high-level data characterization}---\textit{``the ability to identify groups of similar items''}~\cite{sarikaya2018design}. Clustering and segmentation of data points in a given dataset reveal characteristics of data and allow visualization designers and practitioners to explore more about the data~\cite{sarikaya2018scatterplots}.
   \end{minipage}

   \subsubsection{Visual Encoding} \ 

    \vspace{-10pt}
    \begin{wrapfigure}[6]{L}{0.09\linewidth}         \vspace{-10pt}         \begin{minipage}[t]{1.2\linewidth}
        \includegraphics[width=28pt]{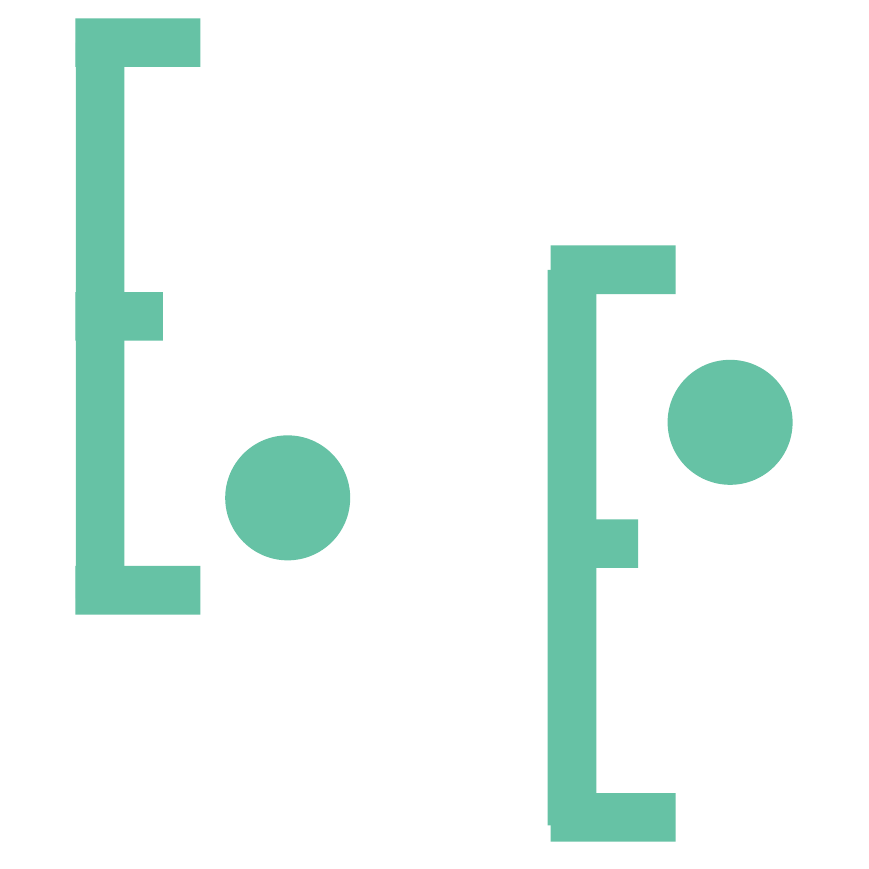}
        
        \vspace{3pt}
        \includegraphics[width=28pt]{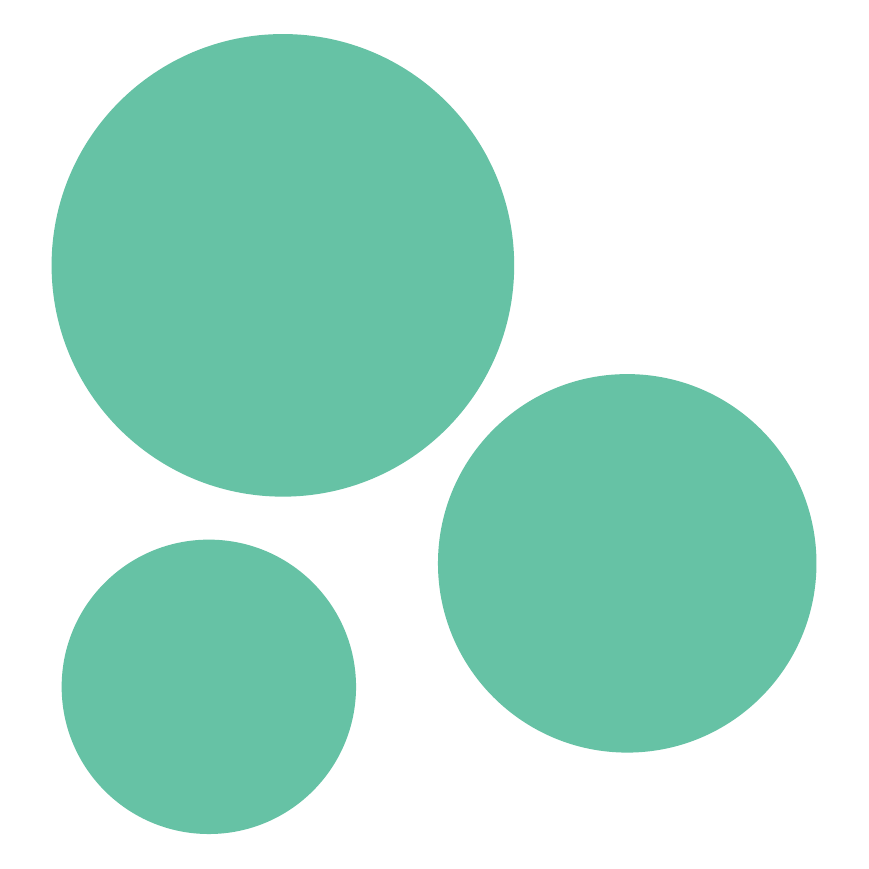}
        \end{minipage}
    \end{wrapfigure} 
    \noindent
    \para{Spatial Position, Shape, \& Size} 
    Symbol or mark size is an influencing visual encoding that affects the density and concentration of point clustering. 
    Additionally, symbol size has a direct influence on identifying the clusters in data, and studies have demonstrated that their discriminability is task-dependent. Li et al.\ demonstrated the effect of mark size and lightness perception on viewers’ ability in multi-class scatterplots for clustering-based tasks~\cite{li2010size,li2010lighntness}. Separability between the symbols or groups of symbols is an important factor in identifying clusters based on the encoding marks. For example, mark shape significantly affects the perception of both size and color, and separability among the three encodings function asymmetrically~\cite{smart2019measuring}.

    Since concentrations of density influence cluster perception, as the size of data points increases, so does the concentration and density. Sadahiro developed a mathematical model to represent cluster perception in point distributions based on proximity, concentration, and density change, and he suggested perception was significantly influenced by the concentration and density change~\cite{sadahiro1997cluster}. 
    Varying densities and gaps between groups of points influence the pattern of data points, potentially forming clusters of points~\cite{o1974human}. 
    A study focusing on the perceptual optimization of scatterplot design studied standard design parameters, including mark size, opacity, and aspect ratio, and it demonstrated that effective choices of the variables enhanced class separation~\cite{micallef2017towards}.

    \begin{wrapfigure}[3]{L}{0.09\linewidth}   
    \vspace{-10pt}        
    \begin{minipage}[t]{1.2\linewidth}
        \includegraphics[width=28pt]{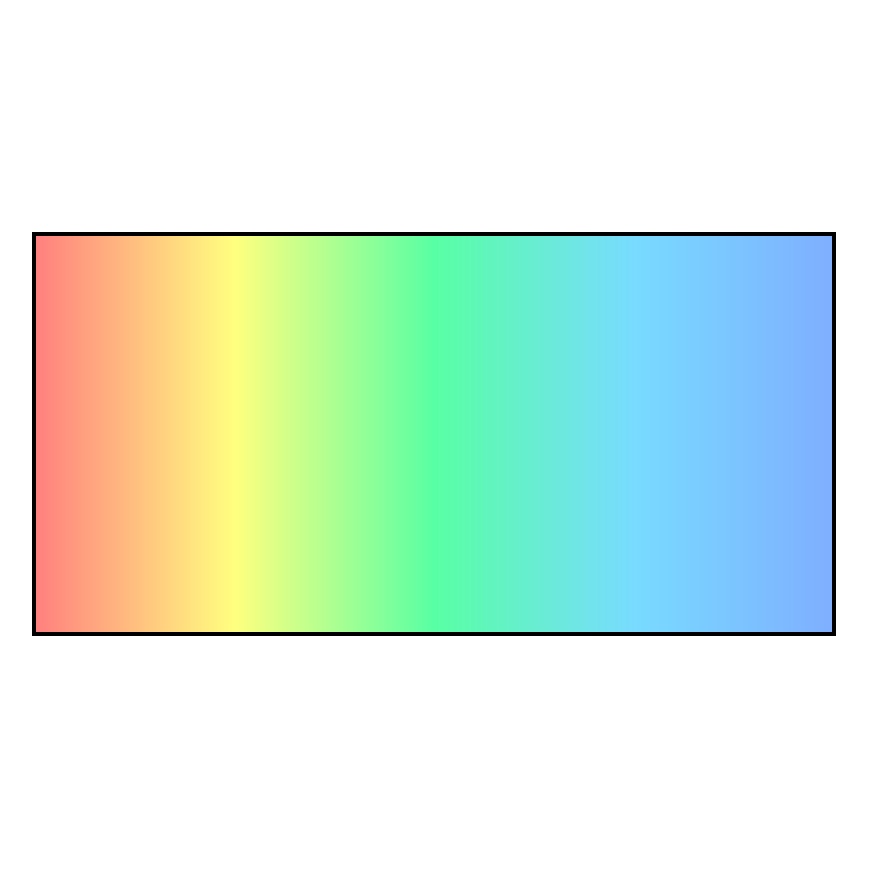}
        \end{minipage}
    \end{wrapfigure}
\begin{figure}[!t]
        \centering
      \includegraphics[ width=0.975\linewidth]{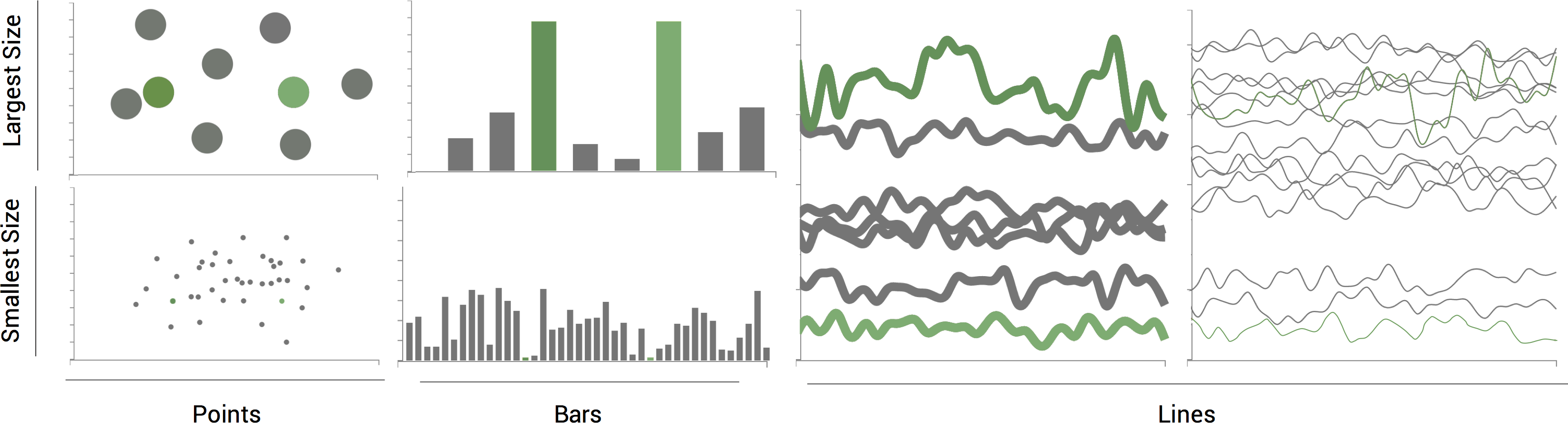}
         \caption{The perceived color difference varies inversely with size, and elongated marks provide significantly higher discriminability. Colors on longer marks were also more discernible than shorter bars of equal thickness. The perceptible color differences for lines vary inversely with thickness. Finally, perceptible color differences for points vary inversely with point diameter. Image reproduced with permission~\cite{szafir2018modeling}.}
        \label{fig:szafir_modelling_2018}
\end{figure}

 
    \noindent
    \para{Color Hue} Color is another important visual encoding of a scatterplot that influences the visual task of segmentation or clustering and grouping, but how we measure the color difference perception varies inversely with mark diameter~\cite{szafir2018modeling} (see \autoref{fig:szafir_modelling_2018}). Furthermore, the shape of the symbol significantly affects how well we perceive the color difference~\cite{smart2019measuring}.
    Hence, an optimized choice of colors aids users in efficiently understanding separability in multi-class scatterplots. They used a method of color assignment to design scatterplots that optimized class separability perception taking into account density-related factors, such as spatial relationship, density, degree of overlaps between points and cluster, and background color~\cite{wang2018optimizing}, which could not be achieved by the default colormapping.

    \begin{wrapfigure}[3]{L}{0.09\linewidth}         \vspace{-10pt}     
    \begin{minipage}[t]{1.2\linewidth}
        \includegraphics[width=28pt]{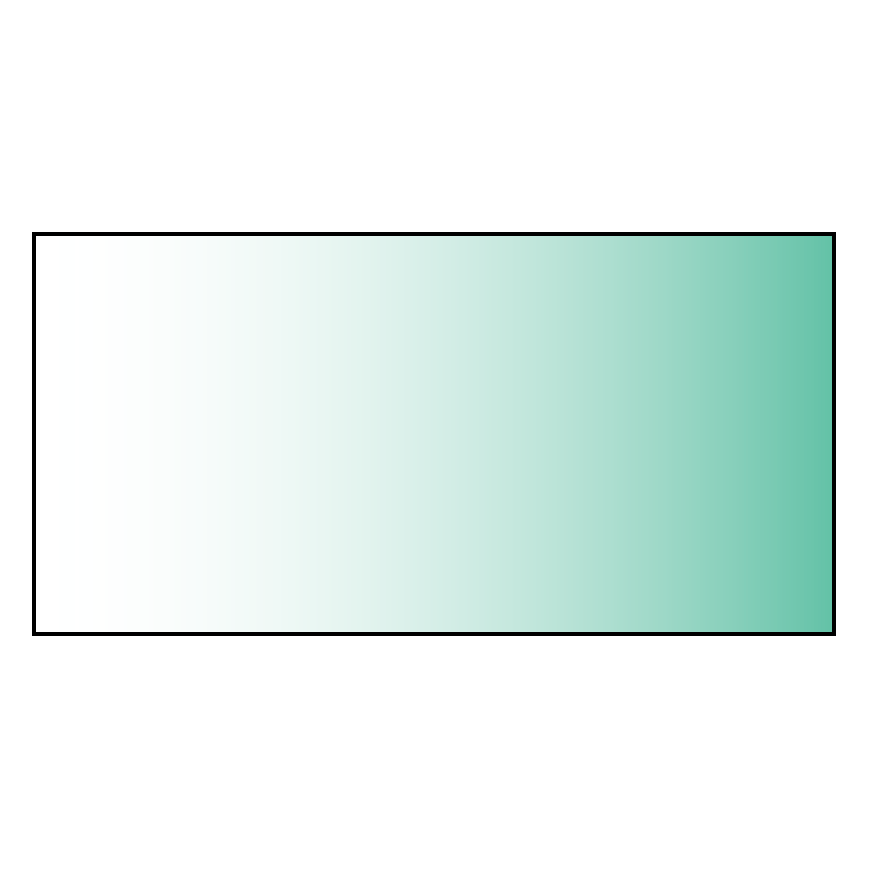}
        \end{minipage}
    \end{wrapfigure} 
    \noindent
    \para{Color Intensity} Reducing mark opacity can alleviate overplotting to aid various visual analytics tasks, e.g., cluster perception or identification~\cite{sarikaya2018scatterplots, matejka2015dynamic,quadri2020modeling} while preserving the spatial information.
    Different opacity levels aid in enhancing class separation---while low opacity benefits density estimation for extensive data, it also makes locating outliers more difficult~\cite{micallef2017towards}.
    Of course, these solutions have their limits, e.g., when opacity is below available precision, or points are their smallest possible size~\cite{few2008solutions}.

   \para{Bias} Priming and anchoring effect distorts the user's decision-making process. Deciding the separability of the two clusters depends not only on how far they are apart but also on previously seen stimuli. Valdez et al.\ elaborated on how repeated exposure to a visualization impacts our interpretation~\cite{valdez2018priming}. While the authors presented the biases caused by these effects, at the same time, the results came with the caveat that judgments were not normally distributed, which further indicating an overestimation of the effect size. The effects can be accidentally caused by chromostereopsis~\cite{allen1981chromostereopsis} and should be studied on monochrome colors.

\subsubsection{Visualization}
    
    Saket et al.'s study found that, with small data, bar charts and pie charts outperformed tables, scatterplots, and line charts in clustering tasks~\cite{saket2018task} (see \autoref{fig:saket2018task}). The performance in cluster perception in pie charts can be traced back to its effectiveness in facilitating proportional judgments through a part-to-whole relationship~\cite{eells1926relative,spence1991displaying}.

    \begin{wrapfigure}[3]{L}{0.09\linewidth}         \vspace{-10pt}         \begin{minipage}[t]{1.2\linewidth}
        \includegraphics[width=28pt]{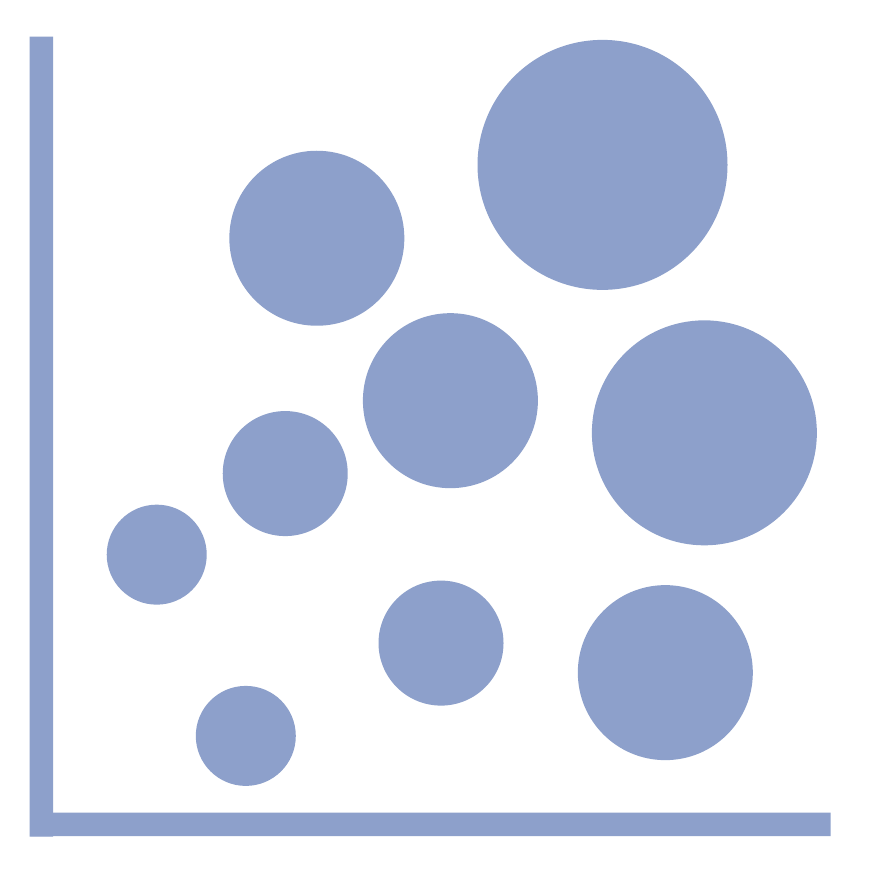}
        \end{minipage}
    \end{wrapfigure} 
    \noindent
    \para{Scatterplot}
    Scatterplots are widely used to visualize data to reveal patterns of data characteristics such as class segmentation or clusters. Numerous studies on visual cluster separation have been conducted to identify the effects of various design factors and visual encodings.

    A qualitative evaluation of cluster separation measures study suggested taxonomy of four factors---scale, point distance, shape, and position--that influence separation perception~\cite{sedlmair2012taxonomy}. With the focus class separation, Sedlmair and Aupetit's extended their prior work and evaluated 15 state-of-the-art class separation measures in a study aimed at mitigating human judgment impacts. They rely on human ground truth as input to a machine learning framework that was used for evaluating the quality of dimension reduction~\cite{sedlmair2015data}. A further continuation of the work included even more measures for improved matching to human perception~\cite{aupetit2016sepme}. The ScatterNet method captured perceptual similarities between scatterplots by applying a deep learning model that was designed to emulate human clustering decisions~\cite{ma2018scatternet}.  The scagnostics technique focused on identifying the patterns in scatterplots, including clusters~\cite{dang2014transforming,matute2017skeleton}. However, a recent study showed that scagnostics do not reliably reproduce human judgments~\cite{pandey2016towards}. The commonality in all of these studies is that they are algorithmically oriented, and most of their evaluations did not consider visual channels.

    Sadahiro presented a mathematical model that suggested perception is significantly influenced by the concentration and density change~\cite{sadahiro1997cluster}. 
    ClustMe used visual quality measures to model human judgments to rank scatterplots~\cite{abbas2019clustme}.  ClustMe performed well in reproducing rational decisions for cluster patterns. Quadri and Rosen built and tested a topology-based model of human perception of scatterplots that considered data distribution, number of data points, size of data points, and opacity of data points in cluster perception~\cite{quadri2020modeling}. Their model demonstrated the strong relationship between all of these factors and the perception of cluster separation.

    \begin{wrapfigure}[3]{L}{0.09\linewidth}         \vspace{-10pt}         \begin{minipage}[t]{1.2\linewidth}
        \includegraphics[width=28pt]{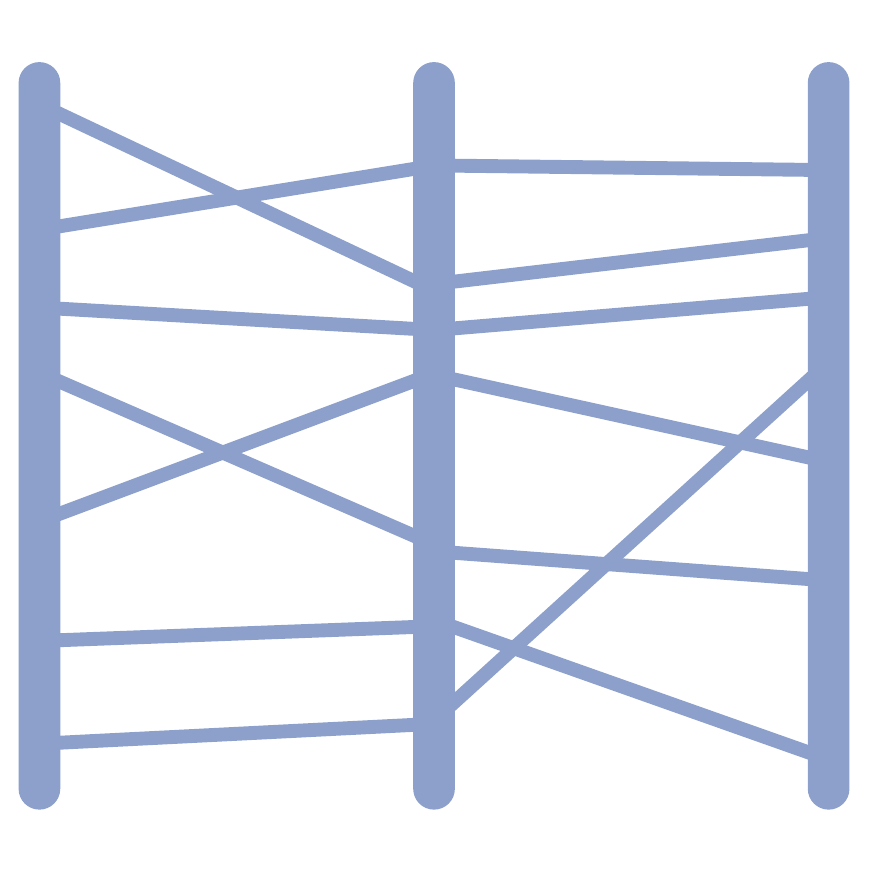}
        \end{minipage}
    \end{wrapfigure} 
    \noindent
    \para{Parallel Coordinates} Parallel coordinates are useful for a variety of tasks, including clustering in real-world applications~\cite{nguyen2018dspcp}. One study that measured the participants' learning outcomes used clustering as a primary task~\cite{kwon2016comparative}. The results showed a more engaging experience for interactive parallel coordinates than static, and the users did not find it challenging to learn the data item mapping to the parallel axes. 
    
    Orientation-enhanced parallel coordinates were developed especially for large datasets to improve the display of emphasizing the underlying data structure, such as allowing the discernibility of clusters~\cite{raidou2015orientation}. The evaluation demonstrated that orientation made it easier to identify the data clusters between the two data dimensions and allowed multiple small clusters between the first two dimensions to be visually enhanced. Ultimately, there remains some question as to whether scatterplots or parallel coordinates are better for identifying clusters, as one study found that parallel coordinates better showed the actual shape of clusters~\cite{holten2010evaluation}.

    \begin{wrapfigure}[3]{L}{0.09\linewidth}         \vspace{-10pt}         \begin{minipage}[t]{1.2\linewidth}
        \includegraphics[width=28pt]{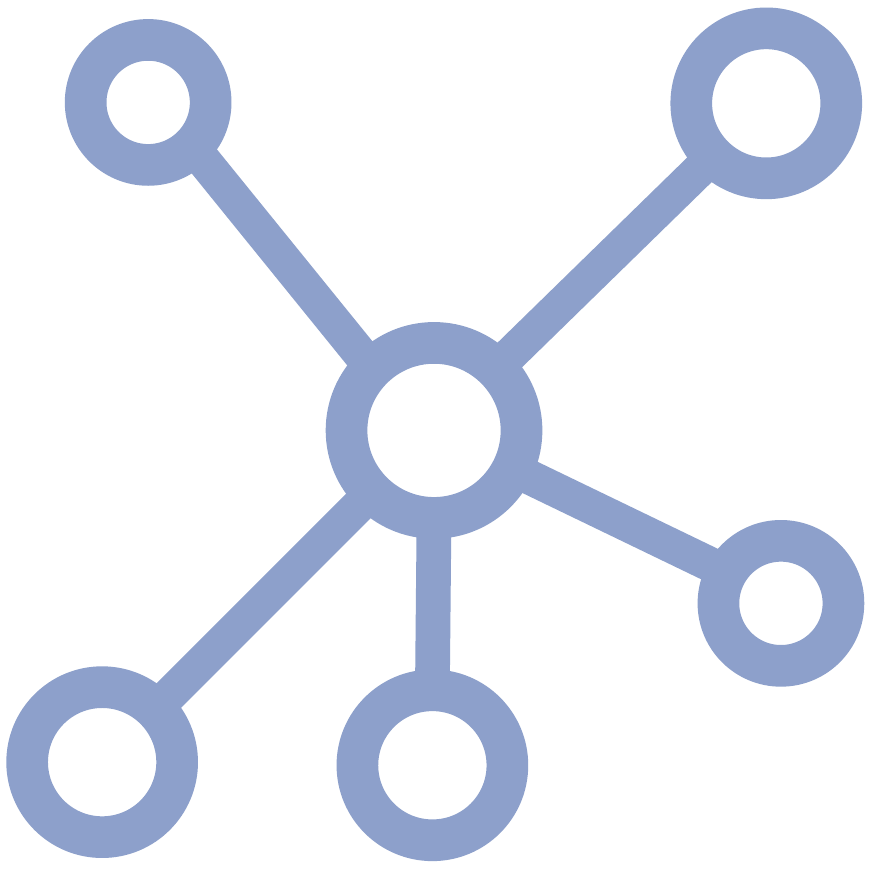}
        \end{minipage}
    \end{wrapfigure} 
    \noindent
    \para{Networks} Graph layout algorithms optimize visual characteristics of visual encodings to create intuitive visualizations. 
    The layout of graphs was investigated in an evaluation study where participants produce their graph from a graph shown earlier to depict clusters~\cite{van2008perceptual}. Users reach confidence and higher overall task accuracy in visualization during the interaction when rapidly adjusting the visual encodings.

     \begin{wrapfigure}[3]{L}{0.09\linewidth}         \vspace{-10pt}         \begin{minipage}[t]{1.2\linewidth}
        \includegraphics[width=28pt]{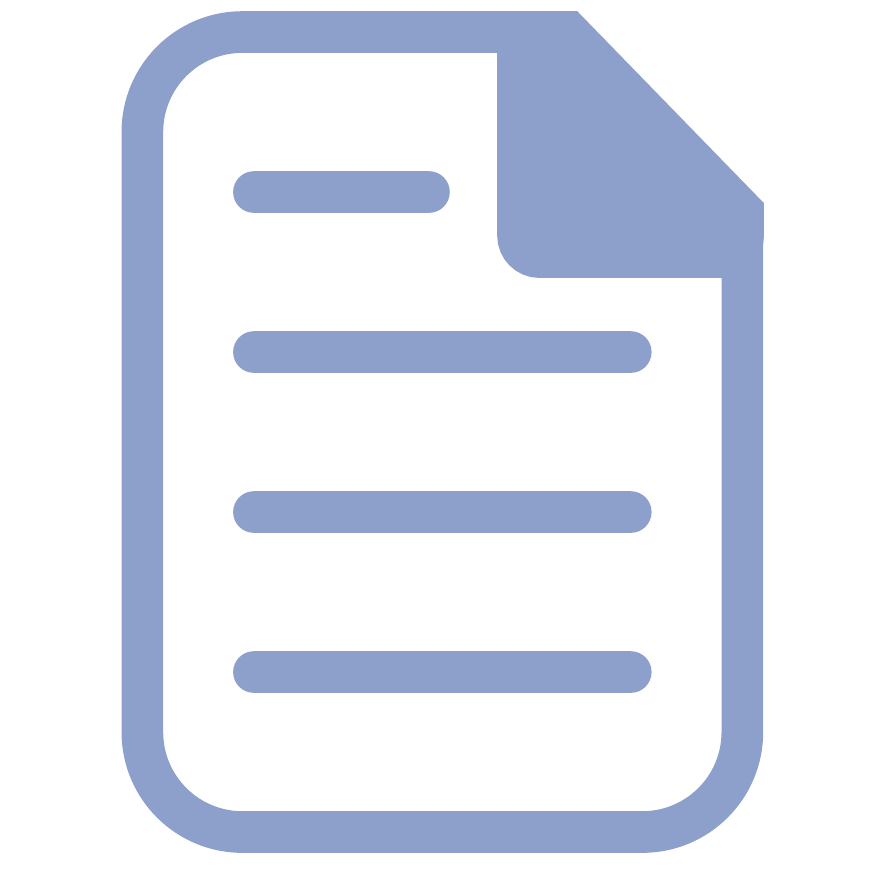}
        \end{minipage}
    \end{wrapfigure} 
    \noindent
     \para{Text} A study to investigate word recall memory, based on classic word clouds layout and font size, stated that word placement (i.e., sorting, clustering, spatial arrangement) is an important consideration as it affects the users' recall potential~\cite{rivadeneira2007getting}.

    \subsubsection{Summary} As scatterplots demonstrate the clustering of bivariate data effectively, most studies we identified used scatterplots. In the case of multivariate data, parallel coordinates and scatterplot matrices can be used to visualize data.

\subsection{Correlate}
\label{sec:task-correlation}

 \vspace{-5pt}
    \noindent
    \begin{minipage}[t]{\linewidth}
    \begin{wrapfigure}[4]{L}{0.145\linewidth}
        \vspace{-10pt}
        \begin{minipage}[t]{1.2\linewidth}
            \includegraphics[width=45pt]{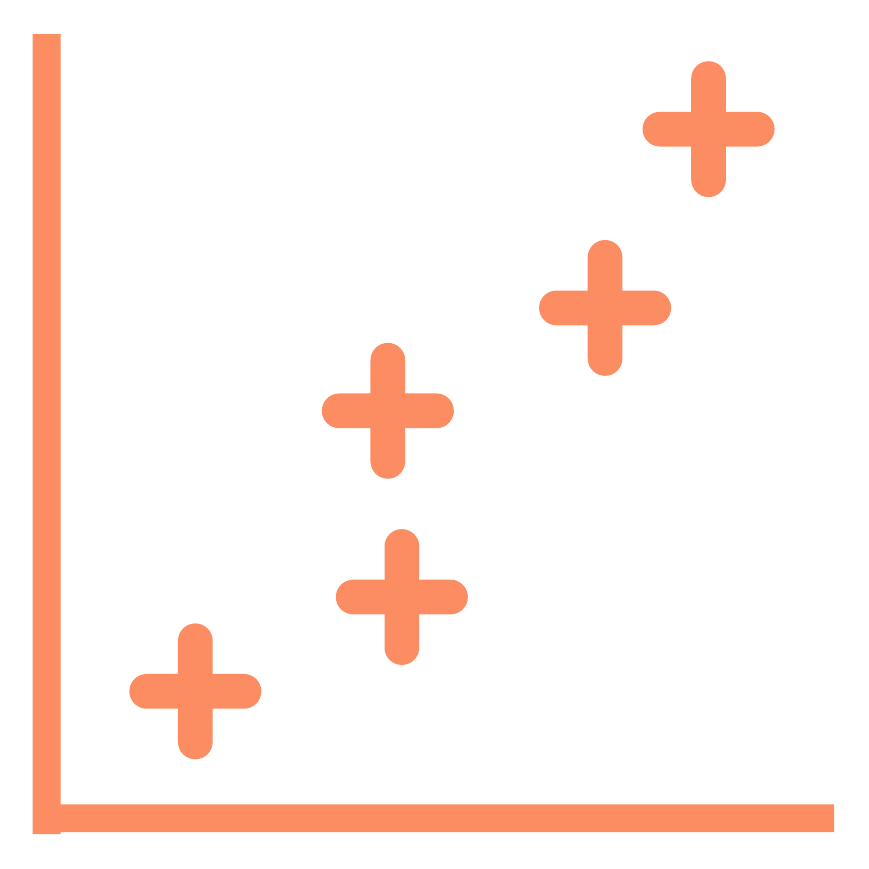}
        \end{minipage}
    \end{wrapfigure} 
    Correlation, in general, is a relationship between values of two or more attributes in a dataset. Formally, correlation is a statistical measure of the linear relationship between two quantitative variables, represented by a correlation coefficient. A positive correlation indicates the extent to which those variables increase in parallel, and a negative correlation indicates the extent to which one variable increases as the other decreases. 
    \end{minipage}

    \subsubsection{Visual Encoding}
    
     Evaluating the correlation perception helps identify people's abilities to perceive and judge differences in visualizations. Numerous studies have focused on identifying linear correlation in a data distribution with visual encodings, including the slope of the points, marker size, opacity, and color~\cite{rensink2010perception, harrison2014ranking, cleveland1982variables, li2010judging, micallef2017towards}. Rensink and Baldridge showed that perception of correlation in scatterplots could be mathematically modeled using the perceptual laws of {Weber's law}~\cite{rensink2010perception}. Further, Chang et al.~\cite{chang2016vision} also use Weber's law to provide a guide for practitioners to select a visualization.

\begin{figure}[!b]
    \centering
    
    \includegraphics[width=0.975\linewidth]{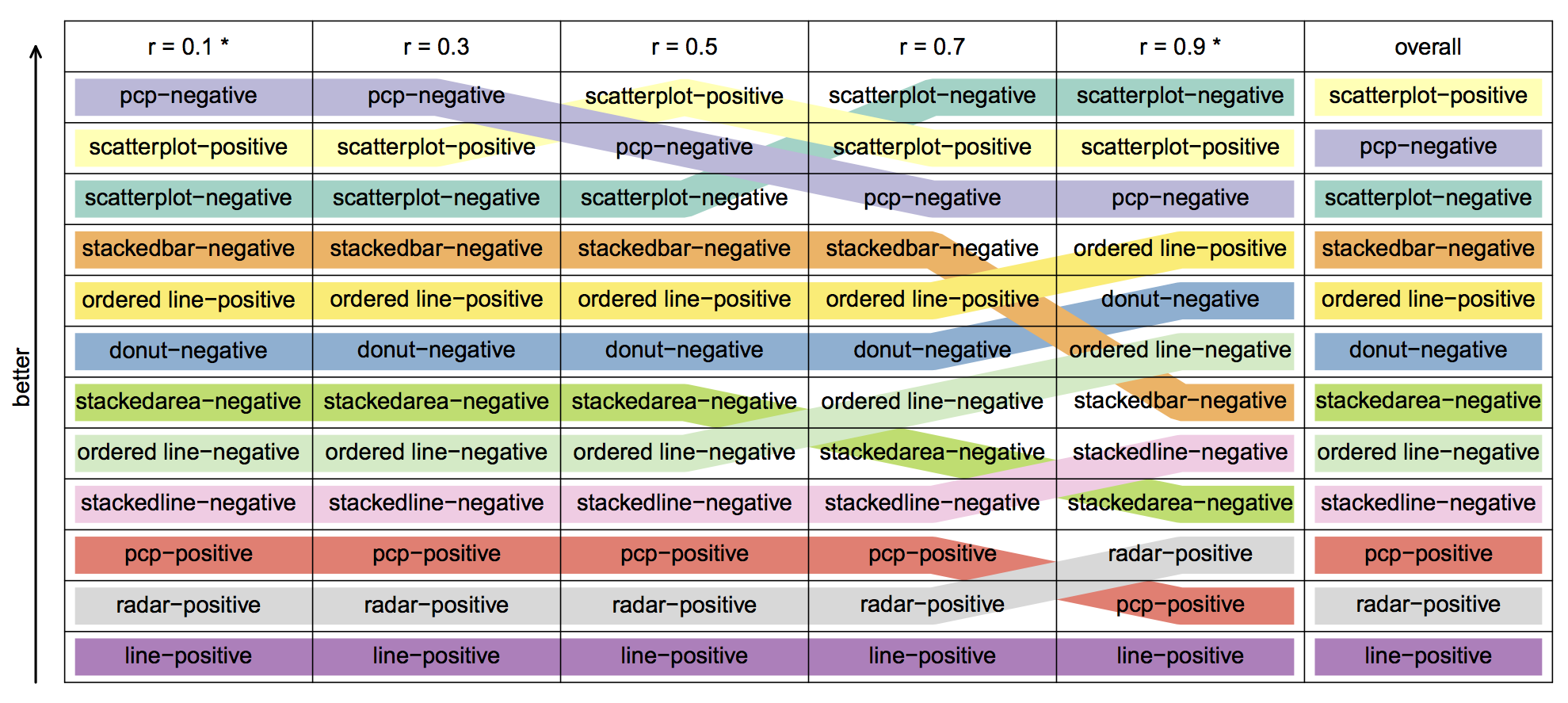}

    \caption{Harrison et al.\ leveraged perceptual laws to evaluate and rank the effectiveness of visualizations for representing correlation. One interesting finding is that judgment precision had a striking variation between negatively and positively correlated data on certain visualizations, e.g., parallel coordinates. Image reproduced with permission~\cite{harrison2014ranking}.} 
    \label{fig:harrison2014ranking}
\end{figure}

\begin{figure}[!t]
    \centering
    
    \includegraphics[width=0.975\linewidth]{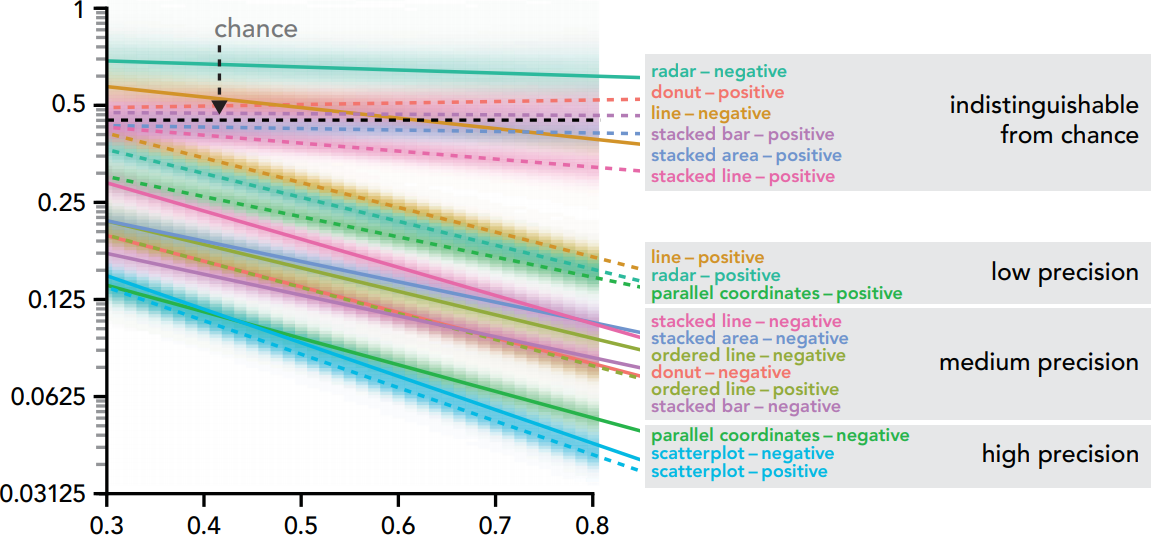}
    
    \caption{The study by Kay and Heer presented a series of refinements to the model presented by Harrison et al.~\cite{harrison2014ranking} (see \autoref{fig:harrison2014ranking}), including the incorporation of individual differences, log transformation, and Bayesian modeling. The left side shows the posterior probability distribution over the mean log (JND) for each value of $r$ using the Bayesian censored log-linear model. The right side shows the ranking and grouping of visualizations based on how precise people’s estimations of correlations are (lower JND implies higher precision). The new model demonstrated notable differences, e.g., parallel coordinates–negative and scatterplot–negative swap positions. Image reproduced with permission~\cite{kay2016beyond}.}
    \label{fig:kay2016beyond}
\end{figure}

    \subsubsection{Visualization}
    
    Identifying correlation supports the decision-making process in multivariate data visualization. A comparative study of parallel coordinates, tabular visualization, and scatterplot matrix reported tabular visualization and the scatterplot matrix have high accuracy, whereas parallel coordinates stand out in terms of completion time~\cite{dimara2018conceptual}.

    The Harrison et al.\ study compared and ranked scatterplots to other visualization methods ({parallel coordinates, stacked area charts, stacked bar charts, stacked line charts, line charts, ordered line charts, radar charts, and donut charts}) for correlation perception tasks and stated that Just-Noticeable Difference (JND) in correlation could be modeled by Weber's law~\cite{harrison2014ranking} (see \autoref{fig:harrison2014ranking}). Kay and Heer proposed a log-linear-based re-analysis of the results of Harrison et al.'s work~\cite{kay2016beyond} (see \autoref{fig:kay2016beyond}). The critical finding of these works was that scatterplots stand above all other tested visualizations for precision on both positive and negative correlations.
    
    As a counterpoint, Saket et al.'s reported that line graphs stand out for detecting correlation in terms of accuracy, time, and user preference, followed by scatterplots~\cite{saket2018task}. However, the experiments only used a small number of points, adding further complexity to the discussion of which method is truly most effective for identifying correlations. Nevertheless, the results are supported by earlier research reporting the effectiveness of line charts for trend-finding tasks~\cite{zacks1999bars}.

    \begin{wrapfigure}[3]{L}{0.09\linewidth}         \vspace{-10pt}         \begin{minipage}[t]{1.2\linewidth}
        \includegraphics[width=28pt]{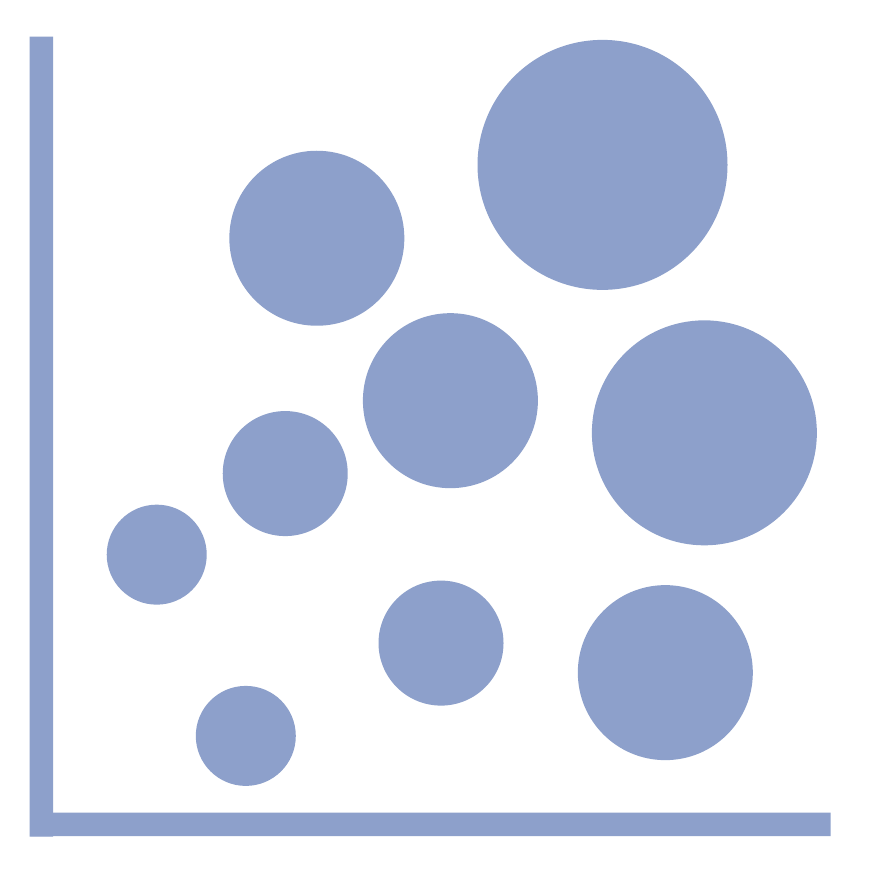}
        \end{minipage}
    \end{wrapfigure} 
    \noindent
   \para{Scatterplot}     
    Correlation is one of the more extensively studied tasks in this taxonomy, particularly concerning scatterplots. One of the first perceptual studies on correlation on scatterplots by Bobkop and Karren~\cite{bobko1979perception} formed the basis for many other experiments. They measured a direct estimation of Pearson's product-moment correlation coefficient, like many other correlation perception studies~\cite{cleveland1982variables, lauer1989density, meyer1997correlation, pollack1960identification, nguyen2016correlation}.  However, Sher et al.\ conducted a study about measuring the offsets of human perception of correlation when changing visual variables, and they found that humans perceive correlations differently compared to the statistical measure of Pearson’s product-moment correlation coefﬁcient~\cite{sher2017empirical}.

    Correlation was also among the first visualization tasks to be studied using JNDs in four experiments~\cite{doherty2007perception}. The study found that users become more confused with low correlation values than with high correlation values. The reason behind the effect was that the user's judgment of discriminability increased with the increased strength in the variable association. 
    Rensink's perception of correlation study opened a new door to effective visualization design by demonstrating a precise way perception of correlation in a scatterplot could be modeled using Weber's law and JND~\cite{rensink2010perception}. Their design studies were focused on the comparison of performance using dot size and density.

    Best et al.\ investigated correlation coefficient estimation in the laboratory-based study where they found that human brain activity during correlation perception increases as correlation is decreased~\cite{best2006perceiving}. Findings indicated that different relationships on a scatterplot are processed differently. However, perceptually, scatterplot processing was similar, and participants used visual features to code the pattern.

    A recent comparative survey included all correlation studies in a timeline taxonomy and investigated a hypothesis---viewers attend to a small number of visual features, e.g., shape, dispersion, and orientation of scatterplots to discriminate correlation in scatterplots~\cite{yang2018correlation}. The report compares the findings with previous pivotal correlation perception studies using {Weber's law}.
    Scaling is another shape property that plays an important role in identifying the correlation between the variables. 
    An early study by Cleveland et al.\ reported that variables in a scatterplot look more correlated when scales are increased~\cite{cleveland1982variables}.

\begin{wrapfigure}[3]{L}{0.09\linewidth}         \vspace{-10pt}         \begin{minipage}[t]{1.2\linewidth}
        \includegraphics[width=28pt]{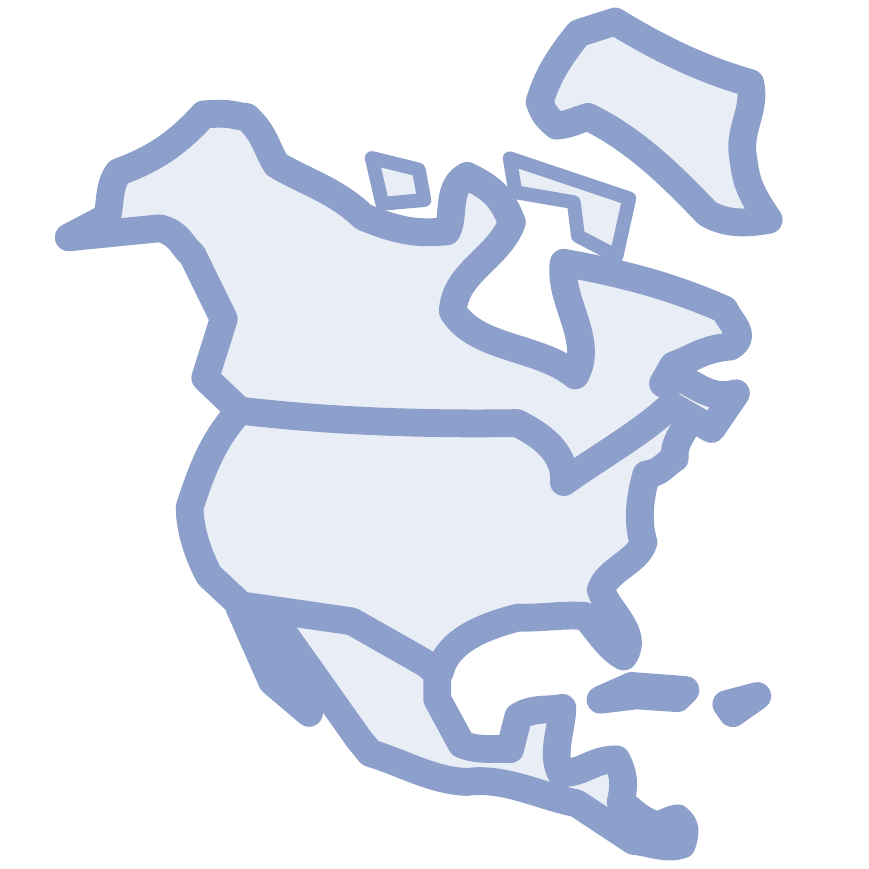}
        \end{minipage}
    \end{wrapfigure}    
    \noindent
    \para{Map} Visualizations on maps show spatial data, and design variations are useful in identifying correlation over geo-temporal data variables. Map lineups used JNDs for evaluating correlation on choropleth maps~\cite{beecham2017map}. They model the differences in visual stimuli of map color and are useful in controlling spatial auto-correlation and increase user's confidence. The comparison of average JNDs measured across all three geographies demonstrated that JND increases and becomes more noticeable when irregularities become more regular.

    The encoding choice on the geophysical maps has a direct influence on correlation judgments. A study on geo-temporal visualization investigated the task of identifying the correlation between two variables that evolve over time and space~\cite{pena2019comparison}. The findings showed that the design choices of geo-temporal multivariate data visualization would impact how users detect a correlation between variables over space and time. The vital design guidelines from this study are: (1)~small multiple visualizations on maps are better for identifying correlation at a specific point in time; and (2)~for identifying correlation for time steps on single maps, bar charts are better than other choices.

    \para{Other} Design variation, such as small multiples, overlay, and mirroring on three visualizations (bar, slope, and donut charts), were shown to influence the correlation task~\cite{ondov2018face}. In the study, participants struggled to use motion animation to extract and compare the correlation between data sets, whereas mirrored arrangements over adjacent arrangements achieved precise results. The comparison arrangement design evaluation confirms the prior hypothesis of better performance in overlaid designs versus small multiples.

    \subsubsection{Summary} The majority of the correlation studies have been conducted on scatterplots, but recent works have also diverged towards other visualization types. Studies have by-and-large shown that either scatterplots or parallel coordinates should be used for correlation tasks. However, their performance may interchange with positive or negative correlation due to the different representations of positive and negative correlations in parallel coordinates. For spatial data correlation choropleth maps can be an effective approach.

    \subsection{Compare}
    \label{sec:task-compare}

    \vspace{-5pt}
    \noindent
    \begin{minipage}[t]{\linewidth}
    \begin{wrapfigure}[4]{L}{0.145\linewidth}
        \vspace{-10pt}
        \begin{minipage}[t]{1.2\linewidth}
            \includegraphics[width=45pt]{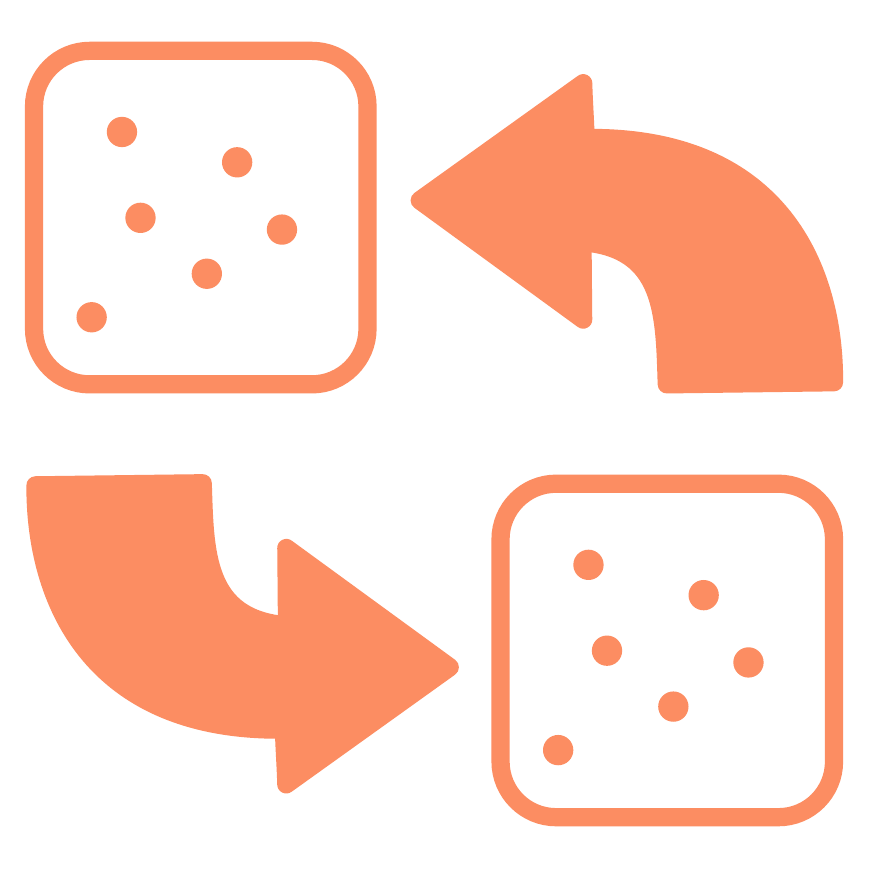}
        \end{minipage}
    \end{wrapfigure}

    \textit{Compare} is a compound task that was mentioned as an intentional ``omission from the taxonomy'' of Amar et al.~\cite{amar2005low}. The task of comparison often involves another subtask, e.g., retrieve a value, compute a derived value, etc., followed by comparison operation. Comparison was implicit in many of the prior tasks. For example, the find extremum task (\autoref{sec:task-extremum}) often requires comparing a set of candidate values to the rest of the data, e.g., ``which cars are more fuel-efficient, Japanese cars or American cars?''~\cite{amar2005low}. This section focuses on the performance of comparison tasks based on different visual features in their design of the same visualizations to analyze the perception judgment. Based on the quantity and importance of studies of this type, we have included it as the eleventh task of the taxonomy. 
     \end{minipage}

     \subsubsection{Visualization} \ 
    
    \vspace{-10pt}
    \begin{wrapfigure}[3]{L}{0.09\linewidth}         \vspace{-10pt}         \begin{minipage}[t]{1.2\linewidth}
        \includegraphics[width=28pt]{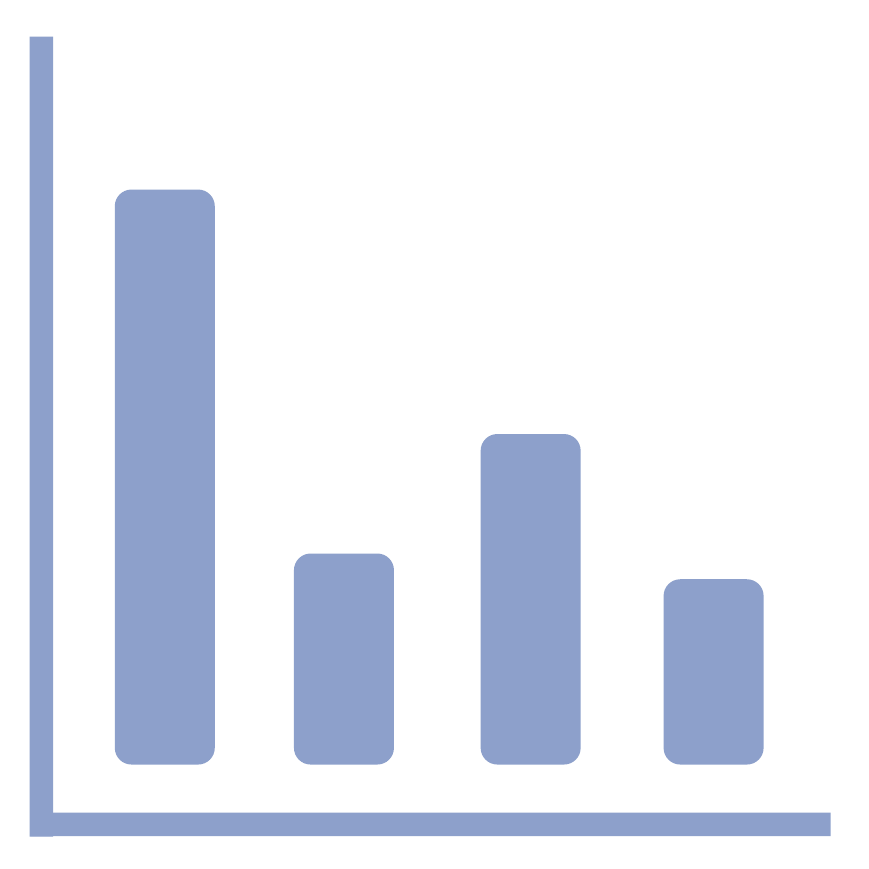}
        \end{minipage}
    \end{wrapfigure} 
    \noindent
    \para{Bar Chart}     The majority of the comparative study performed on bar charts investigate the user performance with design variations.
    The visual comparison task can be traced back to Cleveland and McGill's work, where charts are shown, and different bar chart designs impact the accuracy and comparison of the viewer's perceptual task~\cite{cleveland1984graphical}. 
    The study reported that a comparison between adjacent bars is more accurate than between widely separated bars.

    The findings from that study were extended to four different types of bar charts~\cite{talbot2014four}. Some of the critical findings on bar chart comparisons are: short bars are difficult to compare; the gap between stacked bars can prevent part-to-whole comparison errors; distractors in bar effects unaligned bar comparison; and separating bars in space makes the comparison of their height more difficult. 
    At the same time, comparing the variants of the bar chart design provides different levels of performance on completion time and standard error~\cite{srinivasan2018s}. Variants of bar charts were compared based on the visual features in their designs, and their perception judgments were used to discover error and distraction factors. Charts with different overlays or hybrid designs that combine aspects of juxtaposition and explicit encoding with superposition are just as good or better than sole juxtaposition or explicit encoding-based charts on individual tasks. A comparison helps the designer to identify differences in the representation of data.

    Other forms of design variation in bar charts are embellishments which attach aesthetics to visual display. Different bar charts with visual embellishments were compared to evaluate the accuracy, and the results demonstrate that accuracy was not worse, but at the same time, it did not provide better outcomes~\cite{skau2015evaluation}.

    Comparing a visualization to a mental image is similar to statistical analysis, and thus repeated interpretation of visualization is sensitive to {the multiple comparison problem}---the probability of discovering false insights when visualization is examined more times or compared~\cite{zgraggen2018investigating}. The study was performed on different types of bar charts to measure and compare the accuracy of user-reported insights such as shape, mean, variance, correlation, and ranking.

    Another work on bar charts is compared with Microsoft Excel to show the difference in activities in terms of sequences of action and pipelines~\cite{wun2016comparing}.  Participants tend to follow a linear pattern when using Excel, whereas while using tiles, they followed a cyclical pattern. In a study of five types of ranked list visualization, a comparison of two data items takes the longest time with the lowest accuracy on Zvinca plots~\cite{mylavarapu2019ranked}. A treemap outperformed all other visualizations on accuracy.

   \begin{wrapfigure}[3]{L}{0.09\linewidth}         \vspace{-10pt}         \begin{minipage}[t]{1.2\linewidth}
        \includegraphics[width=28pt]{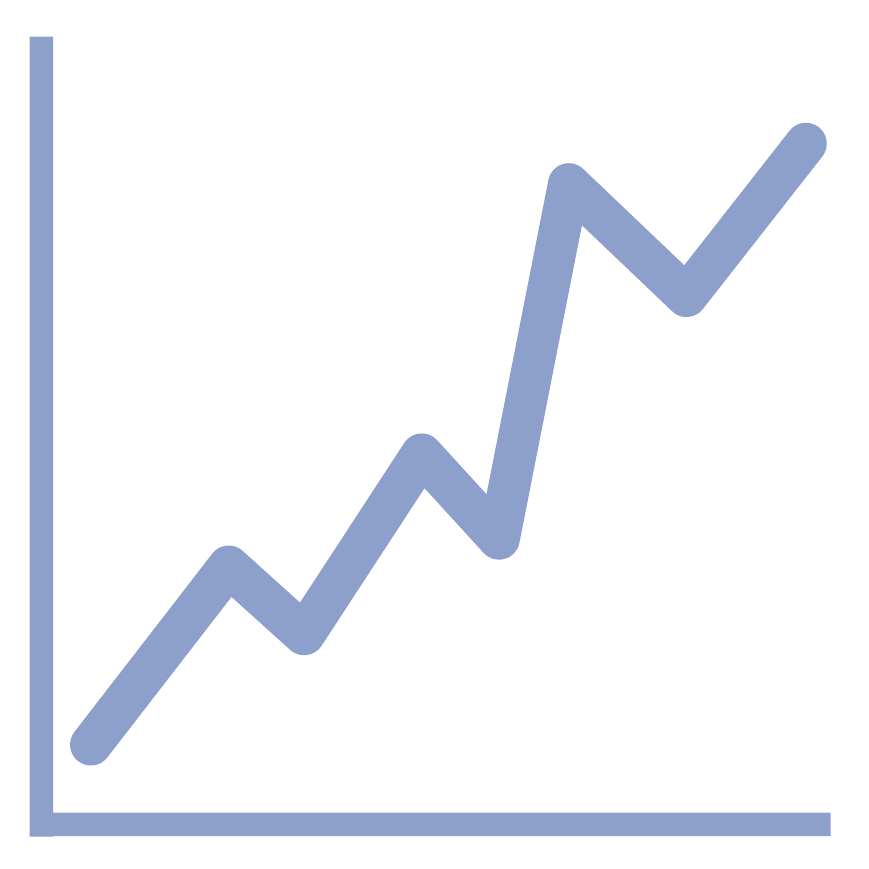}
        \end{minipage}
    \end{wrapfigure} 
    \noindent
     \para{Line Chart} A series of studies have evaluated and compared the performance of line chart variants: {horizon graphs}\footnote{Horizon graphs split the space (mainly vertically) and attempted to optimize the vertical footprint to visualize multiple time series~\cite{javed2010graphical}.}~\cite{reijner2008development,heer2009sizing,perin2013interactive,javed2010graphical}, {colorfields},\cite{gogolou2018comparing,correll2012comparing} {braided graph}~\cite{javed2010graphical}, and {small multiples}~\cite{tufte2001visual}.   A recent study assessed horizon graphs and colorfields, along with a simple line graph on the perception of similarity between time series---two patterns were considered similar irrespective of their amplitudes or their stretching along the time dimension~\cite{gogolou2018comparing}. Layered bands are more useful as chart size decreases.

     In another study, where mirrored and offset horizon graphs were compared, found that estimation error and time increased at four bands in the horizon graphs~\cite{heer2009sizing}. At the same time, the different chart types did not affect the estimation time or accuracy. The effects of chart size of horizon graphs and layering on comparison and estimation showed that horizon graphs performed better than line charts for small chart sizes. 
     A time-series visualization comparison studied line charts, horizon graphs, and color fields in a similarity perception comparison where the choice of visualization affects the viewer's perception of temporal patterns~\cite{gogolou2018comparing}.

     \begin{wrapfigure}[3]{L}{0.09\linewidth}         \vspace{-10pt}         \begin{minipage}[t]{1.2\linewidth}
        \includegraphics[width=28pt]{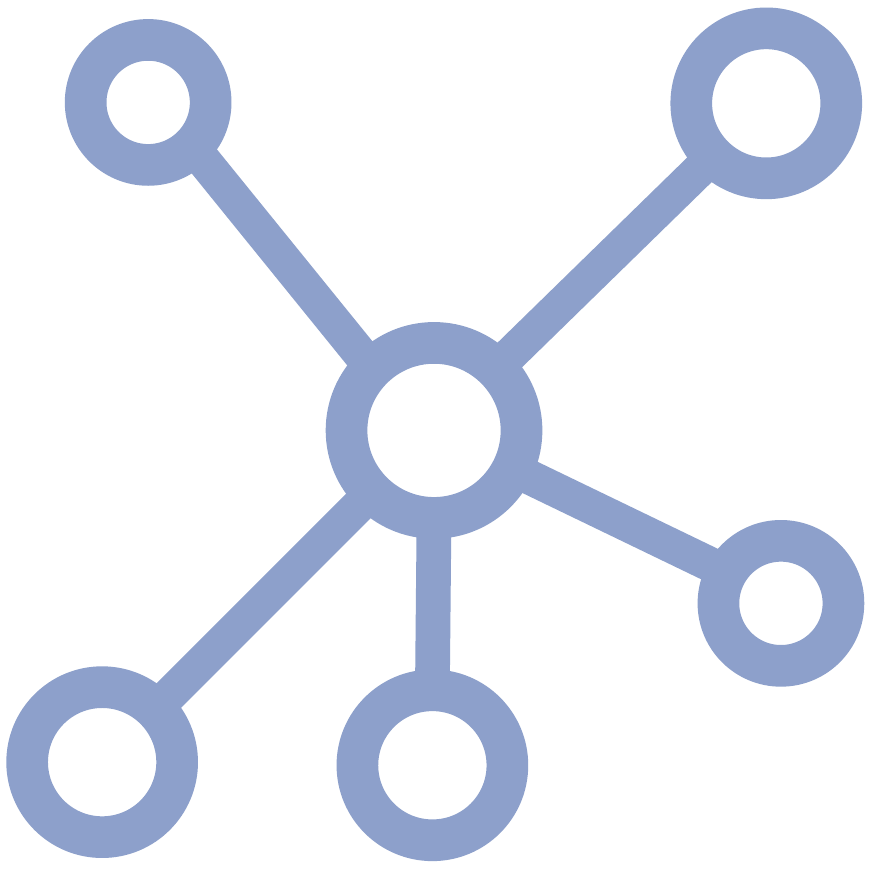}
        \end{minipage}
    \end{wrapfigure} 
    \noindent
    \para{Network} Complementary views are best and increased the accuracy of network exploration tasks. Chang et al.'s work compared matrix diagrams, node-link diagrams, and weighted networks to find effective matrix representations in side-by-side views for network exploration tasks on error, completion time, and user preference~\cite{chang2017evaluating}. Findings state node-link and matrix views are well suited for different visual tasks.
          Another study compared graphs either for different tasks or different datasets to measure their effectiveness. Graph edge attributes with uncertainty were visualized using two separate visual variables. For the task of comparing two graphs on overall strength or certainty showed that lightness was an effective mechanism for uncertainty~\cite{guo2015representing}.

     Readability is another network feature evaluated in a comparison study between node-link diagrams and their matrix-based representations on generic low-level tasks~\cite{ghoniem2004comparison}. They found readability depended upon graph detail, familiarity, graph meaning, and the layout used to visualize them, but the findings also reported that it deteriorated when the size of graphs and link density increased.
     In a recent work on the perception of graph properties, three layout algorithms---{force-directed, multidimensional scaling, and circular}---were modeled and compared using Weber's law to discriminate between graphs~\cite{soni2018perception}. The results showed that Weber's law could be used to model density perception.

     \subsubsection{Summary} For the comparison task, we observed a relatively fair distribution of studies on bar charts, line charts, and scatterplots. Ultimately, the choice of visualization is more subtask-driven and design-dependent.

\section{Discussion}
\label{sec:discussion}

    We have presented a systematic review of research in the perception of visualizations. One important observation from our survey is that much of the research has been pursued through the lens of low-level tasks in terms of efficiency and effectiveness. Early works were broad in nature. For example, Cleveland and McGill's early work demonstrated how the many different types of visual encodings influence perceptual judgments~\cite{cleveland1984graphical}.

\subsection{Visualization Effectiveness Is Task-Dependent} 
    
    One of the important conclusions we have seen time and again in prior work is that low-level task effectiveness varies with the dataset at hand, the visualization used, and specific design variations with the visualization.  While some visualization types tended to perform better than others on average, it seems, from our observation, there is no single visualization that is suitable for all situations. Saket et al.\ seemed to agree in their study of five visualizations on small datasets using ten low-level tasks~\cite{saket2018task}, when they stated that ``No One Size Fits All.'' In other words, depending on the task at hand, various visualizations perform can perform better or worse. In a similar vein, even within a single visualization type, design variations can have a serious effect on performance. Mylavarapu et al.'s study on ranked-list visualizations, which included wrapped bars, packed bars, piled bars, and Zvinca plots, quantified the differences and trade-offs for three tasks. The effectiveness of the representations varied, as each had its own strengths and weaknesses that depended upon the task, data, and user~\cite{mylavarapu2019ranked}.

\subsection{Progress Understanding Graphical Perception}

    Investigating and evaluating the effectiveness of visualizations to optimize their visual design is a perennial topic.  We witnessed the continuously evolving nature of perceptual research, with the majority focusing on visual task judgment. The recent upward trend in perception-based visualization studies, as noted in \autoref{fig:paperCountByyear} and supported by two recent related STAR reports by Borgo et al.~\cite{borgo2018information} and Behrisch et al.~\cite{behrisch2018quality}, show the maturing of this area of research. Many of the works discussed in this report have applied perceptual laws to evaluate visualizations, hopefully leading to better visual design.

    Much of our knowledge about perception in visualization is taken for granted, and despite the diverse set of perceptual research in visualization, many topics we have presumed as ``fact'' have never been sufficiently studied. Take, e.g., parallel coordinates---despite their reputation, studies have shown that parallel coordinates may not be as difficult to use as we think~\cite{siirtola2009visual}. The point here is not entirely to discard the facts but instead to consider that re-evaluating what we know might lead to better design guidelines. This align with some of the work of Kosara~\cite{kosara2016empire}, where the effort was to tease apart what we know and what we only think we know, using examples. The goal is to point out specific gaps in our knowledge and to assist researchers in starting investigations of the underlying theories to systematically build up a better foundation for our field.

    As the research objectives and methodologies have evolved, insights have become more fine-grained and nuanced, e.g., Chung et al.\ observed that color saturation with size could be used as an ordering variable~\cite{chung2016ordered}. Similarly, Szafir's work measured perceptual judgments on color difference that varied with size encodings, e.g., bar-width, circle-radius, line-width~\cite{szafir2018modeling}. These innovations, while enlightening, also stand somewhat in contrast to applied works, e.g., Colin Ware's book~\cite{ware2012information} on visualization design and perception, which provides a practical view of effective design.

\begin{table*}
    \centering

    \caption{Table summarizing the number of studies we reviewed per task, visual encoding, and visualization type. The table reveals which area have received the most attention and those that need more work.}
    \label{fig:synthesis}    
    
    \includegraphics[width=0.915\linewidth]{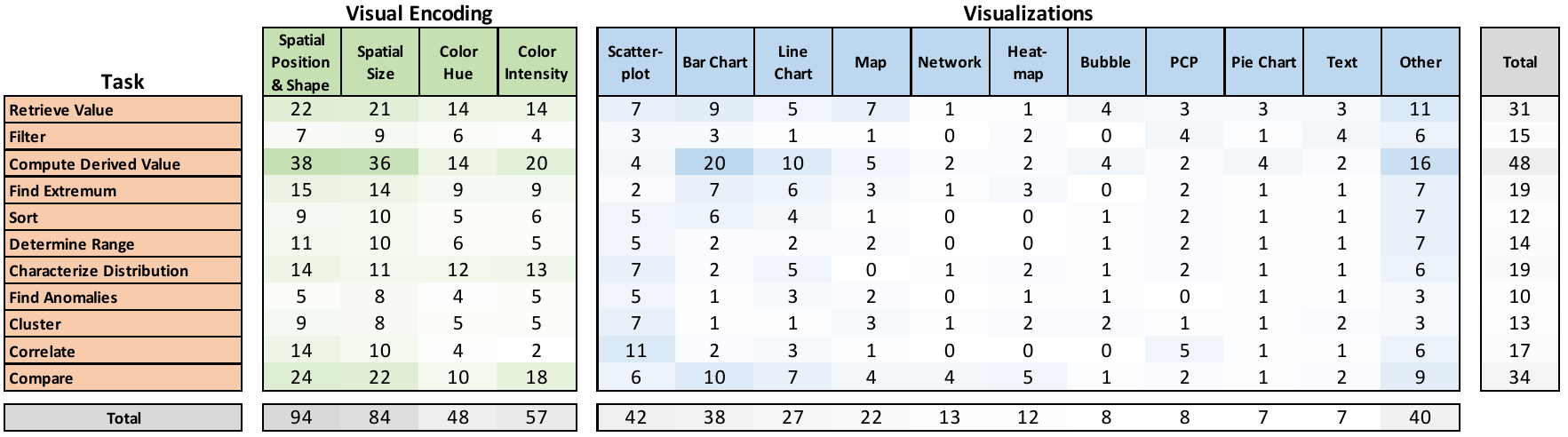}
    
\end{table*}

\subsection{Limitations of Scope and Reproducibility}

    As findings become more nuanced, so too does the scope of the findings. Most individual studies come with limitations in scope, complexity, objective-goal, and sometimes with datasets or demographics. For example, studies that are performed using limited and less diversified subjects, e.g., on computer science students, need to be further investigated before their guidelines could be put into practical use for a broader population. Readers and practitioners need to pay close attention to the limitation studies, as they are critical in effectively applying their conclusions.

    An integral part of understanding the limited scope of studies is considering the issue of reproducibility in studies. Reproducibility is being increasingly encouraged by communities of science to validate conclusions and to extend, re-evaluate, or specialize the original ideas~\cite{quadri2019you,hornbaek2014once}. With a few exceptions, e.g.,~\cite{heer2010crowdsourcing,kay2016beyond}, there has not been significant consideration of replication within the perception studies we have surveyed. Several fields of science are experiencing a ``replication crisis'' that has negatively impacted their credibility. The problem is further exacerbated by the limited scope of visualization studies, specificity of experiments, and difficult logistics required for reproducing a study.

    The lack of sufficient incentives, e.g., generating highly cited publications at high-quality venues, for reproducing studies also stands in the way. Recent suggestions for mitigating this problem revolve around tightly coupling original research with replication studies~\cite{quadri2019you,hornbaek2014once}, though this approach does not address the lack of incentives. One recent step in the right direction has been the trend of encouraging authors to make their research data available. A great set of recommendations are provided for open research practices in~\cite{wacharamanotham2020transparency}.

\subsection{Limitations and Open Questions}

    Throughout this survey, we uncovered various open challenges and areas requiring more study.

    \para{Beyond Scatterplots, Bar Charts, and Line Charts} Many research studies in visualization are driven by popularity, familiarity, and applications. \autoref{fig:synthesis} provides an overview to what extent the various areas have been studied in our survey of papers. Clearly, scatterplots, bar charts, and line charts received the majority of attention in research papers. It is undeniable that bar charts and line charts are widely used to represent univariate data, while scatterplots are used to represent bivariate relationships effectively. Parallel coordinates are an interesting case---even though there has been extensive design-based research work on parallel coordinates, the number of perceptual evaluations is low. One reason may be that it is complex and challenging to perform the perceptual evaluation. Similarly, we observe a bias in \autoref{fig:synthesis} to the tasks of Retrieve Value, Compute Derived Value, and Compare. While it is unlikely that the distribution of studies will balance out, insights gained from some of the less frequently studied techniques can have significant value. Researchers should consider focusing their new research efforts on the less studied tasks and visualization types.

    \para{Limitation of Experiment Data Collection} One topic we did not address in any detail is the collection of human subject data used in these studies. Variations in data collection make meta analyses difficult and can harm reproducibility by fixating on limited subject pools or specific experimental setups. The evaluation and perceptual studies we identified most often used metrics such as accuracy, time to completion, and subjective preference. Data collection methods range from keyboard input to mouse input to eye-tracking (e.g., scanpath and fixation) to voice. The experimental strategies also varied from methods of adjustment or interaction, forced-choice, and think-aloud protocols, which can provide more cognitive and problem-solving insights. Many of the experiments also suffer from low ecological validity, which refers to how closely the experimental setting matches the real-world setting in which the results might be applied~\cite{loomis1999immersive}. There is an explicit trade-off between experimental control and ecological validity~\cite{carpendale2008evaluating}. Studies with high ecological validity closely reflect real-world use, while those with low ecological validity are often highly controlled. Finally, we found variation in the participant pools used, generally consisting of university students or random subjects on crowdsourcing platforms. All of these variations in metrics, apparatus, and subject pools limit the applicability studies. For example, familiarity with the visualization, task, datasets, and design can be extremely biasing to participants' performance. In the same regard, other biases, such as affective priming, can influence perceptual judgment.  These limitations require serious consideration when applying conclusions to designs.

    \para{Relationship to Cognitive Processes} Perception and cognition are tightly linked and difficult to separate. Almost every study we discussed had some cognitive component to it. For example, estimating correlation is both perceptual and a cognitive task. In terms of the efficacy of visualization, cognition plays as important of a role as perception does. For example, one of the metrics on task performance, which was largely ignored or at least largely unreported in experiments, was confidence. Based on the complexity of the task, e.g., interaction versus forced-choice, visualization type familiarity, e.g., parallel coordinates, comfort with the type or size of data, or the type of experimental study being performed, subject confidence can vary widely. There is even a possibility that the user's confidence will vary task-by-task. A related challenge is finding the right set of subjects, which is a particular limitation when recruiting participants through crowdsourcing.  Due to the population diversity, their experience and confidence vary, and it is difficult to check quality. It fits then for researchers to evaluate the participants' level of confidence in task judgment, and accordingly, data quality should be evaluated. New studies should consider the participant's level of confidence and how this affects high- and low-level tasks.

\subsection{Conclusion}  

    We have presented our report with a particular focus on the links between visualization types, visual encodings, and tasks. Through our taxonomy, we wanted to emphasize perception-based research findings and their impact on visual encoding and visualization choices. We believe this report will be a valuable starting point for those designing visualizations and researchers looking to advance the state-of-the-art in perception-based visualization research. It should be noted, however, that the aim of this survey was not to directly summarize the visualization recommendations but instead to provide a broad understanding of the topics that have been evaluated. Furthermore, all of the perceptual studies were studied under limited conditions and come with caveats that we could not enumerate.


\section{Appendix: Perception Fundamentals}
\label{sec.background.fundamentals}

    We provide a brief introduction to some fundamental concepts of perception discussed in the paper. The presented information is concise, and we encourage readers to refer to the related articles for more in depth information.

\subsection{Psychophysical Effects}

    \textbf{Psychophysics} is a set of methods relating sensations to the characteristics of a stimulus. It is used to quantitatively investigate relationships between physical stimuli and the sensation of the perception they produce~\cite{gelfand2017hearing, gescheider2013psychophysics}. The transition of psychology from a philosophical to a scientific discipline was facilitated when G.T.\ Fechner introduced techniques to measure mental events. The attempt to measure sensations through the use of Fechner's procedures was termed psychophysics and primarily investigated the relationships between sensations in the psychological domain and stimuli in the physical domain. Central to psychophysics is the concept of a sensory threshold, that measurement can have a differential and an absolute sensitivity. The \textit{absolute threshold} or \textit{stimulus threshold} is defined as the smallest amount of stimulus energy necessary to produce a sensation. The \textit{differential threshold} was defined as the amount of change in a stimulus required to produce a just noticeable difference (JND) in the sensation.

    \textbf{Weber's law}, or Weber-Fechner's law, relates to {psychophysics} and is used to determine the relationship between the perceived change in a stimulus and the actual change, which has been used to modeling how humans perceive certain features in a visualization. The law states that the change in a stimulus that will be just noticeable is a constant ratio of the original stimulus~\cite{weber1834pulsu, weber1996eh}.

    As an example of Weber's law, consider the act of lifting a 5 kg weight. Adding a small amount of weight, say 0.1 kg, may not make the weight \textit{feel} any heavier. With further additions of weight, the difference will eventually be noticeable. Weber's law is the ratio of change in the stimulus ($\Delta I=0.1$ kg) to the stimulus magnitude ($I=5$ kg), which is $0.02$. Weber's law has been shown to hold for weight discrimination, visual discrimination, and tone discrimination\cite{weberweber}.

    \textbf{Just-Noticeable Difference (JND)}, also known as the difference threshold, is the minimum level of stimulation that a person can detect, usually $50\%$ of the time, though other ratios can be used~\cite{coren1999sensation}. For example, one is asked to hold two objects of different weights---the JND would be the minimum weight difference between the two that one could sense half of the time. JND is used as a component for the perceptual studies, which are required to determine how much a given stimulus must be regulated in order for a human to detect a change reliably. The relation between JND and the stimulus can be represented using Weber's law~\cite{weber1834pulsu,weber1996eh} as follows: $dP= k \frac{dI}{I}$, where, $dP$ is the differential change in perception; $k$ is the Weber fraction; $dI$ is the differential change in stimulus, and $I$ is the actual intensity of the stimulus. With the given $I$ and Weber fraction, JND corresponds to the minimum change of stimulus will produce a noticeable difference in the perception. The application of JNDs with psychophysics evaluation is useful for measuring human judgments in the effectiveness of visual encodings for different tasks or design improvements.

\subsection{Bias and Effect}

    In the process of evaluating visualizations, the effect of bias, whether cognitive or perceptual, is critical to understanding and evaluating experimental results. The primary types of bias observed and studied in visualization are perceptual biases and cognitive biases.

\subsubsection{Perceptual Bias}
 
    \textbf{Perceptual biases} are systematic errors that occur at the perceptual level in perceiving visualization and/or related tasks. Some types of perceptual biases studied include clustering illusions and priming biases~\cite{valdez2017framework}. For example, the clustering illusion is where people underestimate the variance seen in patterns in a small set of random data~\cite{morgan1990biases, seizova2006biases}. Priming relates to associative memory theories where a concept/effect is quickly activated after a similar concept/effect has been activated.

    Perceptual bias has been studied in several regards. Change blindness refers to the inability of humans to recognize large visual changes between images. An optimization-based method introduced and evaluated an approach to generate ``spot-the-difference'' alternatives~\cite{ma13change}. Perceptual biases have been studied in virtual reality platforms, e.g., in the perception of size and stiffness of virtual objects~\cite{wu1999visual} and how photorealism negatively affects our perception of virtual characters~\cite{zibrek2018effect}. Irregularities in data can cause bias, influencing a user's response to the conclusion of analysis~\cite{song2018s, correll2018looks}. One study found that perceptual biases influence a user's awareness of uncertainties, further influencing the user's trust building~\cite{sacha2016role}. Finally, researchers have investigated how bias and perception intersect to create deceptive views. Pandey et al.\ developed a method to quantify and compare the exaggeration caused by misleading representations~\cite{pandey2015deceptive}.

\subsubsection{Cognitive Bias}

    \textbf{Cognitive bias} research has grown considerably at both the cognitive science level and specifically for visual analytic and decision-making tools. Cognitive biases differ from perceptual biases in that they persist even if the information has been correctly processed at a perceptual level. Broadly, collective works on cognitive biases can be found in~\cite{ellis2018so}. The following are examples of cognitive bias topics that have been evaluated in visualization.

\begin{figure}[!b]
    \centering
    \includegraphics[ width=0.8\linewidth]{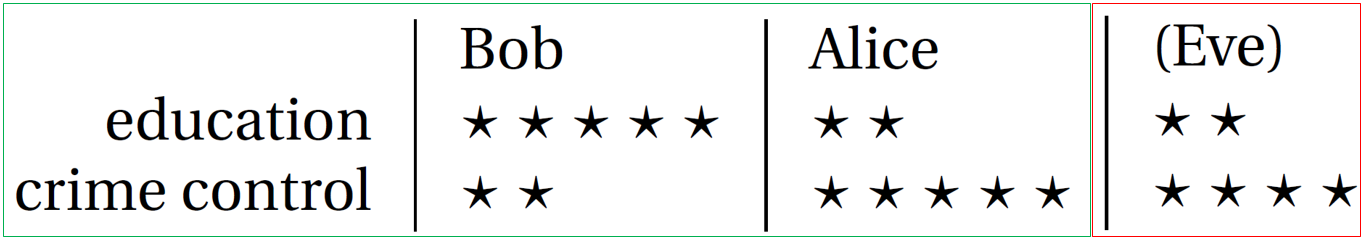}
    \caption{Example of attraction effect in selecting hypothetical election candidates. Image reproduced with permission~\cite{dimara2017attraction}.}
    \label{fig:attraction}
\end{figure}

    \textbf{Attraction effect} is a type of cognitive bias where the presence of irrelevant alternatives influences the choice between two options. When the user's choice between two options is influenced by the presence of an irrelevant (dominated) third alternative. A good example, coming from~\cite{sue1995attraction}, involves selecting between two candidates based on a rating of their education and crime control plans (see \autoref{fig:attraction}). Bob has a solid education plan, while Alice’s strategy for crime control is excellent. The choice is difficult if we consider both the criteria. A third candidate, Eve, called the ``decoy,'' is focused more on crime control than education, though not as good as Alice. The introduction of Eve as alternative biases our preference towards Alice. Awareness of the attraction effect is important, as it may introduce skew into decision-making tasks, as evidenced by Dimara et al.'s study on scatterplots~\cite{dimara2017attraction}.

    \textbf{Anchoring effect} is a type of cognitive bias, where a stimulus might influence human judgment at the perceptual level of the decision-making process. Anchoring effects and ordering effects describe how the order in which information is presented can affect the perceived size of an effect, with subjects across a wide range of domains tending to assign more rhetorical weight to evidence that comes near the beginning of a sequence.

    \textbf{Priming} is another form of stimulus that can influence human judgment. The priming effect is a phenomenon in which an alternative perceptual stimulus influences human responses~\cite{meyer1971facilitation}. Priming effects are seen more frequently than anchoring effects in separability judgment. 
 
    \textbf{Emotions}, in psychology, are defined by two dimensions: \textit{valence}, positive or negative feelings, and \textit{arousal}, the intensity of the feelings. \textbf{Affective priming} is the technique of inducing emotion, also known as \textit{affect}, in a user to study the impact on cognitive tasks. It involves manipulating valence and/or arousal via emotional stimuli. Harrison et al.'s contribution to affective priming suggested that it can influence accuracy in graphical perception tasks~\cite{harrison2012exploring, harrison2013influencing}. The crowdsourced study indicated that affective priming significantly influenced visual judgment, while positive priming improved accuracy.

    The cognitive biases of both anchoring and priming suggest that the decision-making process not only depends on what the visual features currently look like but also on the previous frame of reference.

\ifCLASSOPTIONcompsoc
  \section*{Acknowledgments}
\else
  \section*{Acknowledgment}
\fi

We thank our reviewers for their helpful feedback on the construction of this paper. The project is supported in part by the National Science Foundation (IIS-1845204).

\ifCLASSOPTIONcaptionsoff
  \newpage
\fi



\bibliographystyle{IEEEtran}
\bibliography{main.bib}
%

%

\begin{IEEEbiography}[{\includegraphics[width=1in,height=1.25in,clip,keepaspectratio]{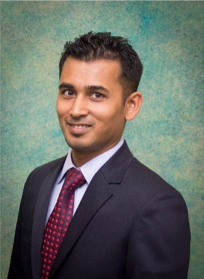}}]{Ghulam Jilani Quadri}
is a fourth-year Ph.D. student at the University of South Florida in the Department of Computer Science and Engineering under the supervision of Dr. Paul Rosen. Before joining the University of South Florida, Ghulam worked for Infosys Limited as System Engineer in Pune. Ghulam and team participated in IEEE VIS 2017 VAST Challenge and awarded Honorable mention. Ghulam received a 2021 Computing Innovation Fellow award. His current research interests are Information Visualization and HCI.
\end{IEEEbiography}

\begin{IEEEbiography}[{\includegraphics[width=1in,height=1.25in,clip,keepaspectratio]{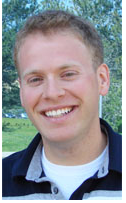}}]{Paul Rosen}
is an Associate Professor at the University of South Florida. He received his Ph.D.\ from the Computer Science Department of Purdue University. His research interests include applying geometry- and topology-based approaches to problems in information visualization. Along with his collaborators, he has received awards for best paper at PacificVis 2016, IVAPP 2016, PacificVis 2014, and SIBGRAPI 2013.  Dr.\ Rosen received a National Science Foundation CAREER Award in 2019.
\end{IEEEbiography}




\end{document}